\documentclass[twoside,11pt]{article}
\usepackage[pdftex]{graphicx}
\usepackage{dcolumn}  
\usepackage{bm}       
\usepackage{color}
\usepackage{amsmath,epsfig,cite}
\usepackage{subfigure}
\usepackage{fullpage}
\begin{document}
\title{Hybrid Mesons}
\author{C.\ A.\ Meyer\,$^{1}$  and E.\ S.\ Swanson\,$^{2}$ \\
$^{1}$\,Carnegie Mellon University, Pittsburgh, PA 15213 USA \\
$^{2}$\,University of Pittsburgh, Pittsburgh, PA 15260 USA}
\maketitle
\begin{abstract}
A review of the theoretical and experimental status of hybrid hadrons is presented. The states $\pi_1(1400)$, $\pi_1(1600)$, and $\pi_1(2015)$ are thoroughly reviewed, along with experimental results from GAMS, VES, Obelix, COMPASS, KEK, CLEO, Crystal Barrel, CLAS, and BNL. Theoretical lattice results on the gluelump spectrum, adiabatic potentials, heavy and light hybrids, and transition matrix elements are discussed. These are compared with bag, string, flux tube, and constituent gluon models. Strong and electromagnetic decay models are described and compared to lattice gauge theory results. We conclude that while good evidence for the existence of a light isovector exotic meson exists, its confirmation as a hybrid meson awaits discovery of its iso-partners. We also conclude that lattice gauge theory rules out a number of hybrid models and
provides a reference to judge the success of others.
\end{abstract}
\clearpage
\tableofcontents
\clearpage

\section{Introduction}

For forty years Quantum Chromodynamics (QCD) has served as one of the pillars of the Standard Model. 
Early doubts about the existence of quarks were dispelled by the discovery of the $J/\psi$, 
while---following a suggestion due to Ellis, Gaillard, and Ross --  gluons were confirmed by the 
discovery of three jet events at DESY in 1979\cite{gluon}.

Although gluons are now firmly established as the carriers of the strong force, their nonperturbative behavior remains enigmatic. This unfortunate circumstance is chiefly due to two features of QCD: the theory is notoriously difficult to work with in the nonperturbative regime, and experimental manifestations of glue tend to be hidden in the spectrum and dynamics of hadrons.  Hadrons that carry valence quark and gluonic degrees of freedom are one such experimental manifestation that has been postulated since the early days of QCD\cite{history}. These states are called {\it hybrids} and a review of their experimental and theoretical status is the purpose of this article.  

That discovering explicit nonperturbative glue is difficult can be gleaned from the history of the development the quark model and QCD. All the well-established mesons have $J^{PC}$ equal to $0^{-+}$, $0^{++}$, $1^{++}$, etc, which can be created with fermion-antifermion pairs in an given orbital momentum state. This strongly suggests that quarks are spin-1/2, while the spectrum ordering suggests that energy increases with orbital angular momentum. Furthermore, the
absence of mesons with isospin or strangeness greater than unity supports the idea that mesons are fermion-antifermion states. Thus the simple quark model of mesons (and baryons) was partly motivated by the {\it absence} of ``exotic" hadrons such as multiquark or gluonic states.  It is thus perhaps no surprise that QCD exotics have been difficult to observe.

The simplest explanation for this observation is that gluonic degrees of freedom are somewhat ``stiff" and therefore difficult to excite. Of course, with the increasing energy, luminosity, and capabilities of modern accelerators and detectors one might hope that this regime can be explored. The interesting question then is: can an analogous spectroscopy of glue be discerned in the data?

A basic question that the data and its interpretation must address is what is meant by the notion of valence glue.
As an example of the importance of choosing correct degrees of freedom, we
mention the simple problem of determining the
number of components of a constituent gluon. For example, one might suppose that 
a massive
constituent gluon (of mass $m_g$) should be transverse so as to maintain consistency with
Yang's theorem. However, this is inconsistent
with the requirements of Lorentz invariance.
Thus, for example, $J = 1$ glueballs are
expected to exist and lattice calculations indicate that they are quite heavy
(roughly 3 GeV)\cite{lattgb}.  Such a state may not be constructed from two
transverse constituent gluons (due to Yang's theorem) and therefore may be expected to
have a mass of roughly $3 m_g$. However massive vector gluons have
no such constraint and one therefore expects them to have a mass of
approximately $2 m_g$. Clearly the assumed active degrees of freedom play a large role in determining the gross features of the spectrum.

There are two broad ideas concerning soft glue: it is some sort of string or flux 
tube or it is an effective constituent confined by a bag or potential. In different language, nonperturbative glue can be thought of as collective, nonlocal degrees of freedom, or as a local quasiparticle degree of freedom. Lattice gauge  theory and experiment are the two main methods which permit resolving this longstanding issue. Of course the primary task is to experimentally verify that such excitations of QCD exist. Then a thorough investigation of the hybrid spectrum and its production and decay mechanisms  must ensue; this must be compared to lattice gauge and model computations, and a qualitative picture must be built before we can claim to fully understand the enigmatic gluonic sector of the Standard Model.

We will start with a review of properties of hybrids and related gluonic systems as determined by lattice gauge computations. This will serve to set terminology and establish a base to which models can be compared. This is followed by brief summaries of the ideas and main results in string-based models, bag models, and constituent glue models. We then summarize the current status of the experimental search for hybrid mesons. Finally, future experimental prospects and an outlook on what is required and expected from theory is presented.

Before starting, it will be useful to note that valence gluonic degrees of freedom increase the
 quantum numbers that are available to fermion-antifermion systems. The parity and charge conjugation that are available to $q\bar q$ systems are specified by $P = (-)^{L+1}$ and $C=(-)^{L+S}$ where $L$ and $S$ are the $q\bar q$ total angular momentum and spin. This means that $J^{PC}= 
0^{--}$, {\tt odd}$^{-+}$, and  {\tt even}$^{+-}$ are not available to simple quark-antiquark systems. Hadrons with these quantum numbers are called ``(quantum number) exotic". A variety of notations have developed over the years for hybrids. Terminology appears to have settled on using standard PDG notation to describe the flavor of a hybrid, a subscript to denote the total spin, with parity and charge conjugation following that of the named hadron. Thus, for example, a $J^{PC} = 1^{-+}$ isovector hybrid would be denoted $\pi_1$.

\section{Lattice Gauge Theory}
\label{sec:latt}

The use of a lattice regulator for Euclidean field theories permits the numerical evaluation of path integrals. Thus all correlation functions can, in principle, be computed under controlled approximations. Historically, this brilliant promise has been compromised by various technical issues including large quark masses, the difficulty in computing ``hairpin" fermion lines, difficulty in computing the excited spectrum, and most prominently, difficulty in computing the determinant of the Dirac operator. The latter issue was handled by ignoring it (this is called the ``quenched approximation"; it is, in fact, not an approximation, but a truncation that renders a quantum field theory inconsistent). The root problem of all of these issues is that they introduce excessive noise into observables -- however, to some extent all of them have been overcome in the past decade. Thus modern lattice computations are made on large lattices, are not quenched (they include the effects of virtual quarks), incorporate large bases of interpolating fields (which permits the extraction of excited states), and employ sophisticated stochastic methods to compute hairpin fermion propagators. The remaining issues are chiefly quark masses that tend to be too large and the necessity of incorporating continuum interpolating operators. Both of these are rapidly becoming things of the past.

\subsection{Adiabatic Gluonic Surfaces}
\label{ad}

Lattice investigations of soft gluonic matter date to an investigation of the adiabatic gluonic spectrum by Griffiths, Michael, and Rakow \cite{GMR}.
The idea is that gluons are a ``fast" degree of freedom with respect to heavy quarks. In the static limit the quark and antiquark serve as a color source and sink at a distance $R$ and the gluonic field arranges itself into configurations described by various quantum numbers. These quantum numbers match those of diatomic molecules: the projection of the gluonic angular momentum onto the $q\bar q$ axis (denoted $\Lambda$), the product of gluonic parity and charge conjugation $\eta \equiv (PC)_g$, and $Y$-parity, which represents the reflection of the system through a plane containing the $q\bar q$ axis. The notation

$$
\Lambda^Y_\eta
$$
will be used in the following, where $\Lambda= 0,1,2, \ldots$ is denoted $\Sigma$, $\Pi$, $\Delta$, $\ldots$; $\eta$ is labelled $u$ (negative) or $g$ (positive); and $Y=\pm$. For $\Lambda \neq 0$ the two $Y$-parity eigenstates are degenerate and the $Y$ quantum number is not given.

Griffiths {\it et al.} sought to compute the adiabatic configuration energies,
$E(\Lambda^Y_\eta; R)$ as a function of the distance between the color source and sink. This computation was made in SU(2) quenched gauge theory on an $8^4$ lattice and rough results for the lowest two surfaces were obtained.

\begin{figure}[ht]\centering
\includegraphics[angle=0,width=8cm]{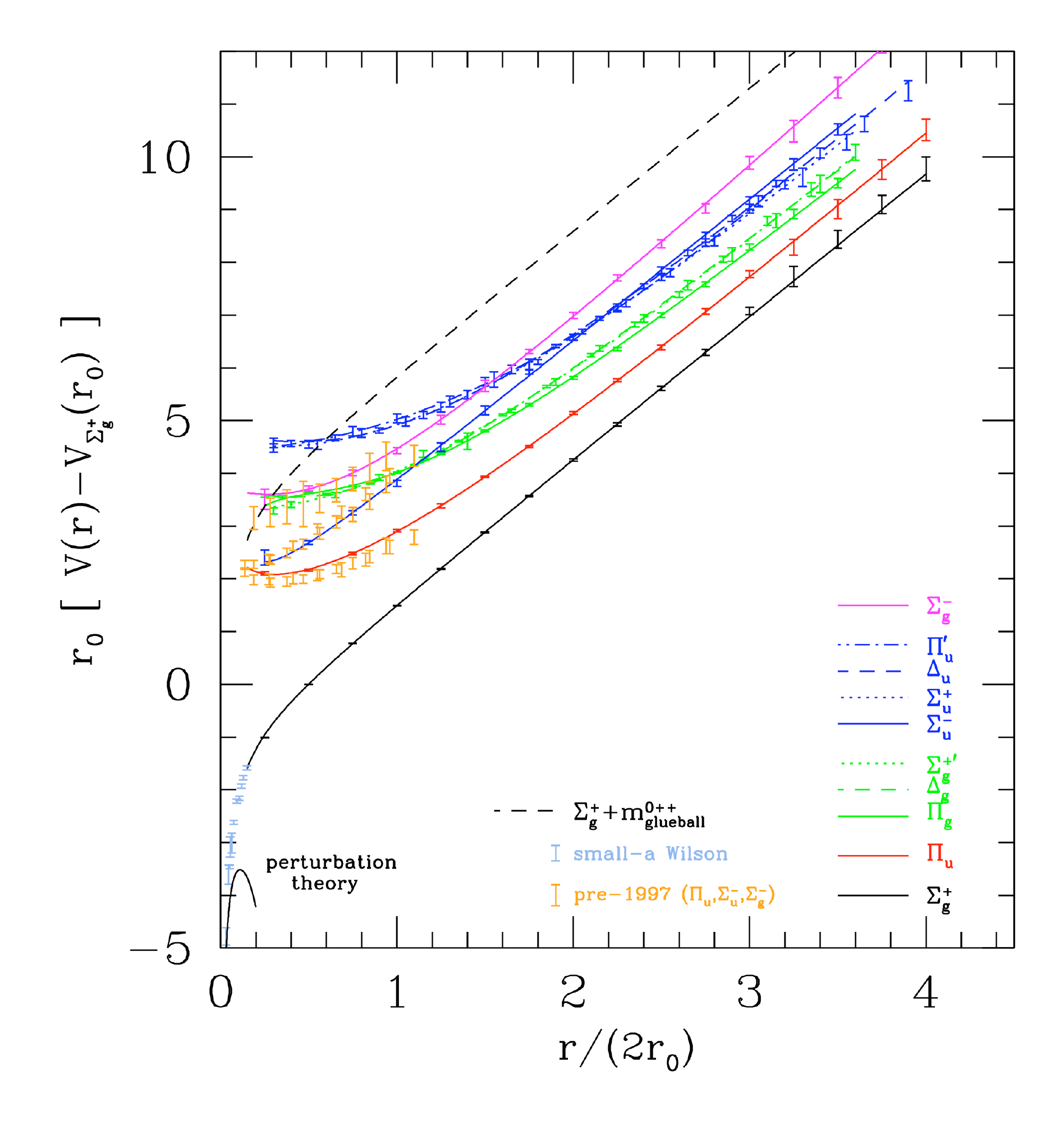}
\caption{Lattice Adiabatic Hybrid Potentials. Figure curtesy of C.J. Morningstar\cite{cjm}.}
\label{VRplot}
\end{figure}

Subsequent work was carried out by the UKQCD collaboration \cite{ukqcd_1,ukqcd_2}
and Juge, Kuti, and Morningstar\cite{JKM1}. Juge {\it et al.} performed a large scale computation with many basis states and were able to obtain many excited states out to large distance. The results are presented in Fig. \ref{VRplot}; here $r_0$ is approximately 0.5 fm and the potentials have been renormalized by subtracting the value of $E(\Sigma_g^+)$ at $r_0$.

An interesting feature of the spectrum is that the adiabatic surfaces tend to finite values at small interquark separation. This may be a surprise because the expected perturbative behavior at small $R$ for a pair of quarks in a color octet is

\begin{equation}
E(\Lambda^Y_\eta;R) \sim \frac{1}{6}\frac{\alpha_s}{R}.
\end{equation}
Alternatively, quarks in a color singlet should have an energy 

$$
E \sim -\frac{4}{3}\frac{\alpha_s}{R}
$$
at very small distances. The figure shows that this distance scale  is evidently smaller than 0.1 fm. 

The actual behavior of the adiabatic potentials at small distance is problematic: lattice gauge computations diverge at the origin and must be regulated in some way. Furthermore, level crossing is expected to occur; namely, it will become energetically favorable for a configuration to emit a scalar glueball and convert to a $q\bar q$ color singlet state at some distance. This surface and the level crossings are indicated by the dashed line in the figure. Of course the quantum numbers must agree for the transition to occur and the mixing matrix element must be large enough to resolve it at the temporal extent used in the lattice computation.

The large distance behavior of the adiabatic surfaces is of interest because it is widely held that this behavior is governed by string dynamics\cite{string_1,string_2}. The simplest geometric string model is described by the Nambu-Goto action

\begin{equation}
\sigma \sqrt{1 + \partial_\mu \vec \xi \cdot \partial^\mu \vec \xi}
\end{equation}
where $\vec \xi$ is a massless vector field with two transverse components. In $D$ dimensions this gives rise to a spectrum (computed with fixed ends) of\cite{ng} (the string tension is denoted $\sigma$)\footnote{Note that $D=26$ is required for consistent quantization, but this problem is not relevant for large $R$.}

\begin{equation}
E(N;R) = \sigma R\, \left(1 - \frac{D-2}{12\sigma R^2} \pi + \frac{2\pi N}{\sigma R^2}\right)^{1/2}
\end{equation}

The fact that this formula works quite well for the ground state surface for $R> 0.5$ fm has been traditionally taken as an indication for the robustness of the string description of gluodynamics\cite{string_1,string_2}. 

At large distances the Nambu-Goto action predicts splittings between adiabatic surfaces that behave as $\pi/R$. Detailed comparison to large distance lattice results by Juge {\it et al.} reveal that the expected universal behavior is achieved, but only for quark separations greater than 2 fm; furthermore residual deviations from string behavior indicate the presence of fine structure in the interaction, so that an effective string interaction is more appropriate for the description of gluonic excitations at large distances\cite{JKM-3}.


\subsection{Gluelumps}
\label{sec:gl}

A ``gluelump" is a hadronic state comprised of a static source in the octet representation with accompanying glue, such that the resulting state is a singlet under gauge transformations. Investigation of gluelumps was motivated by interest in the properties of bound states of massive gluinos and gluons\cite{gluino}. The initial lattice investigation was made by Michael in 1985 in quenched SU(2) gauge theory\cite{mm,fm}.

We note that the gluelump spectrum can be determined up to an infinite self energy due to the static adjoint gluon. Thus energy differences can be unambiguously extracted. Absolute energies can also be obtained under specified renormalization schemes. Here we present the results of Bali and Pineda\cite{bp}, which were obtained in quenched SU(3) gauge theory. Absolute gluelump energies were obtained by employing a renormalon subtraction scheme\cite{P} with a matching scale set to $2.5/r_0 \approx 1$ GeV. Results for the lowest lying eight levels are shown in Table \ref{tab1}. The authors of Ref. \cite{marsh} have examined a  large set of quantum numbers and non-octet sources, with results similar to those presented here.

\begin{table}[h]\centering
\begin{tabular}{l|l}
\hline\hline
$J^{PC}$ $^{\ }$ & mass (GeV) \\
\hline
$1^{+-}$ & 0.87(15) \\
$1^{--}$ & 1.25(16) \\
$2^{--}$ & 1.45(17) \\
$2^{+-}$ & 1.86(19) \\
$3^{+-}$ & 1.86(18) \\
$0^{++}$ & 1.98(18) \\
$4^{--}$ & 2.13(18) \\
$1^{-+}$ & 2.15(20) \\
\hline\hline
\end{tabular}
\caption{Low Lying Gluelump Spectrum\cite{bp}.}
\label{tab1}
\end{table}


HERE

\begin{figure}[h]
\centering
\includegraphics[angle=0,width=8cm]{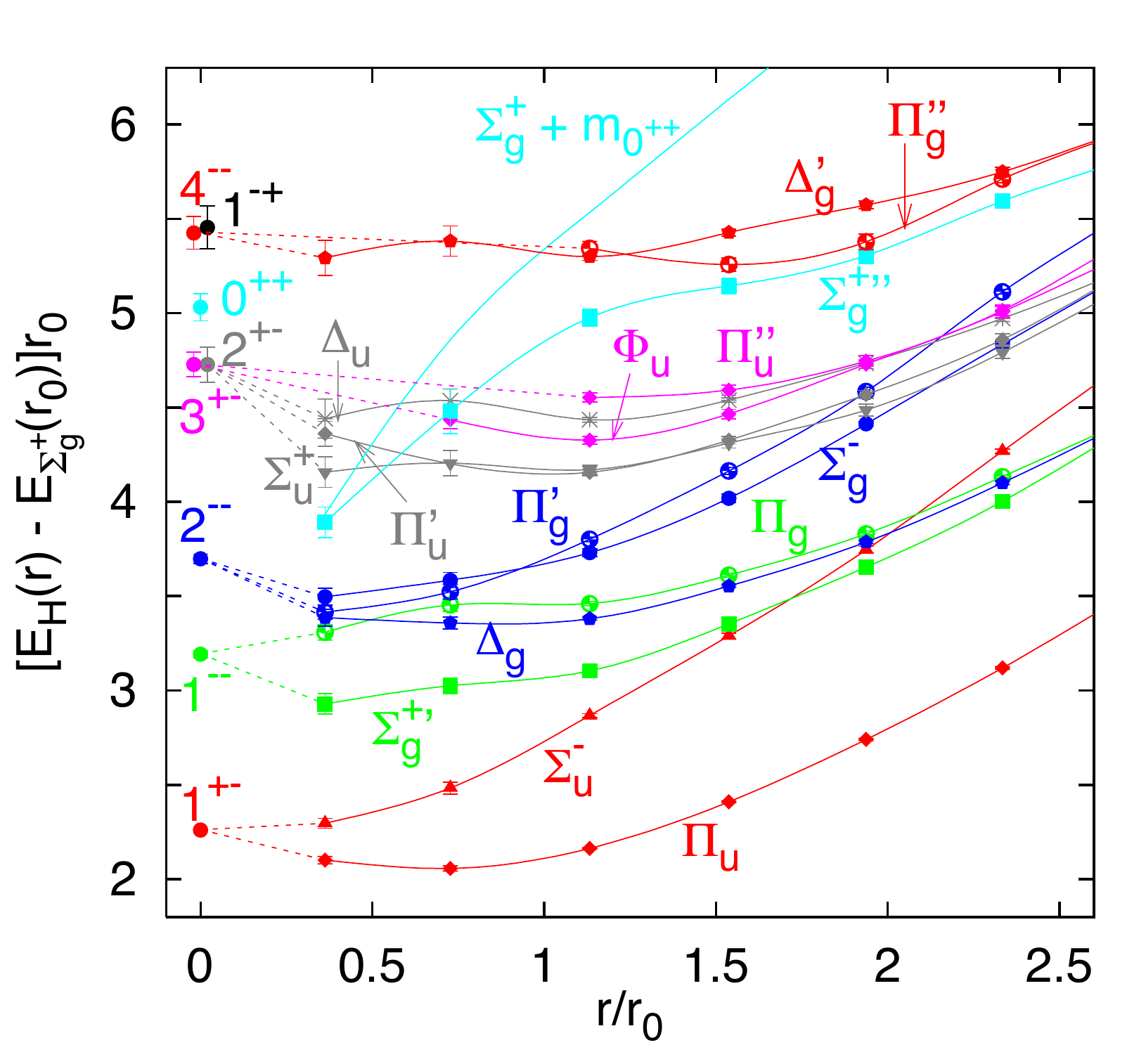}
\caption{Adiabatic Hybrid Potentials with Gluelump Spectrum. (Figure reproduced with
permission from reference \cite{bp}).}
\label{bgfig}
\end{figure}

Because the quark and antiquark can be regarded as merging into a color octet as the separation between them lessens, the spectrum of adiabatic hybrid surfaces is related to the gluelump spectrum as $R\to 0$ \cite{jm}. In this limit greater symmetry is obtained because the gluonic spin, $J_g$, becomes a good quantum number \cite{fm,b} and the gluelump spectrum can be mapped to the adiabatic surfaces as shown in Table \ref{tab2}. The match between the adiabatic and gluelump spectra is displayed in Fig. \ref{bgfig}, where reasonably good agreement is seen.
The exception is the ${\Sigma_g^+}''$, which appears headed towards the $0^{++}$ gluelump energy, but then deviates downwards. This is perhaps an  example of the adiabatic surface crossing mentioned above. In fact, the final point on the $\Sigma_g''$ surface appears to have landed on the surface given by $V(\Sigma_g^+;R) + m_{0^{++}}$, shown in the figure as a cyan line.

\begin{table}[ht]\centering
\begin{tabular}{ll}
\hline\hline
gluelump $J^{PC}$ & adiabatic surface quantum numbers \\
\hline
$1^{+-}$ & $\Sigma_u^-$, $\Pi_u$ \\
$1^{--}$ & $\Pi_g$, ${\Sigma_g^+}'$ \\
$2^{--}$ & $\Sigma_g^-$, $\Pi_g'$, $\Delta_g$ \\
$2^{+-}$ & $\Sigma_u^+$, $\Pi_u'$, $\Delta_u$ \\
$3^{+-}$ & ${\Sigma_u^-}'$, $\Pi_u''$, $\Delta_u'$i, $\Phi_u$ \\
$0^{++}$ & ${\Sigma_g^+}''$ \\
$4^{--}$ & ${\Sigma_g^-}'$, $\Pi_g''$, $\Delta_g'$i, $\Phi_g$, $\Gamma_g$ \\
$1^{-+}$ & ${\Sigma_u^+}'$, $\Pi_u'''$ \\
\hline\hline
\end{tabular}
\caption{Adiabatic Gluon Surface Degeneracies at Small Distance. The $2^{+-}$ and $3^{+-}$ identifications could be reversed.}
\label{tab2}
\end{table}

\subsection{Heavy Hybrids}
The simplest method to access hybrids with heavy quarks is to leverage the adiabatic gluonic potentials of section \ref{ad} by solving the non-relativistic Schr\"{o}dinger equation for quark motion on the relevant surface. This approach has been adopted by Juge, Kuti, and Morningstar\cite{JKM2}, who address a non relativistic Schr\"{o}dinger equation with a centrifugal factor given by

\begin{equation}
\langle L_{Q\bar Q}^2\rangle = L(L+1) - 2 \Lambda^2 + \langle J_g^2\rangle,
\label{eq:LQ}
\end{equation}
where  $L$ is the total angular momentum of the system. The result is a spin-averaged spectrum that depends on the adiabatic surface employed as a potential. For the $\Sigma_g^+$ surface one has $\langle J_g^2\rangle = 0$ and, since the $\Sigma_g^+$ surface mimics the phenomenological Cornell potential, the usual results for heavy quarkonia are recovered. For excited surfaces, $\Pi_u$ and $\Sigma_u^-$, one can approximate $\langle J_g^2\rangle =2$ since this corresponds to the minimum gluonic field angular momentum permitted in these configurations.

Meson quantum numbers are constructed in terms of the total quark spin $S$, the total angular momentum $L$, and the mesonic spin $\vec J=\vec L+\vec S$. In the leading Born-Oppenheimer approximation the eigenvalues $L(L+1)$ and $S(S+1)$ of $\vec L^2$ and $\vec S^2$ are good quantum numbers and the parity $P$ and charge conjugation $C$ of a meson is given in terms of $L$, $S$, and $\Lambda$  by\cite{JKM2}

\begin{equation}
P = \epsilon\, (-)^{L+\Lambda+1}
\end{equation}
 and

\begin{equation}
C=\epsilon \, \eta \, (-)^{L+\Lambda+S}.
\end{equation}
Here $L\geq \Lambda$  $\epsilon = \pm$ for $\Sigma^\pm$ and both signs apply for $\Lambda > 0$. Note that $\eta$ is the product of gluonic charge and parity quantum numbers defined above.  Low lying quantum numbers in the ground state $\Sigma_g^+$ state are thus as given in Table \ref{tab:qnn}.

\begin{table}[ht]\centering
\begin{tabular}{lll}
\hline\hline
$S$ & $L$ & $J^{PC}$ \\
\hline
0 & 0 & $0^{-+}$ \\
1 & 0 & $1^{--}$ \\
0 & 1 & $1^{+-}$ \\
1 & 1 & $(0,1,2)^{++}$ \\
\hline\hline
\end{tabular}
\caption{$\Sigma_g^+$ Meson Quantum Numbers.}
\label{tab:qnn}
\end{table}

\noindent
This is, of course, the usual pattern of the non-relativistic constituent quark model.

For the phenomenologically relevant $\Pi_u$ surface the low lying quantum numbers are (we take $\Lambda=1$, $\eta = -1$, $\epsilon=\pm$) given in Table \ref{tab:piuu}.

\begin{table}[ht]\centering
\begin{tabular}{lll}
\hline\hline
$S$ &  $L$ & $J^{PC}$ \\
\hline
0 & 1 & $1^{--},\  1^{++}$ \\
1 & 1 & $(0,1,2)^{-+},\ (0,1,2)^{+-}$ \\
0 & 2 & $2^{++},\ 2{--}$ \\
1 & 2 & $(1,2,3)^{+-},\ (1,2,3)^{-+}$ \\
\hline\hline
\end{tabular}
\caption{$\Pi_u$ Meson Quantum Numbers.}
\label{tab:piuu}
\end{table}

Because the spectrum only depends on the radial quantum number and $L$ at this order, the leading Born-Oppenheimer multiplets are $(0^{-+},1^{--})$, $(1^{+-},(0,1,2)^{++})$, etc, for the $\Sigma_g^+$ surface and 

\begin{equation}
1^{--},\ (0,1,2)^{-+};\ 1^{++},\ (0,1,2)^{+-}
\end{equation}
for the lowest states on the $\Pi_u$ surface. We remark that these multiplets do \emph{not} agree with those observed in explicit lattice hybrid computations. Rather, the low lying multiplet contains $J^{PC} = 1^{--}$, $(0,1,2)^{-+}$, which corresponds to $\epsilon = +1$.

\begin{figure}[h]
\centering
\includegraphics[angle=0,width=8cm]{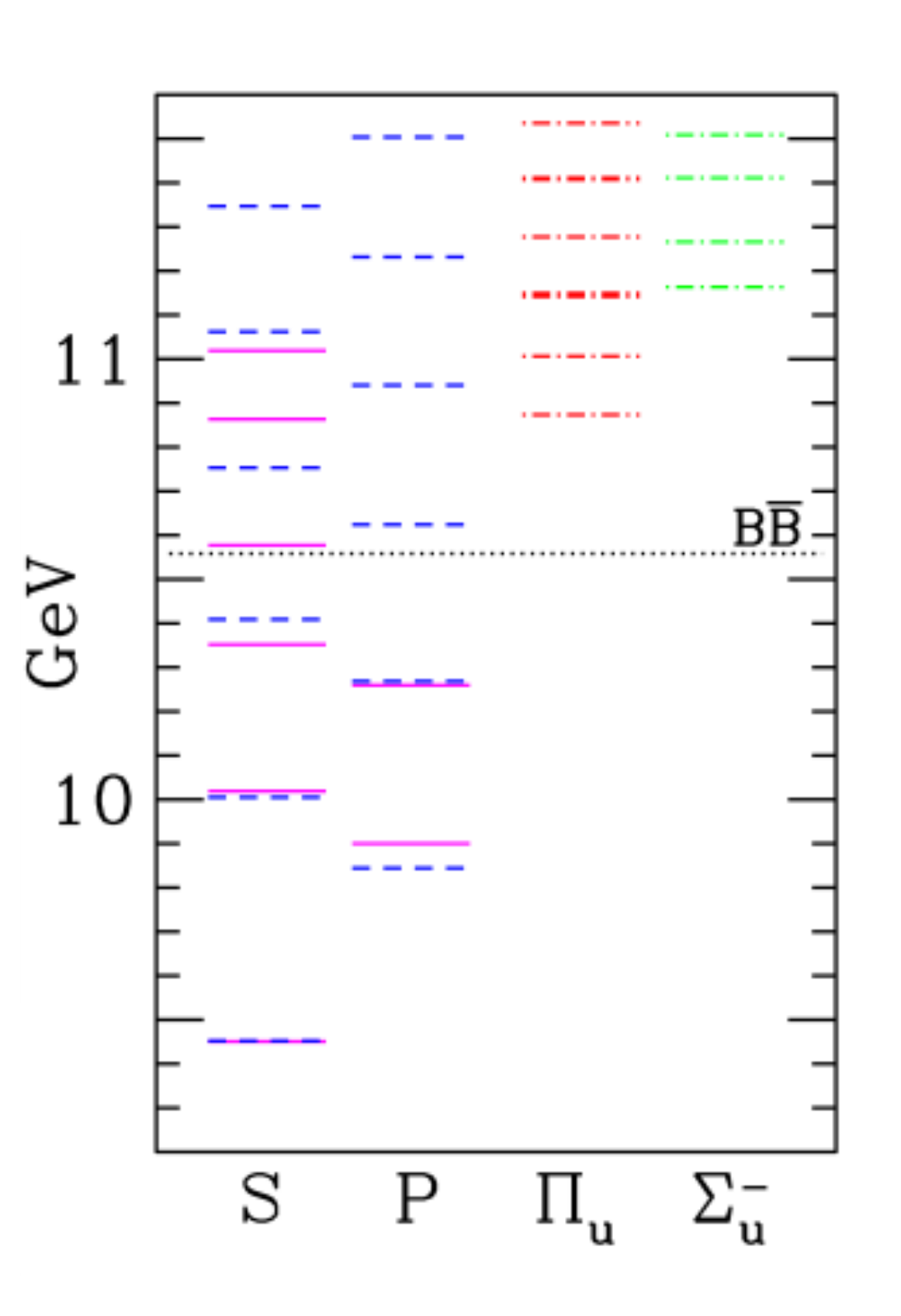}
\caption{Leading Born-Oppenheimer Spectrum for bottomonium (S, P) and Hybrids ($\Pi_u$ and $\Sigma_u^-$). (Figure reproduced with permission from reference \cite{JKM2}).}
\label{bbg}
\end{figure}

Fitting the $S$-wave $\Sigma_g^+$ energy to the spin-averaged $\eta_b$ and $\Upsilon$ mass gave a bottom quark mass of 4.58 GeV. This was then used to obtain the spectrum shown in Fig. \ref{bbg}. Experimental results are given as solid lines in the figure. Notice that the agreement of the conventional spectrum with experiment appears to worsen dramatically once past $B\bar B$ ($B\bar B^*$, $B^* \bar B^*$) threshold. Although similar large shifts due to light quarks may be present in the hybrid spectrum, there is reason to believe that these will be small for the lowest lying states. This is discussed further in section \ref{dec}.

Lattice studies with heavy quarks are difficult because heavy quarks tend to lie above the ultraviolet cutoff on typical lattices and hence are removed from the dynamics of the theory. One way to avoid this problem is to work directly with non-relativistic effective field theory, which removes the heavy quarks in an ordered expansion. An early study of this type by the CP-PACS collaboration neglected all spin-dependent operators in the effective Lagrangian\cite{CP}. The authors employed a ``magnetic" hybrid interpolating fields for a spin singlet $H^1 = \psi^\dagger B_i \chi$ and spin triplet $H^3 = \psi^\dagger \sigma_j B_i \chi$ states ($\psi^\dagger$ and $\chi$ are the heavy quark and antiquark creation operators respectively). Since spin-dependent operators were not included in the Lagrangian, the result was a degenerate multiplet of hybrid states with quantum numbers $1^{--}$, $(0,1,2)^{-+}$.

The computations yielded a charmonium hybrid multiplet 1.323(13) GeV above the spin averaged charmonium ground state (i.e., 3.069 GeV) and a bottomonium hybrid multiplet 1.542(8) GeV above the ground state (at 9.445 GeV).


A calculation of hybrid masses with large lattices in quenched QCD has been performed by Bernard {\it et al.}\cite{b}. They obtained charmonium hybrid masses of

\begin{equation}
1^{-+} \ 4.39(8)\ {\rm GeV} \qquad 0^{+-} \ 4.61(11)\ {\rm GeV}.
\end{equation}
The error in these results is statistical only -- additional truncation and quenching errors also exist.

%
%

Advances in computer technology, especially leveraging the power of graphical processing units, has permitted the recent computation of charmonium states with the full QCD Lagrangian. In particular, the Hadron Spectrum Collaboration has performed a large scale unquenched calculation that employs a large variational basis, a fine temporal lattice spacing, two light dynamical quarks, a strange dynamical quark, and improved lattice actions to obtain a comprehensive charmonium spectrum\cite{liu}. Despite these  technical advances, the dynamical quarks are still heavy, yielding a pion mass of 396 MeV, and a $J/\psi - \eta_c$ splitting of 80(2) MeV -- too small compared to the experimental value of 117 MeV.

\begin{figure}[h]
\centering
\includegraphics[angle=0,width=12cm]{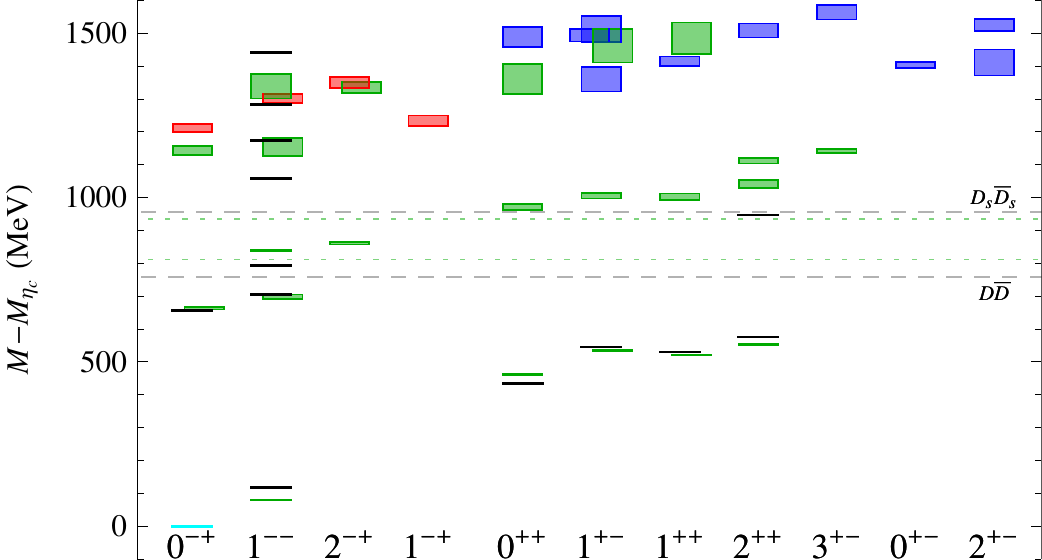}
\caption{A lattice QCD calculation of charmonium States. (Figure reproduced with permission from reference \cite{liu}).}
\label{ccfig}
\end{figure}

Despite these shortcomings, the spectrum, shown in Fig. \ref{ccfig}, is of relatively high quality. The figure shows charmonium masses  as boxes (with vertical extent indicating errors) and experimental states as lines. One sees good agreement for low lying states, with diminishing accuracy higher in the vector channel. Numerical values for the computed masses are presented in Table \ref{tab:ccspec}.

The authors of Ref. \cite{liu} also measured state overlaps with various operators, $\langle M|{\cal O} |0\rangle$, as a probe of the internal structure of the state $|M\rangle$.  Thus, for example, some vectors have large overlaps with a quark-antiquark pair in a ${}^3S_1$ state, while others have larger overlap with ${}^3D_1$ operators. These overlaps only provide qualitative indications of state configurations because they are scale-dependent and comparison to continuum matrix elements can be confounded by operator mixing. Nevertheless, comparison with conventional quarkonia structure as determined by the non-relativistic quark model~\cite{qm_1,qm_2,qm_3} reveals that the method is reliable.

\begin{table}[ht]\centering
\begin{tabular}{c|llllll}
\hline\hline
$J^{PC}$ &\multicolumn{6}{c}{$(M-M_{\eta_c})$ / MeV}\\
 \hline
$0^{-+}$ & 0 & 663(3) & 1143(13) &1211(13) & & \\
$1^{--}$ &80.2(1) &698(6) & 840(3)   &1154(28) &1301(14) &1339(38) \\
$2^{-+}$ &860(3) &1334(17) &1350(17) & & & \\
$2^{--}$ &859(5) &1333(18) & & & &\\
$3^{--}$ &867(3) &1269(26) &1392(12)  & & &\\
$4^{-+}$ &1444(10) & & & & &\\
$4^{--}$ &1427(9) & & & & & \\
\hline
$0^{++}$ &461.6(7) &972(9) &1361(46) &1488(30)  & & \\
$1^{+-}$ &534(1) &1006(9) &1360(38) &1462(51) &1493(19) &1513(39)  \\
$1^{++}$ &521.6(9) &1002(10) &1415(14) &1484(48) & &\\
$2^{++}$ &554(1) &1041(12) &1112(8)  &1508(21) & & \\
$3^{+-}$ &1142(6) &1564(22) & & & &\\
$3^{++}$ &1130(9) & & & & & \\
$4^{++}$ &1129(9) & & & & & \\
\hline
$1^{-+}$ &1233(16) & & & & &\\
$0^{+-}$ &1402(9) & & & & & \\
$2^{+-}$ &1411(40) & 1525(18) & & & & \\
\hline\hline
\end{tabular}
\caption{Charmonium Spectrum\cite{liu}.}
\label{tab:ccspec}
\end{table}

This method can be used to determine  states with large overlaps with operators with gluonic content. The resulting states are indicated with red and blue boxes in Fig. \ref{ccfig}. As can be seen, an approximate multiplet forms with quantum numbers
$1^{--}$ and $(0,1,2)^{-+}$.
This structure can easily be obtained if the effective gluonic degrees of freedom have $(J^{PC})_g = 1^{+-}$. Combining this with a $q\bar q$ pair in a ${}^1S_0$ state yields the $1^{--}$ component of the hybrid multiplet, while combining it with $q\bar q$ in ${}^3S_1$ yields the remaining members. Higher quantum numbers obtained in this manner are listed in Table. \ref{tab:multi}. Note that the higher hybrid multiplet in Fig. \ref{ccfig} contains
$J^{PC} = $ $(0,1,2,)^{++}$ and $(0,1,2,3)^{+-}$ states, which maps very well to the expected $P$-wave  multiplet.

\begin{table}[ht]\centering
\begin{tabular}{llll}
\hline\hline
$(J^{PC})_g$ & $L$ & $S$ & $J^{PC}$ \\
\hline
$1^{+-}$ & 0 & 0 & $1^{--}$ \\
$1^{+-}$ & 0 & 1 & $(0,1,2)^{-+}$ \\
$1^{+-}$ & 1 & 0 & $(0,1,2)^{++}$ \\
$1^{+-}$ & 1 & 1 & $(0,1,1,1,2,2,3)^{+-}$ \\
\hline\hline
\end{tabular}
\caption{Hybrid Multiplets}
\label{tab:multi}
\end{table}

%
%
%
%
%
%
%

\subsection{Light Hybrids}

Computations with light quarks are more difficult than those with moderate mass quarks because the Dirac matrix that must be inverted becomes rapidly ill-conditioned and larger (a lighter pion requires a larger lattice to maintain constant physics). Furthermore, light quarks mean that decays to multiple pion states are possible, greatly complicating the extraction of observables.

%
%
%

The earliest computations of light hybrid masses were made in the quenched approximation which effectively ignores internal $q\bar q$ loops. These calculations  all predicted that the $1^{-+}$ nonet of hybrid mesons was the lightest, with a masses in the 1.8 to 2.1 GeV mass region\cite{early_1,early_2,early_3,early_4,early_5}.

\begin{figure}[ht]
\centering
\includegraphics[width=0.6\textwidth]{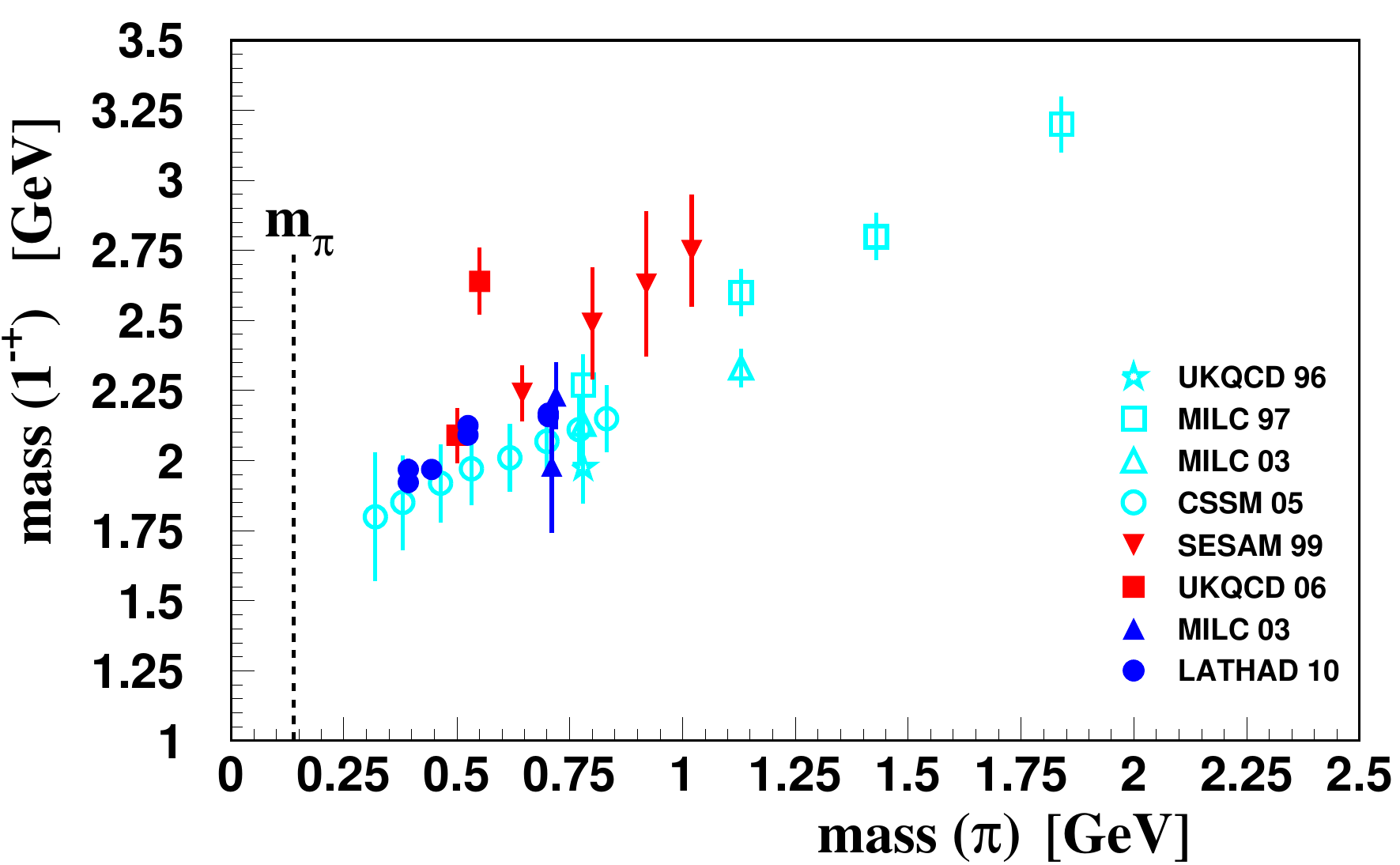}
\caption{The mass of the $J^{PC}=1^{-+}$ exotic hybrid
as a function of the pion mass from lattice calculations. The open (cyan) symbols correspond to
quenched calculations, while the solid (red and blue) symbols are dynamic (unquenched)
calculations: open (cyan) star~\cite{early_1}, open (cyan) squares~\cite{bernard-97}, open (cyan) upright
triangles~\cite{bernard-04}, open (cyan) circles~\cite{hedditch-05}, solid (red)
downward triangles~\cite{lacock-98}, solid (red) squares~\cite{McNeile:2006bz}, solid (blue)
upright triangles~\cite{bernard-04} and solid (blue) circles~\cite{Dudek:2010wm}.}
\label{fig:exotics}
\end{figure}

A summary of $1^{-+}$ hybrid mass computations is shown in Fig. \ref{fig:exotics}. Open symbols represent quenched computations, while filled symbols are unquenched. General agreement in the data is evident; and a naive extrapolation to the physical pion mass gives an unquenched $1^{-+}$ mass of approximately 1.6 GeV.

Finally, we present the results of a recent exhaustive computation of the light meson spectrum by the Hadron Spectrum Collaboration\cite{dudek-2013}. This computation was made with a large operator basis, on lattice of size $16^3\times 128$ up to $24^3\times 128$ lattice, a temporal lattice spacing around 0.034 fm and a spatial lattice spacing of approximately 0.12 fm, and  pion masses of 702, 524, and 391 MeV. Hairpin (disconnected) diagrams were included with the aid of the ``distillation" method\cite{dist}. The computation did not include glueball or multihadron operators, and thus the extracted ``masses" are only approximations to the resonance mass parameters. 

\begin{figure}[ht]
\centering
\includegraphics[angle=0,width=14cm]{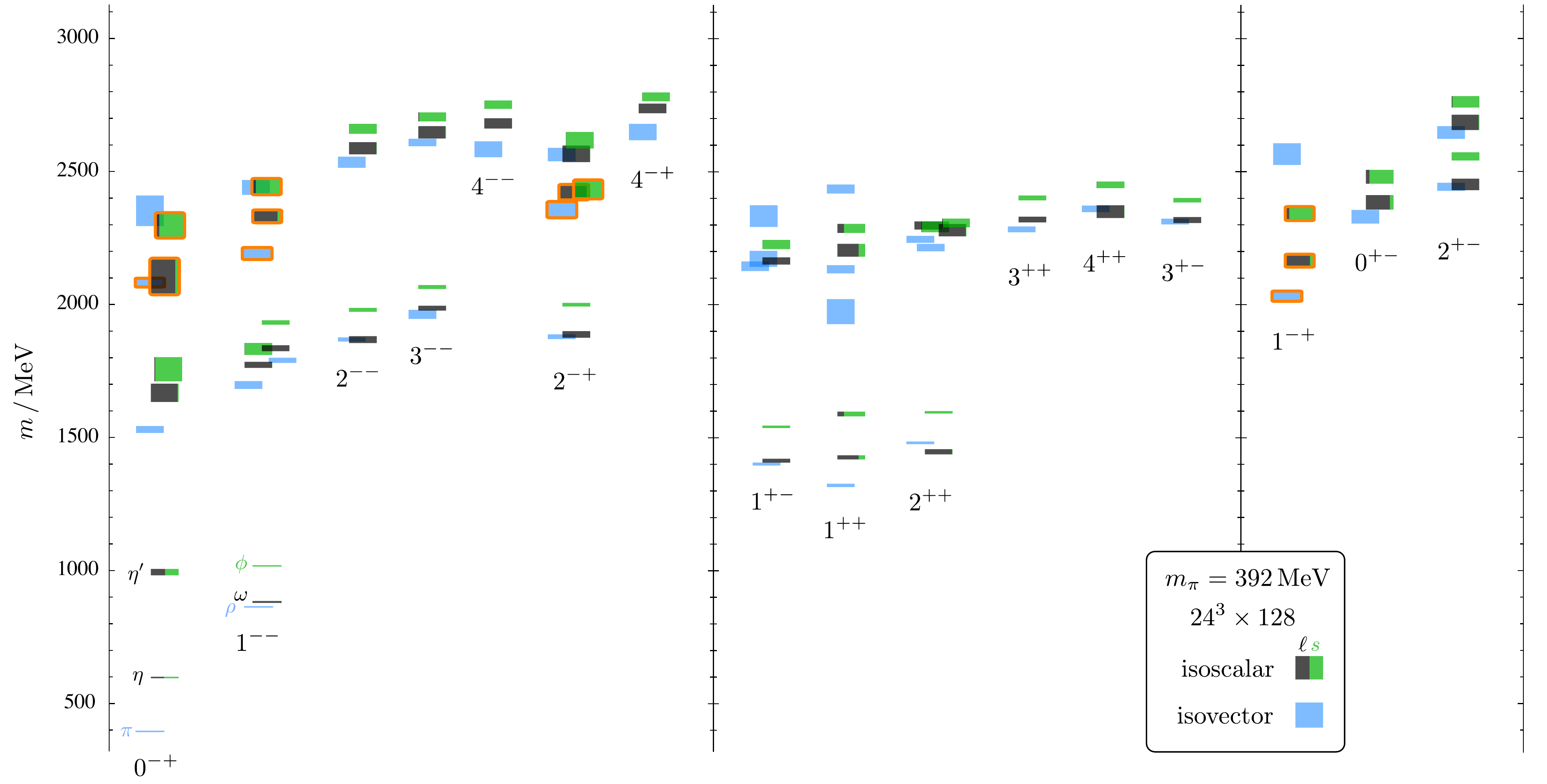}
\caption{Isoscalar and Isovector Hybrid Spectrum. States outline in orange have large gluonic content, (Figure reproduced with permission from reference \cite{dudek-2013}).}
\label{fig:iso}
\end{figure}

Results for the isovector and isoscalar spectra are shown in Fig. \ref{fig:iso} for pion masses of 392 MeV. Notice that mixing between light and strange quarks is represented (in green and black) in the figure. States outlined in orange have large overlap with gluonic operators. Notice that the quantum number-exotic states (to the right) are all predicted to be approximately ideally mixed.

\begin{figure}[h]
\centering
\includegraphics[angle=0,width=12cm]{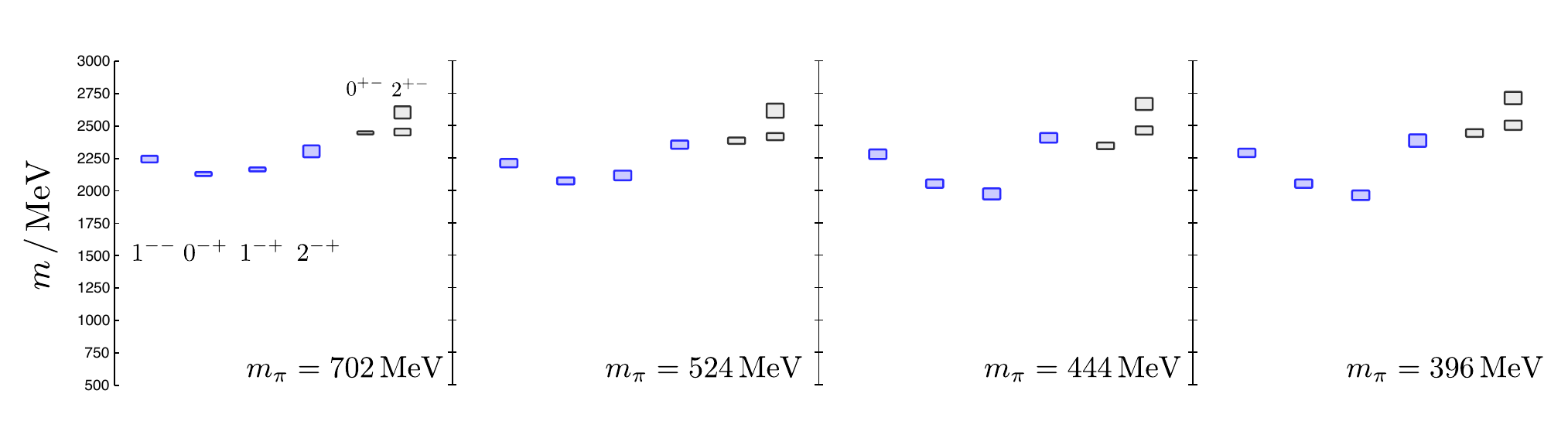}
\caption{The Lightest Hybrid Multiplet as a function of Quark Mass. (Figure reproduced with permission
from reference \cite{dudek-2011}).}
\label{fig:splittings}
\end{figure}

The quark mass-dependence of the lightest hybrid multiplet is of phenomenological interest. This has been evaluated by the Hadron Spectrum Collaboration and is discussed by Dudek in Ref. \cite{dudek-2011}. Fig. \ref{fig:splittings} shows the multiplet for four different pion masses. It is apparent that the P-wave $0^{+-}$ and $2^{+-}$ are approximately independent of quark mass, implying that a short range potential dominates the effective hybrid spin-dependent potential. Furthermore, the $1^{--}$ state is also largely spin-independent, which implies that quark spin-triplets are required in the spin-dependent potential. Finally, it appears that the spin-triplet $J^{-+}$ multiplet slowly splits as the quark mass is reduced, with the $J=0$ component decreasing slowly, the $J=1$ decreasing more rapidly, and the $J=2$ increasing slowly. Developing a phenomenological model for these observations is an interesting task.

\subsection{Hybrid Baryons}
\label{sec:hybbar}

All quantum numbers are available to baryons, hence `exotic' quantum number baryons do not exist (perhaps explaining the relative lack of interest in these states). Furthermore, conventional and hybrid baryons will mix to form the physical spectrum, which can seriously affect the ability to compute the properties of these states and to discover them experimentally.

\begin{figure}[h]
\centering
\includegraphics[angle=0,width=10cm]{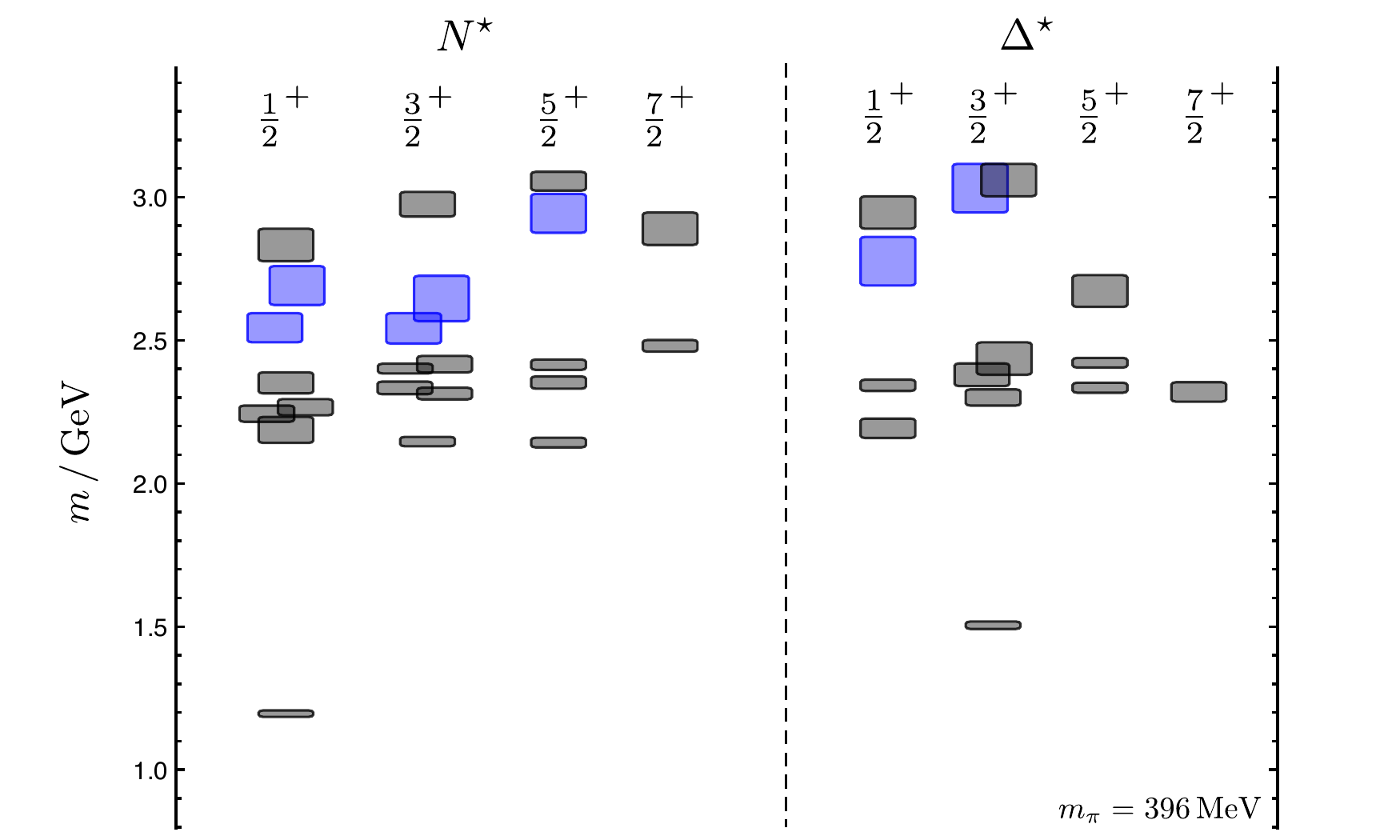}
\caption{Nucleon and Delta States. Grey boxes are conventional baryons; blue boxes are hybrid baryons. (Figure reproduced with permission from reference \cite{Dudek:2012ag}).}
\label{fig:lattbar}
\end{figure}

We are aware of only one lattice computation of the hybrid baryon spectrum, which was carried out by the JLab group\cite{Dudek:2012ag}. The authors considered the spectrum of nucleons and deltas at several quark masses and found a set of positive parity hybrid baryons with quantum numbers $2[N_{1/2+}]$, $2[N_{3/2+}]$, $N_{5/2+}$, $\Delta_{1/2+}$, and $\Delta_{3/2+}$ above the first band of conventional excited positive parity baryons. Results are shown in Fig. \ref{fig:lattbar}. These have been obtained on a $16^3\times 128$ lattice with 2+1 dynamical quarks and a pion mass of 396 MeV.

The low lying hybrid baryons appear as a band with positive parity that lie above the first-excited positive parity conventional states. This pattern of states is compatible with a color octet gluonic excitation with $J^P = 1^+$, in keeping with hybrid mesons. Furthermore, the hybrid excitation scale is approximately 1.3 GeV for both mesons and baryons, which indicates a common mechanism in both of these systems.
It is notable that hybrid states appear to mix weakly with conventional states, which encourages the idea that they might be experimentally detectable.

\subsection{Transitions}

Relatively little lattice work has been done on hybrid transitions. 
Indeed, the difficult task of evaluating two-point correlation functions such as $\langle {\cal O}_2(t_2) {\cal O}_1(t_1)\rangle$ and extracting physical states is replaced by the daunting task of evaluating three point functions $\langle {\cal O}_2(t_2) j_\mu(t_i) {\cal O}_1(t_1)\rangle$, extracting physical states and normalizing correctly.

\subsubsection{Vector Hybrid Mixing}

Computing an overlap matrix element is a somewhat simpler task than measuring a transition matrix element. For example, the MILC collaboration has examined the effects of moving beyond the static limit in the heavy quark expansion of quenched NRQCD by evaluating the matrix element $\langle Q\bar Q | {\cal O} | Q\bar Q g\rangle$\cite{milc}. The operator is the leading order correction in the NRQCD Lagrangian:

\begin{equation}
{\cal O} = -c  g \frac{\sigma\cdot B}{2M}.
\end{equation}
where $M$ is the heavy quark mass, $B$ is the chromomagnetic field, and $c$ is a coupling that is unity at tree level.

The resulting matrix elements can be written in terms of admixture fractions as

\begin{equation}
|\langle \Upsilon|Q\bar Q g (1^{--})\rangle|^2  \approx 0.4\%,
\end{equation}

\begin{equation}
|\langle \eta_b|Q\bar Q g (0^{-+})\rangle|^2  \approx 1\%,
\end{equation}

\begin{equation}
|\langle J/\psi|Q\bar Q g (1^{--})\rangle|^2  \approx 2.3\%,
\end{equation}

\begin{equation}
|\langle \eta_c|Q\bar Q g (0^{-+})\rangle|^2  \approx 6\%.
\end{equation}
The authors caution that the results for charmonium are less reliable than those for bottomonium since higher order operators can contribute substantially. Furthermore, unquenching effects can be significant. Neglecting these issues permits a simple estimate for the rate of charmonium vector hybrid production in electron-positron collisions:

\begin{equation}
\Gamma(c\bar c g \to e^+e^-) \approx 0.023(2) \cdot \Gamma(J/\psi \to e^+e^-) \approx 0.12(1) \ {\rm keV}.
\end{equation}

\subsubsection{Heavy Hybrid Hadronic Transitions}

The first lattice computation of a hadronic transition was made by the UKQCD collaboration for the case of heavy hybrids\cite{mcn}. The static quark limit imposes  important constraints on the decay process since the quark-antiquark configuration must remain unchanged. The authors focus on the decay of the exotic $1^{-+}$ state and determine that decay into S-wave mesons is forbidden (since production of the light quark pair in a spin triplet is forbidden by $\eta$ while a spin singlet is forbidden by $\Lambda$).  Furthermore, decay to an S-wave $(Q\bar q)$ + P-wave $(q\bar Q)$ configuration is forbidden because the P-wave excitation energy is typically greater than the hybrid excitation energy. Thus the only allowed transition in the heavy quark limit is a hidden flavor process, $Q\bar Q g \to Q\bar Q + S$, where $S$ is a light flavor singlet meson.   This amounts to a de-excitation of the excited gluonic string by emission of a light quark-antiquark pair.

The authors computed two such transitions using unquenched QCD with light quark masses near the strange quark mass. Heavy valence quarks were fixed in place and integrated out, so that they did not appear explicitly in the computation. Results (applied to bottomonium) were

\begin{equation}
b\bar b g(1^{-+}) \to \eta_b \, \eta(s\bar s) \sim 1\ {\rm  MeV}
\end{equation}
and
\begin{equation}
b\bar b g(1^{-+}) \to \chi_b \, \sigma(s\bar s) \sim 60\ {\rm  MeV}.
\end{equation}

\subsubsection{Light Hybrid Hadronic Transitions}
\label{sect:lht}

The decays of light hybrids are much less constrained than those with heavy quarks, thus lattice computations become even more important to guiding experiment and phenomenology.  The UKQCD collaboration has examined the decay of a light $1^{-+}$ exotic (termed the $\pi_1$) with two dynamical quarks\cite{mcn2}. They determined the $\pi_1$ mass to be 2.2(2) GeV and obtained effective couplings for two decay modes as follows

\begin{eqnarray}
\Gamma (\pi_{1}\rightarrow b_{1}\pi)/k & = & 0.66\pm 0.20 \\
\Gamma (\pi_{1}\rightarrow f_{1}\pi)/k & = & 0.15\pm 0.10,
\end{eqnarray}
where $k$ is the relative momentum in the final state.
Assuming that these results determine a constant effective coupling then permitted the authors to obtain partial widths of

\begin{eqnarray}
\Gamma (\pi_{1}\rightarrow b_{1}\pi) & = & 400 \pm 120 \ {\rm MeV} \\
\Gamma (\pi_{1}\rightarrow f_{1}\pi) & = & 90 \pm 60 \ {\rm MeV}.
\end{eqnarray}

As a check of their procedure, they also carried out a similar calculation for
$b_{1}\rightarrow \omega \pi$, obtaining  $\Gamma/k\sim 0.8$, which
leads to $\Gamma(b_{1}\rightarrow\omega\pi)\sim 220$ MeV. This is about a
factor of $1.6$ larger than the experimental width. Finally, the authors noted that the large hybrid widths arose chiefly due to the large available phase space in the decay. We remark that in general this expectation is false because hadronic form factors cause effective couplings to diminish rapidly with increasing momentum.

The latter point was subsequently pursued by Burns and Close, who compared the lattice results for the transition element near threshold to those of the flux tube model (see section \ref{sec:ftm})\cite{Burns:2006}. The two approaches were found to be in rough agreement near threshold, see Table \ref{tab:ftm}. Thus, if the flux tube model can be trusted to extrapolate to the physical recoil momentum, one obtains substantially reduced partial widths of $\Gamma(\pi_1 \to b_1\pi) \approx 80$ MeV and $\Gamma(\pi_1 \to f_1 \pi) \approx 25$ MeV. The lattice results also suggest that the light quark creation vertex has spin triplet quantum numbers.

\begin{table}[ht]\centering
\begin{tabular}{llll}
\hline\hline
   & $b_1 \to \omega \pi$ (GeV${{}^{-1/2}}^{\ }$) & $\pi_1 \to b_1\pi$ (GeV$^{-1/2}$) & $\pi_1 \to f_1 \pi$ (GeV$^{-1/2}$) \\
\hline
Lattice (UKQCD) & 2.3(1)  & 2.9(4) & 1.5(4) \\
Lattice (CP-PACS) & 3.4(2) & 2.9(3) & 1.1(4) \\
\hline
flux tube (a) & 2.7 & 2.9 & 1.4 \\
flux tube (b) & 3.3 & 3.9 & 1.9 \\
\hline\hline
\end{tabular}
\caption{Comparison of lattice and flux tube model transition amplitudes. UKQCD and CP-PACS refer to different gauge configuration ensembles.}
\label{tab:ftm}
\end{table}

\subsubsection{Hybrid Radiative Transitions}
\label{sect:hrt}

More recently, the Hadron Spectrum Collaboration has computed charmonium hybrid radiative transitions\cite{dudek-trans}. The calculation was made with a  large operator basis in the quenched approximation. The renormalization constant required to compare the lattice matrix elements to physical ones was determined nonperturbatively by conserving charge at zero recoil. The resulting widths are presented in Table \ref{tab:cctrans}, where one sees quite good agreement with experiment, where available. Notice that the process $c\bar c g(1^{-+}) \to J/\psi \gamma$  is a magnetic dipole transition. With conventional charmonia, these require a spin flip and therefore are suppressed for heavy quarks. In this case, however, the extra gluonic degrees of freedom can permit the transition, and hence it can be large.

\begin{table}[ht]\centering
\begin{tabular}{lll}
\hline\hline
transition & $\Gamma_{\rm lattice}$ (keV) & $\Gamma_{\rm expt}$ (keV) \\
\hline
$ \chi_{c0} \to J/\psi \gamma$ & 199(6) & 131(14) \\
$ \psi' \to \chi_{c0} \gamma$ & 26(11) & 30(2) \\
$ \psi'' \to \chi_{c0} \gamma$ & 265(66) & 199(26) \\
$ c\bar c g(1^{--}) \to \chi_{c0} \gamma$ & $<$ 20  &  \\
\hline
$J/\psi \to \eta_c \gamma$  & 2.51(8) & 1.85(29) \\
$\psi' \to \eta_c \gamma$  & 0.4(8) &  0.95 -- 1.37 \\
$\psi'' \to \eta_c \gamma$  & 10(11) &   \\
$c \bar c g(1^{--})  \to \eta_c \gamma$  & 42(18) &   \\
\hline
$c \bar c g(1^{-+})  \to J/\psi \gamma$  & 115(16) &   \\
\hline\hline
\end{tabular}
\caption{Quenched Lattice Charmonia Radiative Decays\cite{dudek-trans}.}
\label{tab:cctrans}
\end{table}

\section{Hybrid Models}

The construction of a reliable model of hybrid meson structure and dynamics is important for the interpretation of experimental results. At the simplest level, this is because it is expensive to compute large numbers of experimentally relevant quantities on the lattice. It is also likely that the computation of complicated amplitudes involving hybrids will remain out of reach of lattice methods for a long time.

In this section we discuss the salient features of bag models, string models, constituent glue models, and attempted computations with QCD sum rules and the Schwinger-Dyson formalism. The prime difference between models is the assumed form that the gluonic degrees of freedom take on: broadly, these are quasiparticle or collective in nature.

\subsection{Bag Models}

A detailed phenomenology of hybrids was first developed with bag models, both in the MIT bag 
model\cite{hybag-1,hybag-2,hybag-3,hybag-4,hybag-5,hybag-6}, 
and in the ``Budapest variant"\cite{HHKR}. The idea was to place a gluonic field in a vacuum `bubble' with appropriate boundary conditions. These imposed either transverse magnetic ($1^{--}$, TM) or transverse electric ($1^{+-}$, TE) solutions, with the TE modes being the lightest. The predicted result was four low lying nonets with, in order, quantum numbers $0^{-+}$, $1^{-+}$, $1^{--}$, and $2^{-+}$ in the mass range 1.2 -- 2.5 GeV.

The Budapest variant was designed to deal with heavy quarks for which a spherical bag was unrealistic (as it would not naturally yield linear confinement). Thus the bag was allowed to deform due to the presence of fixed heavy quark and antiquark sources. The resulting adiabatic energy surface was used in a two-body Schr\"{o}dinger equation to give mass estimates for hybrids. 
Masses found for the lightest hybrids were $\approx 3.9$ GeV
for $c\bar c$ and 10.45 GeV for $b\bar b$. Readers interested in examining this model closely should be aware that the formulas for quantum numbers reported therein are incorrect\cite{horgan}.

The bag model has also been applied to hybrid baryons\cite{bc-baryon}. In this case a TE gluonic field was combined with three quarks in an overall color octet to produce a set of hybrid states: $2[N_{1/2+}]$, $2[N_{3/2+}]$, $N_{3/2+}$, $N_{5/2+}$, $\Delta_{1/2+}$, and $\Delta_{3/2+}$. Notice that these quantum numbers agree with those reported for the lattice in section \ref{sec:hybbar}.

The overall mass scale of these states is not firmly established; for example in Ref. \cite{bc-baryon} the scale was set assuming that the $\pi(1300)$ was a hybrid meson, yielding hybrid baryons near 1.6 GeV. These days the  $\pi(1800)$ is a more plausible hybrid candidate, which places the low lying bag model hybrid baryons above 2 GeV in mass.

%
%
%
%
%
%
%
%
%
%
%

We remark that although bag models provide a compelling qualitative picture of confined quarks and gluons, they suffer from a number of ambiguities which makes their application less than ideal. Among these are difficulties in determining the response of the bag boundary when quarks and gluons are present, ambiguities in gluon self-energies, and the existence of spurious degrees of freedom associated with the center of mass.

A more subtle problem concerns the presence of both a bag pressure and valence gluons. This issue is illustrated with a statement by Jaffe and Johnson, who discuss exotic quantum numbers in the MIT model and note that they can occur due to ``type II exotic quark states", 
which are $q\bar q$  or $qqq$ states with quantum numbers that cannot be obtained in the ``conventional" quark model\cite{jj}. They contrast these with ``mixed states of quarks and glue" and go on  to explain that it is the bag surface that permits the extra quantum numbers.

The problem is that the degrees of freedom in QCD are quarks and gluons, and it is natural to associate the bag pressure with gluonic properties of the medium. Thus the extra quantum numbers are due to glue, and ``type II" and ``$q\bar q g$" states are really the same. In short, an intrinsic ambiguity exists between bag and gluonic degrees of freedom.

\begin{figure}[h]
\centering
\includegraphics[angle=0,width=8cm]{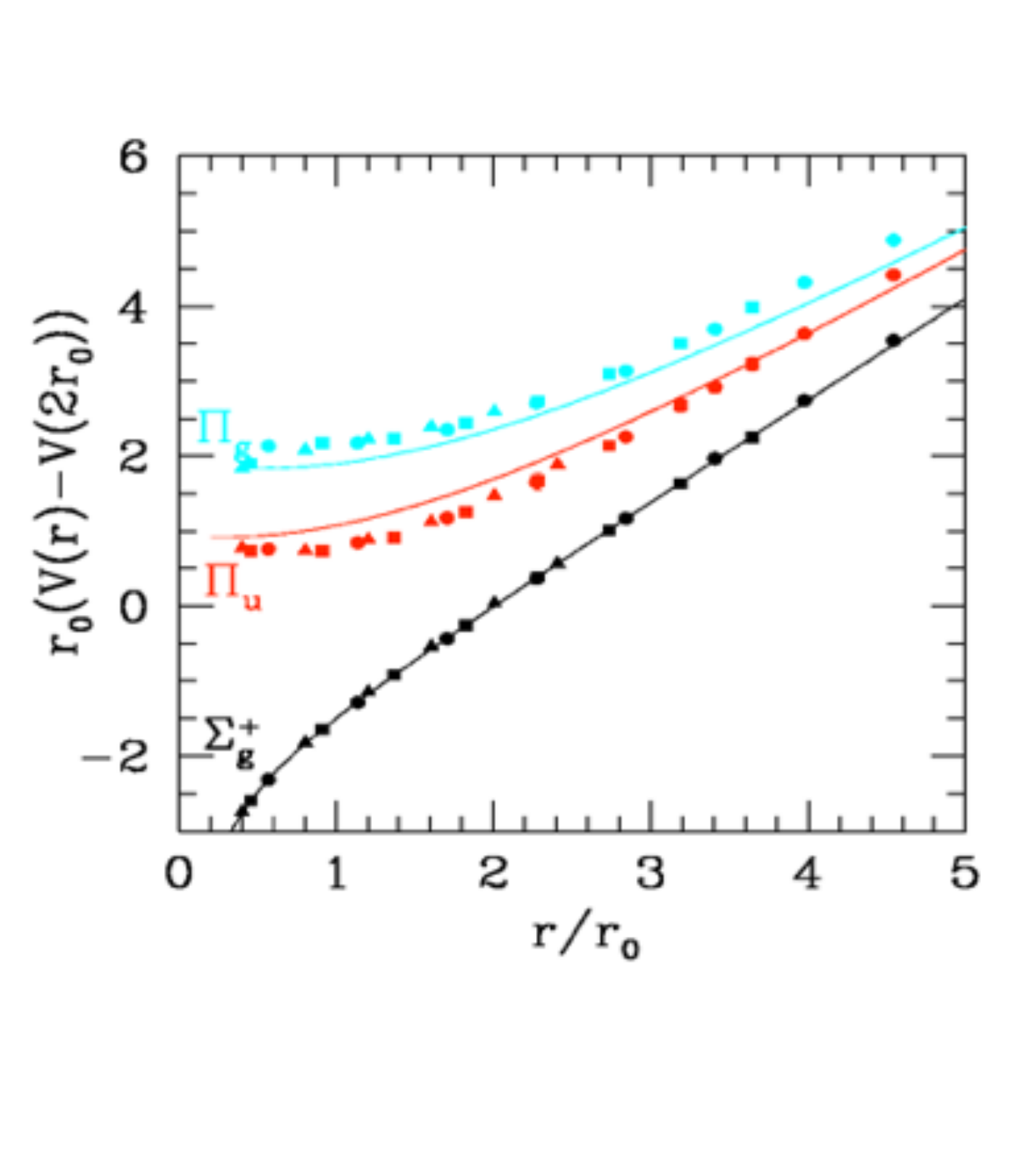}
\caption{Lattice and Bag Model Adiabatic Potentials. (Figure reproduced with permission from 
reference \cite{Juge:1997nd}).}
\label{fig:hhkr}
\end{figure}

More modern applications of bag models have focussed on reproducing lattice results. For example, Juge, Kuti, and Morningstar have resurrected the Budapest variant and compared its adiabatic surfaces to those of the lattice\cite{Juge:1997nd}. The results, some of which are given in Fig. \ref{fig:hhkr}, show quite good agreement between the approaches.

Finally, an attempt at reproducing the gluelump spectrum in the bag model has been made by Karl and Paton\cite{Karl:1999wq}. The calculation draws on the mapping between adiabatic surfaces at zero quark separation and the gluelump spectrum. Computed single gluon modes in a spherical bag do not agree with the lattice gluelump ordering; including corrections due to the Coulomb interaction and adjusting the strong coupling gave levels as reported in Table \ref{tab:kp}. For comparison we display the lattice results of table \ref{tab1} with energies shifted to match 1.43 GeV for the $1^{+-}$ state. As can be seen, there is broad agreement, while detailed agreement cannot be claimed.

\begin{table}[ht]\centering
\begin{tabular}{llllll}
\hline\hline
$J^P$: & $1^+$ & $2^-$ & $1^-$ & $3^+$ & $2^+$ \\
mass (GeV): & 1.43 & 1.97 & 1.98 & 2.44 & 2.64 \\
lattice (GeV): & -- & 2.01 & 1.81 & 2.42 & 2.42 \\
\hline\hline
\end{tabular}
\caption{Gluelump Bag Model Predictions and Shifted Lattice Spectrum.}
\label{tab:kp}
\end{table}

\subsection{Relativistic String Models}

Models of soft glue as collective, `stringy', degrees of freedom have a long history, dating from early ideas contained in dual resonance models. 
The formal development of string theory veered off into mathematical details once it was realized that substantial formalism was required for consistent quantization. However, variant string models were presented as explicit models of mesons. Thus, for example, Andreo and Rohrlich noted that dual strings and gauge theories are related, and thus, ``one is ... led to believe that in a certain approximation a meson can be regarded as two point-like quarks confined to each other by a string"\cite{ar}. Once this idea is in place it is  natural to examine the adiabatic potentials of the model and note that ``daughter trajectories" correspond to additional mesonic states.

To our knowledge, the first to draw this connection were Giles and Tye in 1976\cite{GT}. These authors coupled quarks to a relativistic two-dimensional sheet in what they called the 
``Quark Confining String Model" and noted that 
``The presence of vibrational levels  gives ... extra states in quantum mechanics ... that are absent in the charmonium model".

The quark-string system was assumed to obey an action 

\begin{equation}
S = \int d^2u\, (-{\rm det}(g))^{1/2} \, \bar \psi i \gamma_\mu \tau^\mu_\alpha \partial^\alpha \psi + \ldots
\end{equation}
where $\bm{\tau}$ is a collection of tangent vectors to the string surface. Vibrational energies were crudely estimated and a Born-Oppenheimer Schr\"{o}dinger equation was solved for the hadronic states:

\begin{equation}
H = -\frac{1}{m}\frac{\partial^2}{\partial R^2} + 2m + V_n(R) + \frac{\ell(\ell+1)}{m R^2}
\end{equation}
with

\begin{equation}
V_n = k R \left(1 - \frac{2 n \pi}{2n\pi + k ( (R-2d)^2 + 4 d^2)}\right)^{-1/2}.
\end{equation}
Here $d$ is a correction due to the finite quark mass.
Vibrational levels (hybrids) in the charmonium sector were predicted to lie at 3.96, 4.21, 4.41, 4.45, and 4.46 GeV; although no quantum numbers (other than the angular momentum) were given.

Similar models have been developed by a variety of groups: Ref. \cite {bt} extended the work of Giles and Tye and predicted a vector bottomonium hybrid with mass $m_\Upsilon + 990$ MeV and a photocoupling of $\Gamma(b\bar b g (1^{--}) \to e^+e^-) = 0.20(15)$ keV. An early application of string models to baryons is contained in Ref. \cite{ckp} where adiabatic potentials relevant to the three quark system are developed. Another application has been to the adiabatic surfaces of section \ref{ad}.  The authors of Ref.\cite{Allen:1998wp} used a simple quantized string model with fixed ends to obtain an expression for the adiabatic surfaces

\begin{equation}
V_N(R) = \sqrt{(aR)^2 + 2\pi a (N+c)} +C
\end{equation}
where $N$ labels the string excitation and is given in terms of phonon occupation numbers as

\begin{equation}
N = \sum_{m=1}^\infty m(n_m^+ + n_m^-).
\label{eq:N}
\end{equation}
The authors added $-4/3 \cdot \alpha_s/R$ to this expression, set $c=0$, and fit to the adiabatic $\Sigma_g^+$ surface to obtain $C$ and $a$.  The resulting excited surfaces agreed reasonably well with the lattice computations. 

The extension of effective string models to $1/M^2$ corrections to the ground state adiabatic potential has been explored by Brambilla \emph{et al.}\cite{Brambilla:2014eaa}. Explicit expressions for a variety of corrections were obtained, but a detailed comparison to quarkonium phenomenology was not made.

Recently, the AdS/QCD hypothesis of Maldacena\cite{mald} has been used to extract the ground state $\Sigma_g^+$ surface\cite{ads} and the $\Sigma_u^-$ hybrid surface\cite{andreev}. Although this effort is in its infancy, the agreement with lattice results is quite good.

\subsection{Flux Tube Model}

\subsubsection{Model Development}

The flux tube model was developed by Isgur and Paton as a response to bag models. In particular, Isgur and Paton objected to the use of perturbative gluons in constructing a hadron, and argued that nonperturbative degrees of freedom, such as in a string model, should be used\cite{jp}. Noting that the QCD lattice Hamiltonian also is not written in terms of perturbative gluons, Isgur and Paton built a simple model by truncating the Hamiltonian with a series of approximations\cite{IP-1,IP-2}. In particular, the lattice degrees of freedom are quarks fields on lattice sites and gluonic `link variables' $U_\ell = \exp(-iag A_\mu(x))$ where $\ell$ represents the link ($x,\hat \mu$). In the strong coupling limit the Hamiltonian is given by

\begin{equation}
H_{scQCD} = \frac{g^2}{2a} \sum_\ell E_\ell^aE_{a\ell} + m\sum_n\bar \psi_n \psi_n
\end{equation}
where $g$ is the strong coupling, $a$ is the lattice spacing, 
and  $n$ is a lattice site. The velocity variables $\dot U_\ell$ have been replaced by
electric field operators $E_\ell$. 
Gauge invariant pure glue states are formed by closed (possibly
multiply connected) loops of link operators.
The commutation relation $[E^a,U] = T^a U$ then implies that the energy of these states is 
simply the sum of the quadratic color charges of each link:

\begin{equation}
E_{\rm loop} = \frac{g^2}{2a}\sum_{\ell \in {\rm loop}} {\cal C}_\ell^2
\end{equation}
where ${\cal C}^2 = 4/3$ for a field in the 3 or $\bar 3$ representations, 10/3 for 6 or $\bar 6$, etc.
The presence of quarks permits gauge invariant states with open flux strings which terminate 
on quark color sources or sinks. 
Perturbations to these states are provided by subleading quark hopping and magnetic
terms. The former allow flux tube breaking via quark pair production or quark motion.
The latter can change link color representations, cause link hopping, or change 
loop topology.

Isgur and Paton simplified the dynamics by (i) assuming an adiabatic separation of
quark and gluon degrees of freedom (ii) neglecting `topological mixing' such
as loop breaking or loop Euler number changing transitions (iii) working in the non-relativistic
limit.  The model is meant
to be applied to the `intermediate regime' where  $1/a \sim \sqrt{b}$.
They then modelled
link variable dynamics in terms of spinless colorless particles (`beads') of mass $ba$ where $b$ is
the string tension in the static quark potential. Finally these particles are assumed
to interact via a linear potential and perform small oscillations about their
resting positions. The result is a simple discrete string model for glue described
by the Hamiltonian:

\begin{equation}
H = b_0 R + \sum_n\left[ \frac{p_n^2}{2b_0a} + \frac{b_0}{2a}(y_n-y_{n+1})^2\right],
\end{equation}
where $y_n$ is the transverse displacement of the $n$th of $N$ string masses, $p_n$ is its momentum, $b_0$ is a bare string tension, and $R=(N+1)a$ is the separation between the static quarks. This Hamiltonian may be diagonalized in the usual way yielding

\begin{equation}
H_{FTM} = b_0R + \left( \frac{4}{\pi a^2}R -\frac{1}{a} - \frac{\pi}{12R} + \ldots\right) + \sum_{n\lambda} \omega_n \alpha_{n\lambda}^\dagger \alpha_{n\lambda}
\end{equation}
where $\alpha_{n\lambda }^\dagger$ creates a phonon in the $n$th mode with polarization $\lambda$. Notice that the string tension has been renormalized by the first term in the brackets. 
The last term in the brackets is the L\"uscher term of string phenomenology\cite{L}. The
mode energies are given by $\omega_n = 2/a \sin[\pi n / 2(N+1)]$. Thus 
$\omega_1 \to \pi/R$ as $N\to \infty$ is the splitting between the ground state
Coulomb+linear potential and the first gluonic excitation surface at long range.
The energy of a given phonon state is approximately

\begin{equation}
E = E_0 + N \frac{\pi}{R}
\label{E}
\end{equation}
with $N$ given by Eq. \ref{eq:N}.

Hybrid mesons are constructed by specifying the gluonic states via phonon
operators and combining these with quark operators with a Wigner rotation matrix:

\begin{equation}
|LM_L; s \bar s; \Lambda, \{n_{m+},n_{m-}\}\rangle \propto \int d^3 r \varphi(r) D^L_{M_L\Lambda}(\hat r) \, b^\dagger_{r/2,s} d^\dagger_{-r/2,\bar s} \prod_m (\alpha_{m+}^\dagger)^{n_{m+}} (\alpha_{m-}^\dagger)^{n_{m-}} | 0 \rangle.
\end{equation}
The projection of the total angular momentum on the $q\bar q$ axis is denoted by $\Lambda = \sum_m (n_{m+}-n_{m-})$.
The parity and charge parity of these states are given by

\begin{equation}
P|LM_L; S M_S; \Lambda, \{n_{m+},n_{m-}\}\rangle = (-)^{L+\Lambda+1}|LM_L; S M_S; -\Lambda, \{n_{m-},n_{m+}\}\rangle,
\end{equation}

\begin{equation}
C|LM_L; S M_S; \Lambda, \{n_{m+},n_{m-}\}\rangle = (-)^{L+S+\Lambda+N}|LM_L; S M_S; -\Lambda, \{n_{m-},n_{m+}\}\rangle.
\end{equation}
States of good parity are thus formed as 
\begin{equation}
|LM_L; S M_S; \zeta; \Lambda, \{n_{m+},n_{m-}\}\rangle = \frac{1}{\sqrt{2}}\Big(|LM_L; S M_S; \Lambda, \{n_{m+},n_{m-}\}\rangle + \zeta |LM_L; S M_S; -\Lambda, \{n_{m-},n_{m+}\}\rangle\Big).
\end{equation}

Possible single phonon $(m=1)$ mesons are listed in table \ref{tab:ftmqn}, where underlined quantum numbers represent quantum number exotic hybrids.

\begin{table}[h]\centering
\begin{tabular}{cccc}
\hline\hline
$\zeta$ &  $L$  &  $S$   &  $J^{PC}$ \\
\hline
+  & 1  &  0  & $1^{++}$ \\
+  & 1  &  1  & $(\underline{2},1,\underline{0})^{+-}$ \\
+  & 2  & 0  &  $2^{--}$ \\
+  & 2  & 1  &  $(\underline{3},2,\underline{1})^{-+}$ \\
\hline
-  &  1 &  0  & $1^{--}$ \\
-  & 1  & 1  & $(2,\underline{1},0)^{-+}$ \\
-  & 2  &  0  & $2^{++}$ \\
-  & 2 &  1  &  $(3,\underline{2},1)^{+-}$ \\
\hline\hline
\end{tabular}
\caption{Flux Tube Model Single Phonon Mesons}
\label{tab:ftmqn}
\end{table}

Isgur and Paton obtained hybrid meson masses by solving a model Hamiltonian of quark
motion on the single-phonon excited surface:

\begin{equation}
H_{IP} = -\frac{1}{2 \mu} \frac{\partial^2}{\partial R^2} + \frac{L(L+1) - \Lambda^2 }{2 \mu R^2} - \frac{4 \alpha_s }{3 R} + \frac{\pi}{R}(1 - {\rm e}^{-f \sqrt{b} R}).
\label{HIP}
\end{equation}
The interaction term incorporates several important additional assumptions. Namely
the $\pi/R$ phonon splitting is softened at short range. The parameter $f$ which
appears in the softening function was estimated to be roughly unity.
Furthermore, it was assumed that the attractive Coulomb ($1/R$) potential  remains
valid for hybrid mesons. 

%

Finally the quark angular momentum operator is now complicated by the presence of 
gluonic/string
degrees of freedom. One may write

\begin{equation}
L_{q\bar q} = L - L_{S_{||}}- L_{S_\perp}
\end{equation}
where $L$ ($L_S$) is the total (string) angular momentum. Note that
$L_{S_\perp}$ mixes adiabatic surfaces. Using $L_{S_{||}} = \Lambda \hat R$ and
neglecting surface mixing  yields the centrifugal term  of Eq. \ref{HIP} (notice that this is \emph{not} the same as used by Juge \emph{et al.}).

The hybrid masses obtained by solving Eq. \ref{HIP} are labelled $E_{IP}$ in table \ref{tab:ipr}.
Isgur and Paton also estimated
the effects of adiabatic surface mixing and used these as their final mass estimates
(labelled $E_{IP}'$). The column labelled KW lists hybrid masses obtained when the adiabatic surfaces of section \ref{ad} are used along with the centrifugal term of Eq. \ref{eq:LQ}\cite{KW}.

\begin{table}[h]\centering
\begin{tabular}{cccc}
\hline\hline
flavor & $E_{IP}$ (GeV)  & $E_{IP}'$  (GeV) &  $E_{KW}$  (GeV)\\
\hline
I=1  &  1.67 &  1.9  &  1.85 \\
I=0  &  1.67  & 1.9  &  1.85  \\
$s\bar s$  & 1.91  &  2.1 & 2.07  \\
$c\bar c$  & 4.19 &  4.3  & 4.34 \\
$b \bar b$ & 10.79  & 10.8 & 10.85 \\
\hline\hline
\end{tabular}
\caption{Hybrid Mass Predictions}
\label{tab:ipr}
\end{table}

\subsubsection{Further Developments}
\label{sect:fd}

In developing the flux tube model, Isgur and Paton assumed adiabatic separation of bead and quark motion and that the string executes small oscillations. Both of these assumptions were
tested by numerically solving a bead-quark-antiquark system with a quantum Monte Carlo algorithm\cite{BCS}. The results indicate that the small oscillation approximation is accurate
for long strings but overestimates gluonic energies by an increasing amount as
the interquark distance shrinks. Typical energy differences are order 100 MeV at
1 fm. Similarly, the adiabatic approximation
underestimates true energies by roughly 100 MeV, with slow improvement as the quarks
get very massive. It thus appears that these approximation errors tend to cancel 
each other, leaving the mass estimates of Isgur and Paton largely intact.

The effects of adiabatic surface mixing were examined by Merlin and Paton by considering the full quark-bead system\cite{MP}.
Although the effects can be quite complicated, with mixing between all surfaces possible,
they found that the majority of the effects can be absorbed in a redefinition of the
hybrid potential by including the rigid body moment of inertia of the string in the
centrifugal term:

\begin{equation}
\frac{1 }{2 \mu R^2} \to \frac{1}{2\mu R^2 + \frac{1}{6} b R^3},
\end{equation}
and with a more important effect that increases the strength of the $\pi/r$ splitting
as $r$ becomes larger than $m_q/b$.
An explicit computation revealed
found mass shifts of order -100 MeV for conventional S-wave light quark mesons and +200
MeV for light quark hybrids.

Merlin and Paton also examined spin orbit forces in the context of the flux tube mode\cite{MP2}. The idea was to map the operators of the leading spin orbit term in the heavy quark
expansion  of QCD, namely $V_{SO} = g/2m\cdot {\bm \sigma}\cdot {\bm B}$, onto phonon degrees of freedom. Explicit computations revealed that spin orbit splittings due to $V_{SO}$ are
small and that the majority of the splittings arise from Thomas precession,
$V_{Th} = 1/4  (\ddot {\bm r}_q \times \dot {\bm r}_q)\cdot {\bm \sigma}$. This was  modelled by including the effects of phonons on the quark coordinate.  The resulting mass splittings for light hybrids are listed in table \ref{tab:mp}.
One sees that the lowest member of the octet of light hybrids is predicted to be
the $2^{+-}$ while the heaviest is the $0^{+-}$. These results do not agree with those of lattice gauge theory.

\begin{table}[h]\centering
\begin{tabular}{l|llllllll}
\hline\hline
$J^{PC}$ & $2^{+-}$ & $2^{-+}$ & $1^{-+}$ & $0^{-+}$ & $1^{+-}$ & $0^{+-}$ & $1^{++}$ & $1^{--}$ \\
\hline
$\delta M$ (MeV) & -140 & -20 & 20 & 40 & 140 & 280 & 0 & 0 \\
\hline\hline
\end{tabular}
\caption{Spin Orbit Hybrid Mass Splittings.}
\label{tab:mp}
\end{table}

Isgur pointed out that the energy carried by the flux
tube will change several features of the naive quark  model\cite{I} (see also \cite{MP}).
For example, zero point oscillation of the flux tube about the interquark axis will induce transverse fluctuations in the quark positions, something which is not present when the flux tube is treated as a potential. The additional fluctuations have the effect of increasing the
 charge radius of a heavy-light meson (qQ):

\begin{equation}
r_Q^2 = \left[ \left( \frac{m_q }{ m_q+m_Q}\right)^2 + \frac{2 b }{\pi^3m_q^2}\zeta(3)\right] \langle r^2\rangle
\end{equation}
where the second term in the bracket is the new contribution. Isgur estimated this to give rise to a 50\% increase in charge radii of light quark hadrons.

\subsection{Hybrid Baryons in the Flux Tube Model}

Capstick and Page have made a detailed study of baryon flux tube dynamics\cite{CapPage-1},\cite{CapPage-2}.
This is a technically challenging problem due to the multitude of vibrational and rotational
modes that are available to a Y-shaped string system. However, they have found that the
problem simplifies considerably because the string junction decouples to good accuracy 
from the rest of the bead motion. Thus a hybrid baryon can be approximated by three 
quarks coupled via linear potentials to a massive ``junction bead". The dynamics of this
system are completely specified by the flux tube model and variational calculations indicate that the
lowest  lying hybrids are $J^P = \frac{1}{2}^+$ and $\frac{3}{2}^+$ states at approximately
1870 MeV. Notice that the lattice confirms these lightest states, but that the lattice mass is somewhat higher at 2550 MeV (section \ref{sec:hybbar}).

We remark that lattice investigations of the static baryon interaction have been carried 
out\cite{BFT}. The chief point of interest is whether the expected flux tubes form into a `Y' shape or a `$\Delta$' shape.  This may be addressed by carefully examining the baryonic energy in a variety of quark configurations.  
Current results are mixed, with some groups claiming support for the two-body hypothesis
\cite{Alexandrou} and some for the three-body hypothesis\cite{Suganuma}.  Finally, a strong
operator dependence in the flux tube profiles has been observed\cite{OW}, which clearly 
needs to be settled before definitive conclusions can be reached.

\subsection{Constituent Glue Models}

Constituent gluon models for hybrids were introduced by Horn and Mandula 
\cite{HM} and were subsequently developed by 
Iddir {\it et al.} and Ishida {\it et al.} \cite{hycg,conDecay,conDecay-2}.
Since these models assume a diagonal gluon angular momentum $\ell_g$ their
predictions for quantum numbers differ somewhat from other models. 

Horn and Mandula assumed a massless $J^P=1^-$ gluon interacting via linear potentials with quarks. The short range repulsive interquark potential was absorbed into a constant with the argument that repulsive forces do not produce vacuum polarization instabilities. Their model thus took the form:

\begin{equation}
H = 2m_c + \frac{p_q^2}{m_c} + p_g + G (|r_q - r_g| + |r_{\bar q}-r_g|) + C
\end{equation}
where the string tension was taken to be $G= 0.30$ GeV$^2$.

For the lightest hybrid states (with $\ell_g=0$) Horn and Mandula predicted
nonexotic quantum numbers equivalent to $P$-wave $q\bar q$ states, since
the gluon has $J^P=1^-$. Exotic quantum numbers including $1^{-+}$ are predicted in the 
higher-lying $(\ell_{q\bar q}, \ell_g ) = (1,0)$ and $(0,1)$ multiplets.
Detailed spectroscopic predictions for hybrids have not been published using these constituent gluon models, and the estimated masses are assigned large uncertainties. A typical result, due to Ishida {\it et al.}, is 1.3-1.8 GeV for light nonexotic hybrids and 1.8-2.2 GeV for light exotics.

%
%
%
%
%
%
%
%
%

More recently, the idea of constituent glue models has been revived in the context of QCD in Coulomb gauge\cite{SS2}. The model takes the Hamiltonian of QCD in Coulomb gauge as its starting point. A nontrivial mean field vacuum Ansatz is used to construct self-consistent constituent quarks and gluons, and these are used to construct hadrons in the Tamm-Dancoff approximation. In this picture the gluon remains transverse, but has a dynamically generated mass. Interactions were truncated at the two-body level. An eigenvalue equation was derived for the case of static quarks and the adiabatic energy surfaces were obtained. Although they mimic those of section \ref{ad} reasonably well, the level ordering was incorrect. This problem can be traced to the $J^P=1^-$ assignment for the constituent gluon, and hence applies to all models that make this assumption.

It is also possible to reproduce the gluelump spectrum with this method by setting the interquark distance to zero. In this limit the gluonic angular momentum becomes a good quantum number and $Y$-parity is simply given by $P (-)^{j_g}$. Thus degeneracies appear in the gluelump spectrum, precisely as shown in section \ref{sec:gl} (although the level ordering remains incorrect). With this method the splitting between the lowest two levels was predicted to be 500 MeV, which should be compared to the 360 MeV measured in lattice gauge theory (see table \ref{tab1}).

\begin{figure}[h]
\centering
\includegraphics[angle=0,width=8cm]{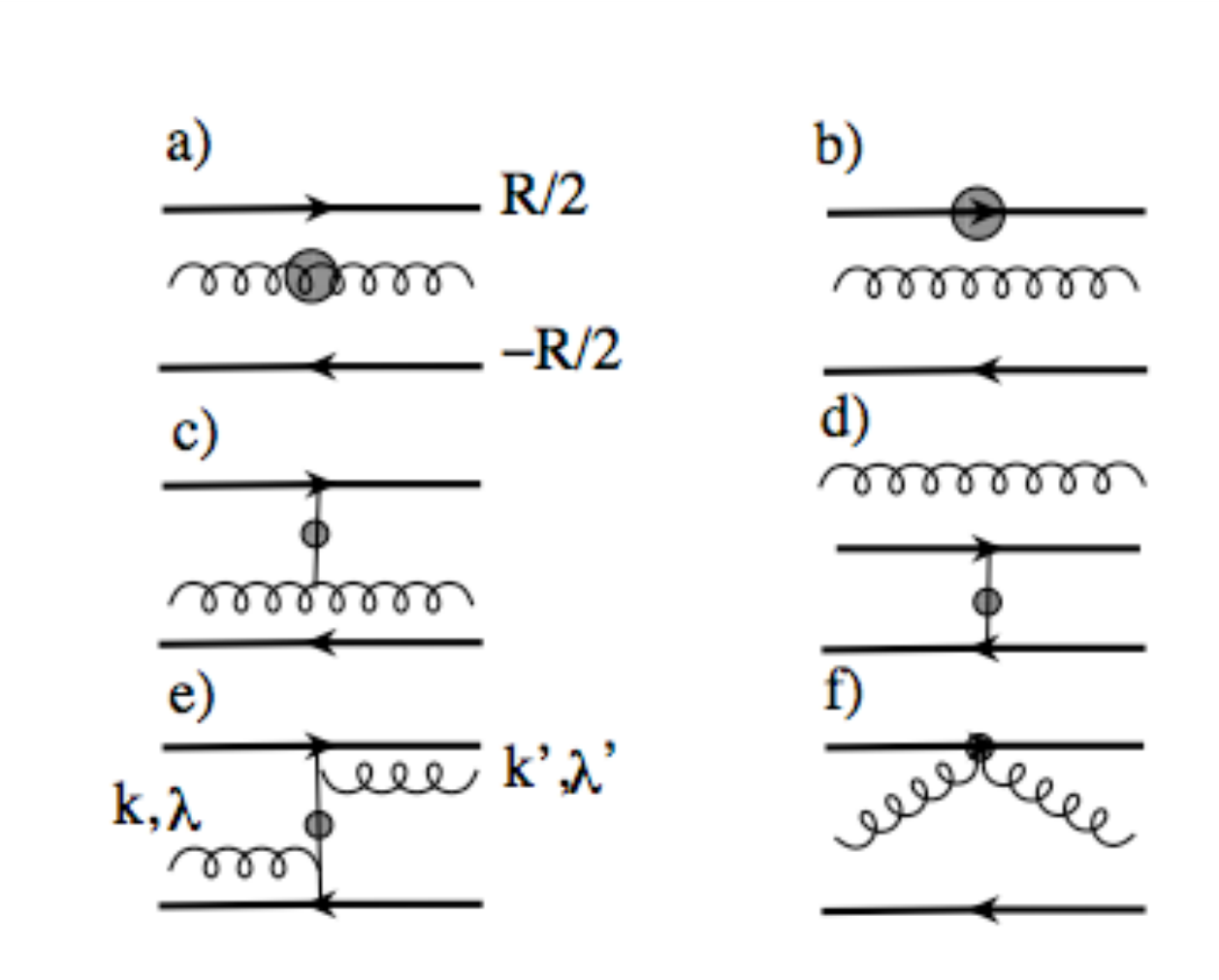}
\caption{Interactions in a Constituent Hybrid Model.}
\label{fig:hyb-int}
\end{figure}

Despite these failings, it would be premature to dismiss a constituent gluon picture of hybrids because important three body interactions exist in the system and these cannot be neglected in a nonperturbative context. Fig. \ref{fig:hyb-int} shows some of these interactions; the first two are self-energies, the second pair were considered in older constituent models, and (e) and (f) represent important additional interactions. In fact, Szczepaniak and Krusinski have shown that the three-body interactions are sufficient to invert the naive parity ordering in the gluelump and adiabatic surface spectra\cite{Szczepaniak:2006nx}. Detailed comparison to lattice results show reasonable agreement, although the model levels tend to be several hundred MeV high.

A follow-up paper generalized the Coulomb gauge model by fitting the gluonic quasiparticle dispersion relation to the gluelump spectrum (rather than obtaining it by minimizing the vacuum energy)\cite{Guo:2007sm}.  The result and lattice data are shown in table \ref{tab:lattMod}.
 
\begin{table}[ht]\centering
\begin{tabular}{l|lllllll}
\hline\hline
$J^P$ & $1^+$ & $1^-$ & $2^-$ & $2^+$ & $3^+$ & $0^+$ & $4^-$ \\
\hline
model (GeV) &  0.87 & 1.27 & 1.47 & 1.59 & 1.99 &-- &  2.49 \\
lattice (GeV) & 0.87(15) & 1.25(16) & 1.45(17) & 1.86(19) & 1.86(18) & 1.98(18) & 2.13(18) \\
\hline\hline
\end{tabular}
\caption{Lattice and Constituent Gluon Model Gluelump Spectra.}
\label{tab:lattMod}
\end{table}

The modified Coulomb gauge model was also employed in a computation of the heavy hybrid spectrum\cite{Guo:2008yz}. General agreement with lattice results were found, although the predictions appear to be systematically high. The $1^{-+}$ charmonium hybrid was predicted to lie at 4.47 GeV.

\subsection{Sum Rule and Schwinger-Dyson Calculations}

QCD sum rules have been used since their inception for the study of nonperturbative properties of QCD\cite{sumrules}. Their application to hybrid mesons dates to the beginnings of the field. A collection of results for the light quantum number exotic hybrid are presented in table \ref{tab:QCDSR}. Evidently mass estimates have been climbing over the years, with more recent results tending to agree with lattice computations.

\begin{table}[h]\centering
\begin{tabular}{llll}
\hline\hline
$1^{-+}$ mass (GeV) & Ref  & year\\
\hline
1.3 & \cite{g-tab} & 1983 \\
$\sim 1$ & \cite{b-tab} & 1986 \\
1.7(1) & \cite{l-tab} & 1987 \\
1.65(5) & \cite{c-tab} & 2000 \\
1.81(6) & \cite{n-tab} & 2009 \\
\hline\hline
\end{tabular}
\caption{Sum Rule Hybrid Mass Estimates}
\label{tab:QCDSR}
\end{table}

Sum rules have also been applied to hybrid baryons in Ref. \cite{kiss}. The authors considered the possibility of a hybrid baryon as the first excited state above the nucleon with $N_{1/2+}$ quantum numbers and concluded that such a state should have a mass near 1.5 GeV. This is 1 GeV lighter than current lattice results.

Finally, exploratory computations of hybrid masses have been carried out with the Schwinger-Dyson and Bethe-Salpeter formalism\cite{SD-hyb-1},\cite{SD-hyb-2}. Although the authors solve a two-body Bethe-Salpeter equation, exotic quantum numbers are accessible because the formalism is relativistic and the Bethe-Salpeter kernel can contain gluonic degrees of freedom.  Of course the dynamics and eigenvalues of hybrids then depends crucially on the assumed form of the scattering kernel. In this case, a separable kernel Ansatz of the form

\begin{equation}
K(p,q) = -\frac{4}{3} g^2 \gamma_\mu[ G(p^2)G(q^2) + p\cdot q \, F(p^2)F(q^2)]\gamma^\mu
\end{equation}
was used. The functions $F$ and $G$ are related to the full quark propagator via the Schwinger-Dyson equations for the gluon propagator. Results for a selection of states are presented in table \ref{tab:BS-hyb} along with possible resonance assignments suggested by the authors. The authors point out that the $1^{-+}$ states carry negative norm, which is normally regarded as a sign of a spurious state\cite{nakanishi}, although they argue that this is not enough to dismiss the states as physical.

\begin{table}[h]\centering
\begin{tabular}{lll}
\hline\hline
$J^{PC}$ & mass (GeV) & assignment \\
\hline
$0^{++}$ & 749 & $\sigma(540)$, $a_0(980)$ \\
$0^{+-}$ & 1082 & exotic  \\  
$0^{--}$ & 1319 & exotic  \\ 
\hline
$1^{--}$ & 730 & $\rho(770)$ \\
$1^{+-}$ & 1244 & $h_1(1170)$ \\
$1^{++}$ & 1337 & $a_1(1260)$ \\
$1^{-+}$ & 1439, 1487 & $\pi_1(1400)$, $\pi_1(1600)$ \\
\hline\hline
\end{tabular}
\caption{Bethe-Salpeter Hybrid Masses}
\label{tab:BS-hyb}
\end{table}

\subsection{Model Comparison}

It is instructive to confront the models of the past section with the lattice data for the gluelump spectrum, adiabatic surfaces, and hybrid masses.

With regards to the general features of the adiabatic surfaces, it is possible for spherical bag models to reproduce the lattice calculations at small quark separation, but they fail at large $R$. Furthermore, flux tube model or Nambu-Goto string models reproduce the lattice reasonably well for intermediate to large quark separations, but do not perform well at small $R$ or in detail at large $R$.

The flux tube model fits the first excited state, $\Pi_u$, quite well over a wide range of the quark separation (one must ignore the Coulomb term present in Eqn. 25 of Ref. \cite{IP-1}\cite{IP-2}). It is, however, the only surface to do so at small distance. Furthermore,
this may be a fluke due to the particular choice of the short distance cutoff of the 
$\pi/R$ term employed in Ref.~\cite{IP-1},\cite{IP-2}. 

Of course, $\pi/R$ splittings are expected to be universal behavior of string systems. 
Juge, Kuti, and Morningstar have carried out a detailed analysis of the relationship of the hybrid surfaces of Fig. \ref{VRplot}  to string excitations.  They have found that $\pi/R$ splittings are manifest only for very large source separation
(roughly 4 fermi or greater).  This is something of a surprise since one
expects a phonon-like excitation spectrum on general grounds. 

The lightest flux tube multiplet: $(0,1,2)^{-+}$, $1^{--}$, $(0,1,2)^{+-}$, and $1^{++}$ is larger than that observed in lattice computations. Taken together, it appears that the flux tube model is not an effective model of gluonic excitations. Alternatively, bag models that have a lowest energy TE mode yield multiplets in agreement with the lattice. Yet these models suffer from internal consistency issues (center of mass motion, bag distortion, double counting degrees of freedom) that render their detailed predictions inaccurate or suspect.

Simply achieving the quantum numbers of the adiabatic potentials is not possible in many models.  Thus, for example, models that employ single spinless gluons
do not contain sufficient degrees of freedom to reproduce the adiabatic potential spectrum.
In particular, transverse ``single bead" flux tube models (Ref. \cite{BCS}) cannot make $\Pi_g$ or $\Sigma_u$ states while three dimensional ``bead" models (Ref. \cite{HM}) cannot make $\Sigma^+$ states. Thus including gluon spin is a minimal necessity in this class of models (although the level ordering problem must be overcome as discussed above). Furthermore, including this spin is not sufficient to obtain the correct level ordering\cite{SS2}, as it appears that three-body interactions are required, at least in some models\cite{Szczepaniak:2006nx}.
Alternatively, models with constituent gluons with $J^P=1^-$ in relative S-waves with $q\bar q$ systems fail to describe the observed lattice states. However, if the dynamics favors a P-wave coupling then the lowest hybrid multiplet can be explained if the $q\bar q$ system is in an S-wave and the heavier lattice multiplet can be explained if the quark pair is in a P-wave. If this promise is fulfilled, constituent glue models must still strive to incorporate the effects of multiple gluons, which are presumably required to obtain string-like large distance behavior\cite{chain}.

\section{Hybrid Decay Models}
\label{dec}

Hybrid decay mechanisms were considered from the beginnings of the development of hybrid models. Tanimoto categorized these in terms of three possibilities\cite{bagDecay}: (i) decays of the type {\tt hybrid} $\to$  {\tt hybrid} + {\tt meson}; these can occur via a mechanism for quark pair production, such as the $^3P_0$ decay model\cite{3p0-1},\cite{3p0-2}; (ii) {\tt hybrid} $\to$ {\tt hybrid} + {\tt meson} $\to$ {\tt meson} + {\tt meson}, a two-step process in which the virtual hybrid undergoes a gluonic de-excitation to a (conventional) meson state; (iii) {\tt hybrid} $\to$ {\tt meson} + {\tt meson}, where the valence gluon produces a quark pair.

An important early observation was that a $1^{-+}$ hybrid cannot decay to identical S-wave mesons\cite{bagDecay,conDecay,conDecay-2}. This result was obtained in the non-relativistic limit employing a decay vertex of the form $\bar \psi {\rlap A}{\,/} \psi$.  Early computations of this sort found that TM hybrids tended to have typical hadronic widths, while TE hybrids tend to be narrow\cite{bagDecay}.

\subsection{Flux Tube Decay Model}
\label{sec:ftm}

The difficulties bag models had in predicting detailed properties of the conventional mesons led to a waning of interest in them and an increasing reliance on flux tube models. In fact, 
shortly after its introduction, the
flux tube model of meson structure was extended by Isgur, Kokoski, and Paton (IKP) to
provide a description of meson\cite{IK} and hybrid\cite{IKP} decays. The 
transition operator was envisioned as arising due to the quark hopping term of the
lattice QCD Hamiltonian. The lowest terms in the expansion of this operator are 

\begin{equation}
H_{hop} = \sum_{n,\mu} \psi_n^\dagger \alpha\cdot \mu \psi_n  + a \sum_{n,\mu}\psi_n^\dagger\alpha\cdot\mu\nabla\cdot\mu \psi_n,
\end{equation}
where $\mu$ is a unit vector along the string.
If one assumes a smooth string then the first term dominates as the lattice spacing gets
small and one has a $^3S_1$ strong decay operator. Alternatively, if the string is rough
then the first term averages to zero upon summing over all local string orientations
and the second term dominates, yielding a  $^3P_0$ strong decay operator. The authors
of Ref. \cite{IK,IKP} assumed the second scenario since it has a long history of phenomenological success.

Flux tube degrees of freedom were incorporated by assuming factorization:

\begin{equation}
\langle \{\ldots\} b d; \{\ldots\} b d | {\cal O} | \{\ldots\} b^\dagger d^\dagger\rangle \approx \langle bd; bd| ^3P_0 | b^\dagger d^\dagger\rangle \cdot
\langle \{\ldots\}; \{\ldots\} | \{\ldots\}\rangle.
\end{equation}
Here $\{\ldots\}$ refers to the collection of (phonon) quantum numbers required to describe the flux tube and ${\cal O}$ is a transition operator.
The first matrix element on the right hand side is a typical $^3P_0$ mesonic
decay overlap. The second represents the overlap of the gluonic/flux tube degrees
of freedom. Assuming that the quark pair creation occurs at a transverse distance
$y_\perp$ from the interquark axis of the parent meson yields the results

\begin{equation}
\langle \{0 \ldots \};\{0 \ldots \}| \{ 0 \ldots \}\rangle \sim {\rm e}^{-f b y_\perp^2}
\end{equation}
for meson decay and
\begin{equation}
\langle \{0 \ldots \};\{0 \ldots \}| \{ 1 \ldots \}\rangle \sim y_\perp{\rm e}^{-f b y_\perp^2}
\end{equation}
for hybrid decay. The factor  $f$ is a computable constant of order unity.  The extra 
factor of $y_\perp$ in the hybrid decay vertex forces the decay to pairs of identical 
$S$-wave mesons to be zero. This gives rise to a selection rule that states that hybrids cannot decay to meson with identical S-wave spatial wavefunctions.  This is because the string angular momentum cannot be absorbed by the relative coordinate between the final state mesons and therefore must go into a $q\bar q$ angular momentum. We stress that this result hinges on the assumed factorization of the quark pair production operator from the gluonic degrees of freedom. A more general discussion of selection rules in hybrid decays is contained in Ref. \cite{page}

The observation that flux tube motion can affect the center of mass of the quarks that was mentioned in section \ref{sect:fd} was 
expanded upon by Close and Dudek who commented that radiative decays of hybrid mesons can proceed because the recoil of the radiating quark affects the string degrees of freedom giving a nonzero overlap of the flux tube wavefunction  with the ground state flux tube wavefunctions of ordinary mesons\cite{CD}.

A similar scheme involving the emission of pointlike pions may be used to compute hybrid
decays to final states such as $\pi\rho$\cite{CD2}. The most striking result here is that
this decay mechanism evades the selection rule  discussed above.

\subsection{\label{sec:pss}The PSS Hybrid Decay Model}

An alternative hybrid decay model has been developed in which the gauge field in the interaction $-g\int \bar \psi {\bm \alpha} \cdot A\psi$ was mapped to phonon degrees of freedom\cite{PSS-1},\cite{PSS-2}. This model is quite different from that of IKP because it explicitly correlates quark pair production with gluonic degrees of freedom. The resulting decay vertex is given by

\begin{equation}
H_{int} = \frac{i g a^2}{\sqrt{\pi}}  \sum_{m,\lambda} \int_0^1 d \xi
\cos(\pi \xi)  T^a_{ij}\, b^\dagger_i(\xi {\bf r}_{Q\bar Q})
{\bm\sigma}\cdot \hat{\bf e}_{\lambda}(\hat{\bf r}_{Q\bar Q}) \left(\alpha^a_{m \lambda}
 - \alpha^{a \dagger}_{m \lambda} \right) d^\dagger_j(\xi {\bf r}_{Q\bar Q}),
\end{equation}

\noindent
where the $\hat{\bf e}(\hat{\bf r})$ are polarization vectors orthogonal to $\hat {\bf r}$.
The $q\bar q$ creation occurs on a line joining
the original $Q\bar Q$ quarks, smeared over the transverse size of the flux
tube.  The spin operator contracts with the flux tube
phonon polarization vector, which is absent in the IKP model.
Finally, the decay amplitude vanishes when the final mesons are identical
due to the nodal structure in the vector potential. 
This is true for any single-phonon hybrid in an odd mode.
Thus one obtains the selection rule:
low-lying hybrids do not decay to identical mesons. This subsumes the 
selection rule of IKP so that none of their qualitative conclusions are 
changed. However the model predicts, for example, that hybrids do not decay
to pairs of identical P-wave mesons. 

Another rule, the ``spin selection" rule, exists: if the $q\bar{q}$ in either hybrid or 
conventional mesons
are in a net spin singlet configuration then decay into final states consisting only of spin singlet
states is forbidden. This rule follows because pair creation is spin-triplet and appears to be a universal feature in all non-relativistic decay models.

Notice that the spin selection rule can be used to distinguish conventional and hybrid vector mesons.  This is because conventional vector mesons have ${}^{(2S+1)}L_J = {}^3S_1$ or ${}^3D_1$ while hybrid vectors have their quarks coupled to a spin singlet. Thus, for example, 
the decay of hybrid $\rho_H$, is forbidden to 
$\pi h_1$  whereas $\pi a_1$ is allowed; this
is the reverse of the case of $^3S_1$ conventional mesons where the
$\pi a_1$ channel is relatively suppressed and $\pi h_1$
is allowed.

Finally, the IKP and PSS decay models can be compared to the UKQCD lattice results for $\pi_1$ decay with the modification introduced by Burns and Close (discussed in section \ref{sect:lht}). The model widths reported in Table \ref{tab:hlc} are obtained from Ref. \cite{PSS-1},\cite{PSS-2}.

\begin{table}[ht]\centering
\begin{tabular}{llll}
\hline\hline
mode & $\Gamma_{\rm PSS}$ (MeV)  & $\Gamma_{\rm IKP}$ (MeV) & $\Gamma_{\rm lattice}$ (MeV) \\
\hline
$\pi_1 \to f_1 \pi$ & 10-18 & 14 & 25 \\
$\pi_1 \to b_1 \pi$ & 40-78 & 51 & 80 \\
\hline\hline
\end{tabular}
\caption{Comparison of lattice and model hybrid decay modes.}
\label{tab:hlc}
\end{table}


\subsection{Hybrid Photocoupling and Production}

Model hybrid decay computations focus on OZI allowed decays, and thus charmonium or bottomonium decays are to open flavor final state mesons. In the case of light flavors, an OZI allowed decay to a final state such as $\pi\rho$ will permit an estimate of the radiative transition to $\pi\gamma$ via the vector meson dominance model.  The phenomenology of this case was examined in Ref. \cite{Close:1994pr}; the authors noted that hybrid photoproduction could be significant in the case of $\omega$ or $\pi$ exchange because the incoming $\rho$ is replaced by a photon and because the exchanged particles are off-shell.
 
The cross section for deep exclusive electroproduction of an exotic $1^{-+}$ hybrid was estimated in the Bjorken regime by Anikin \emph{et al.} \cite{pire}. This cross section was found to be large, and in fact scales like $1/Q^2$, as is usual for meson electroproduction. The authors also noted that forward-backward asymmetry in the production of $\pi$ and $\eta$ mesons can serve as a useful signal for hybrid meson production.

We are aware of one attempt to compute the hidden flavor decay of hybrid mesons\cite{Guo:2014zva}. The model assumed hybrids constructed with constituent gluons such that the TE mode is the lowest lying. The electromagnetic transition was modelled in bound state perturbation theory with the valence gluon being absorbed by quarks or by the Coulomb interaction with photon emission from a quark line. The authors were not able to predict absolute rates so that comparison to the lattice results of section \ref{sect:hrt} is not possible. The rates for all the low lying TE hybrid multiplet transitions were found to be comparable.

The utility of polarized photoproduction of hybrid mesons was analyzed in Ref. \cite{AS}.
The authors examined the process $\vec \gamma p \to \rho^0 \pi^+ n$ at low momentum transfer via pion exchange. They assumed a simple form for the photon-pion-hybrid vertex and fixed the ratio of partial waves according to the vector meson dominance model. The computed cross sections  indicated that corrections due to absorption are required and that linear photon polarization permits separating the exotic wave, even without partial wave analysis.

\section{Experimental Situation}
Experimental data on exotic hybrid mesons goes back nearly three decades, and continues
until today. An extensive review on spectroscopy in general can be found in reference~\cite{klempt-07},
while a review focussing only on hybrid mesons is given in reference~\cite{Meyer:2010ku}. The
experimental situation has evolved since these previous reviews, with new data from several 
experiments and detailed systematic studies of production mechanisms made possible
by very-large data sets. To date, there is experimental evidence for three exotic quantum number
states~\cite{pdg12}. All have the same quantum numbers, $I^{G} J^{PC}=1^{-}1^{-+}$, the $\pi_{1}(1400)$,
$\pi_{1}(1600)$ and $\pi_{1}(2015)$, and there are experimental and interpretational issues
surrounding all of them. The field is now on the cusp of a new experiment to study the 
photoproduction of hybrids, and a few years away from a new antiproton experiment. Thus
it seems timely to take stock of the current situation.

In order to search for exotic hybrids, it is first necessary to catalog the possible decay modes.
These are given in Table~\ref{tab:exotic-hybrid-widths} where the widths come from model
calculations (see Section~\ref{sec:pss}). The decay modes are obtained by simple quantum
number counting arguments. The list is not exhaustive, but covers the most likely and most
detectable modes.
\begin{table}[ht]\centering
\begin{tabular}{ccccc}\hline\hline
  &  &  &  & \\ 
Name & $J^{PC}$ & \multicolumn{2}{c}{Total Width MeV} & 
Allowed Decay Modes \\ 
         &                   & PSS & IKP & \\ \hline
$\pi_{1}$   & $1^{-+}$ &  $81-168$ & $117$ & 
$b_{1}\pi$, $\pi\rho$, $\pi f_{1}$, $\pi\eta$, $\pi\eta^{\prime}$, $\eta a_{1}$, $\pi \eta(1295)$ \\
$\eta_{1}$  & $1^{-+}$ &  $59-158$ & $107$ &
$\pi a_{1}$, $\pi a_{2}$, $\eta f_{1}$, $\eta f_{2}$, $\pi \pi(1300)$, $\eta\eta^{\prime}$,  
$K K_{1}^{A}$, $K K_{1}^{B}$ \\
$\eta^{\prime}_{1}$ & $1^{-+}$ &  $95-216$ & $172$ &
$K K_{1}^{B}$, $K K_{1}^{A}$, $K K^{*}$, $\eta\eta^{\prime}$ \\ \hline
$b_{0}$     & $0^{+-}$ & $247-429$ & $665$ &
$\pi \pi(1300)$, $\pi h_{1}$, $\rho f_{1}$, $\eta b_{1}$ \\
$h_{0}$     & $0^{+-}$ & $59-262$  & $94$  &
$\pi b_{1}$, $\eta h_{1}$, $K K(1460)$ \\
$h^{\prime}_{0}$    & $0^{+-}$ & $259-490$ & $426$ &
$K K(1460)$, $K K_{1}^{A}$, $\eta h_{1}$ \\ \hline
$b_{2}$     & $2^{+-}$ &    $5-11$ & $248$ &
$\pi a_{1}$, $\pi a_{2}$, $\pi h_{1}$, $\eta\rho$, $\eta b_{1}$, $\rho f_{1}$ \\
$h_{2}$     & $2^{+-}$ &    $4-12$ & $166$ &
$\pi\rho$, $\pi b_{1}$, $\eta\omega$, $\omega b_{1}$  \\
$h^{\prime}_{2}$    & $2^{+-}$ &    $5-18$ &  $79$ &
$K K_{1}^{B}$, $K K_{1}^{A}$, $K K^{*}_{2}$, $\eta h_{1}$ \\ 
\hline\hline 
\end{tabular}
\caption[Hybrid Widths]{\label{tab:exotic-hybrid-widths}
Exotic quantum number hybrid widths and possible  decay modes~\cite{PSS-1}.}
\end{table}

\subsection{Exotic Quantum Number states}
\begin{figure}[h!]\centering
\includegraphics[width=0.25\textwidth]{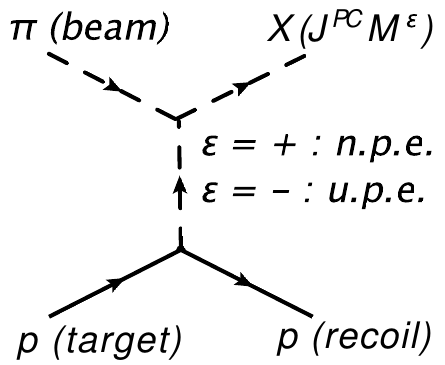}
\caption[]{\label{fig:t-channel}The diffractive process with an incident pion ($\pi$ beam) and
a proton ($p$) target. The exchange has $z$-component on angular momentum $M$ and reflectivity 
$\epsilon$. The recoil proton and a state $X$ of given $J^{PC}$ constitute the final state. For positive 
reflectivity, the $t$-channel is a natural parity exchange, while for negative reflectivity, it is 
unnatural parity exchange.}
\end{figure}

Data on exotic-quantum-number mesons have come from both diffractive production using
incident pion beams, antiproton annihilation on protons and neutrons and charmonium decays. 
Diffractive production is schematically shown in Fig.~\ref{fig:t-channel}. A pion beam is incident 
on a proton (or nuclear) target, which recoils following a $t$-channel exchange.  
The process can be written down in the reflectivity basis~\cite{Chung:1974fq} where the
production factorizes into two non-interfering amplitudes---positive reflectivity ($\epsilon=+$)
and negative reflectivity ($\epsilon=-$). The absolute value of the spin projection along the 
$z$-axis is $M$, and is taken to be either $0$ or $1$ (it is usually assumed that contributions from
$M$ larger than $1$ are small and can be ignored~\cite{chung-99}). It can be shown in this process that 
naturality of the exchanged particle can be determined by $\epsilon$. Natural parity 
exchange (n.p.e.) corresponds to $J^{P}$s of $0^{+}$, $1^{-}$, $2^{+}$, $\cdots$, while unnatural 
parity exchange (u.p.e.) corresponds to $J^{P}$ of $0^{-}$, $1^{+}$, $2^{-}$, $\cdots$.
For a state which is observed in more than one decay mode, one would expect that the production
mechanism ($M^{\epsilon}$) would be the same for all decay modes. If not, this could be indicative 
of more than one state being observed, or possible analysis problems.

In antiproton-nucleon annihilation, there are a number of differences between various annihilation
processes. For the case of $\bar{p}p$, the initial state is a mixture of isospin $I=0$ and $I=1$. For
$\bar{p}n$ annihilation, the initial state is pure $I=1$. For annihilation at rest on protons, the 
initial state is dominated by atomic S-waves. In particular, $^{1}S_{0}$ and $^{3}S_{1}$ atomic states, 
which have $J^{PC}=0^{-+}$ and $1^{--}$ respectively (with a small admixture of $P$ states). For 
annihilation in flight, the number of initial states is much larger and it may no longer make 
sense to try to parametrize the initial system in terms of atomic $\bar{p}p$ states. 

The combination of initial isospin and final state particles may lead to additional selection 
rules that restrict the allowed initial states. In the  case of $\bar{p}p\rightarrow \eta\pi^{0}\pi^{0}$, 
the annihilation is dominated by $^{1}S_{0}$ initial states ($J^{PC}=0^{-+}$). For the case of 
$\bar{p}n\rightarrow \eta\pi^{0}\pi^{-}$, quantum numbers restrict this annihilation to occur from 
the $^{3}S_{1}$ initial states ($J^{PC}=1^{--}$). In addition, the neutron is bound in deuterium, where
the Fermi motion introduces substantial p-wave annihilation. Thus, one may see quite different 
final states from the two apparently similar reactions. The results to date came from the Crystal 
Barrel experiment at CERN. New results are expected from the PANDA experiment at FAIR sometime
after 2020.

A limited number of results on the photoproduction of exotic mesons have been published,
with no reported signals to date. Extensive new results are expected with the advent  of the 
GlueX experiment at Jefferson Lab in 2015, and the CLAS-12 experiment two years later.

The decays of $\chi_{c}$ mesons should also be a good place to search for hybrid mesons. 
Results from CLEO-c report interesting results in the $\eta^{\prime}\pi$ system. More 
extensive results with much higher statistics should be available from BES-III.

\subsection{\label{sec:p1_1400}The $\pi_{1}(1400)$}
\subsubsection{GAMS results on the $\pi_{1}(1400)$}
The GAMS experiment made the first reported observation of an exotic quantum number state,
using a $40$~GeV/c $\pi^{-}$ beam to study the reaction $\pi^{-}p\rightarrow p \eta \pi^{-}$. They 
reported a $J^{PC}=1^{-+}$ state (the $M(1405)$) in the $\eta\pi^{-}$ system~\cite{alde-88}. 
They reported a mass of $1.405\pm 0.020$~GeV and a width of $0.18\pm 0.02$~GeV for this state.
In the neutral channel, $\eta\pi^{0}$, an earlier search found no evidence for an exotic 
quantum-number state~\cite{appel-81}.  

\subsubsection{KEK results on the $\pi_{1}(1400)$}
Somewhat later, an experiment at KEK using a $6.3$~GeV/c 
$\pi^{-}$ beam observed a $1^{-+}$ state in the $\eta \pi^{-}$ system with a mass of 
$1.3431\pm 0.0046$~GeV and a width of $0.1432\pm 0.0125$~GeV~\cite{aoyagi-93}. Since this mass 
and width is very close to that of the strong signal for the $a_{2}(1320)$, the possibility of signal leakage
into the weak exotic wave from the strong $a_{2}$ signal often comes up.

\subsubsection{VES results on the $\pi_{1}(1400)$}
\begin{figure}[t]\centering
\includegraphics[width=0.75\textwidth]{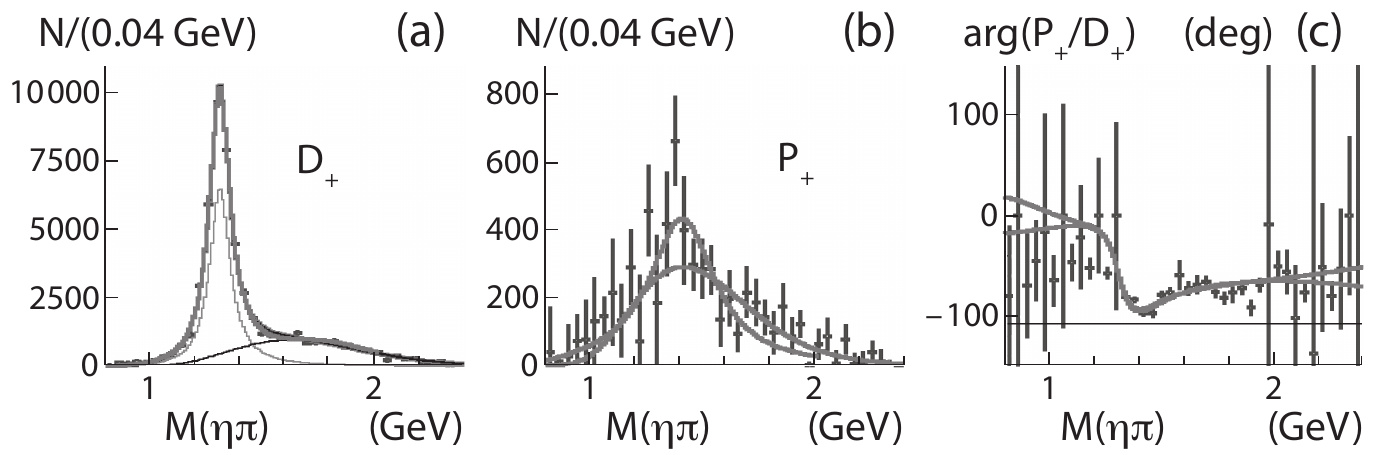}
\caption[]{\label{fig:ves_etapi}The results of a partial-wave analysis of the $\eta\pi^{-}$
final state from VES. (a) shows the intensity in the $2^{++}$ partial wave, (b) shows the 
intensity in the $1^{-+}$ partial wave and (c) shows the relative phase between the waves.
(Figure reproduced from reference~\cite{Amelin:2005ry}.)}
\end{figure}

The VES experiment studied interactions using a 37  GeV/$c$ pion beam. They reported intensity 
in the $1^{-+}$ $\eta\pi^{-}$ exotic wave and saw rapid phase motion between the $1^{-+}$ wave and  
the $a_{2}(1320)$~\cite{Beladidze:93}, as shown in Fig.~\ref{fig:ves_etapi}. 
The exotic wave was present in the $M^{\epsilon}=1^{+}$ (natural parity) exchange, but 
not in the $0^{-}$ and $1^{-}$ (unnatural parity) exchange.  The observed data could be fit using  
$J^{PC}=1^{-+}$ intensity and the phase motion with respect to the $a_{2}(1320)$ using
a Breit-Wigner distribution with a mass of $1.316\pm 0.012$~GeV and width of $0.287\pm 0.025$~GeV.
However, VES stopped short of claiming an exotic resonance, as they could not unambiguously 
establish the nature of the exotic wave~\cite{Dorofeev:02}. In a later analysis of the $\eta\pi^{0}$ 
system, they claimed that the peak near $1.4$~GeV can be understood without requiring an exotic 
quantum number meson~\cite{Amelin:2005ry}.

\subsubsection{E852 results on the $\pi_{1}(1400)$}
\begin{figure}[h]\centering
\includegraphics[width=0.5\textwidth]{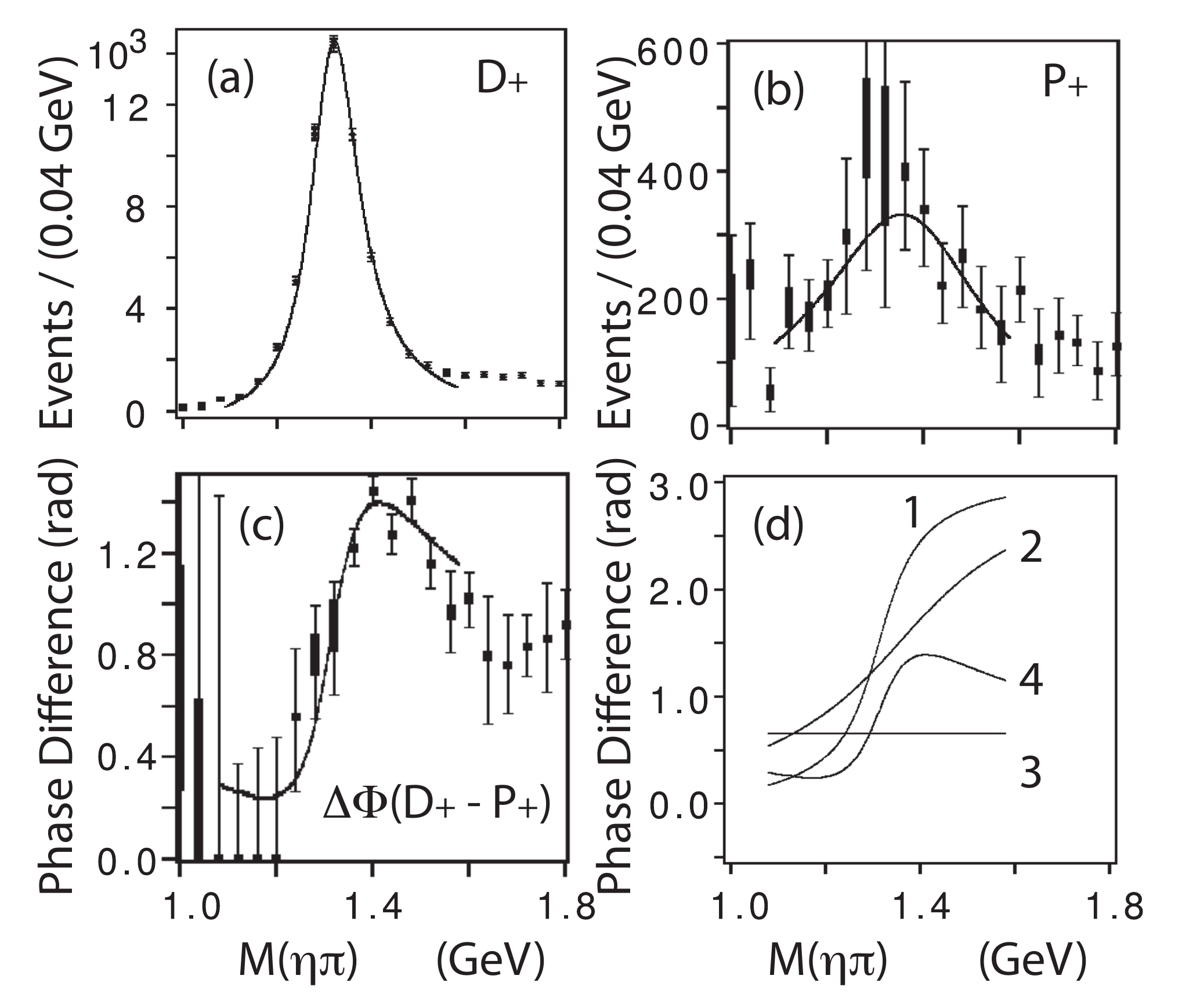}
\caption[]{\label{fig:e852_p1_1400}Data from the E852 experiment showing the 
$\pi_{1}(1400)$~\cite{Thompson:1997bs}. In (a) is shown the intensity of the $J^{PC}=2^{++}$ 
partial wave, dominated by the $a_{2}(1320)$, as a function of $\eta\pi$ mass. In (b) is shown
the intensity of the exotic $1^{-+}$ wave as a function of the $\eta\pi$ mass. Frame (c) shows the 
phase difference between the two partial waves, $2^{++}$ and $1^{-+}$. Frame (d) shows the
evidence for the resonant nature of the $\pi_{1}(1400)$ by looking at the three components that
build the total phase difference. (Figure reproduced with permission from reference~\cite{Thompson:1997bs}.)}
\end{figure}

The E852 collaboration used an $18$~GeV/c $\pi^{-}$ beam to study the reaction 
$\pi^{-}p\rightarrow p \eta \pi^{-}$. They reported the observation of a $1^{-+}$ 
state in the $\eta\pi^{-}$ system~\cite{Thompson:1997bs}. The state was only produced 
in natural parity exchange ($M^{\epsilon}=1^{+}$). Fitting to a Breit-Wigner distribution, they 
found a mass of $1.37\pm 0.016^{+0.050}_{-0.030}$~GeV and a width of 
$0.385\pm 0.040^{+0.065}_{-0.105}$~GeV. The intensity and phase difference plots are shown in 
Fig.~\ref{fig:e852_p1_1400}.  While the observed exotic signal was only a few percent of the 
dominant $a_{2}(1320)$ strength, they noted that its interference with the $a_{2}$ provided clear 
evidence of the existence of the $\pi_{1}(1400)$. This is most clearly shown in plot (d) of the 
figure where curve $1$ is the phase of the $a_{2}$, curve $2$ is the phase of the $\pi_{1}$, $3$
is an assumed flat background phase, and $4$ is the resulting phase difference. When their 
intensity and phase-difference plots were compared with those from VES~\cite{Beladidze:93}, 
they were found to be identical. 

Due to disagreements over the interpretation of the $1^{-+}$ signal, the E852 collaboration
split into two groups. The majority of the collaboration published the resonance interpretation,
$\pi_{1}(1400)$~\cite{Thompson:1997bs}, while a subset of the collaboration did not sign
the paper. As this latter group, centered at Indiana University, continued to analyze data 
collected by E852, we will refer to their publications as E852-IU to distinguish 
the work of the two groups.
An analysis by the E852-IU group of data for the reaction $\pi^{-}p\rightarrow n \eta \pi^{0}$
found evidence for the exotic $1^{-+}$ partial wave, but was unable to describe it as
a Breit-Wigner-like $\pi_{1}(1400)$  $\eta\pi^{0}$ resonance~\cite{dzierba03}. 
However, a later analysis by the E852 collaboration of the same final state and data confirmed 
their earlier observation of the $\pi_{1}(1400)$~\cite{Adams:2006sa}. E852 found a mass of 
$1.257\pm 0.020\pm 0.025$~GeV and a width of $0.354\pm 0.064\pm 0.058$~GeV with 
the $\pi_{1}(1400)$ produced via natural parity exchange ($M^{\epsilon}=1^{+}$).
Much of the discrepancy between these two works arose from the treatment of backgrounds.
The E852 collaboration consider no background phase, and attribute all phase motion to
resonances. The E852-IU group allow for non-resonant interactions in the exotic channel,
these background processes are sufficient to explain the observed phase motion.

\subsubsection{Crystal Barrel results on the $\pi_{1}(1400)$}
\begin{figure}[h!]\centering
\includegraphics[width=0.5\textwidth]{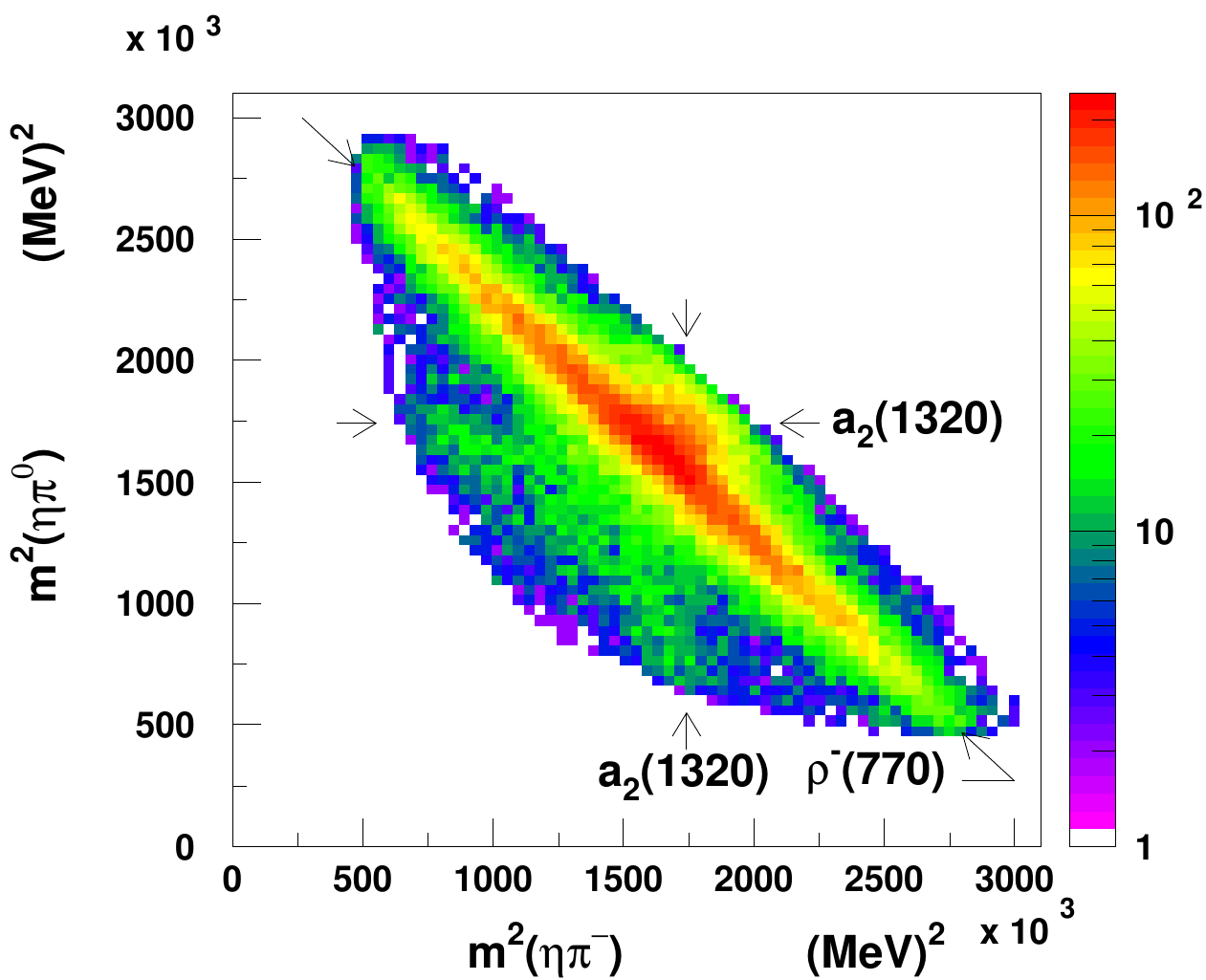}
\caption[]{\label{fig:cbar_p1_1400} (Color on line.) 
The Dalitz plot of $m^{2}(\eta\pi^{0})$ versus $m^{2}(\eta\pi^{-})$
for the reaction $\bar{p}n\rightarrow \eta\pi^{-}\pi^{0}$ from the Crystal Barrel 
experiment~\cite{cbar98}. The bands for the $a_{2}(1320)$ are clearly seen in both $\eta\pi^{0}$ and
$\eta\pi^{-}$, while the $\rho(770)$ is seen in the $\pi^{0}\pi^{-}$ invariant mass.}
\end{figure}

\begin{figure}[h!]\centering
\begin{tabular}{c}
\includegraphics[width=0.5\textwidth]{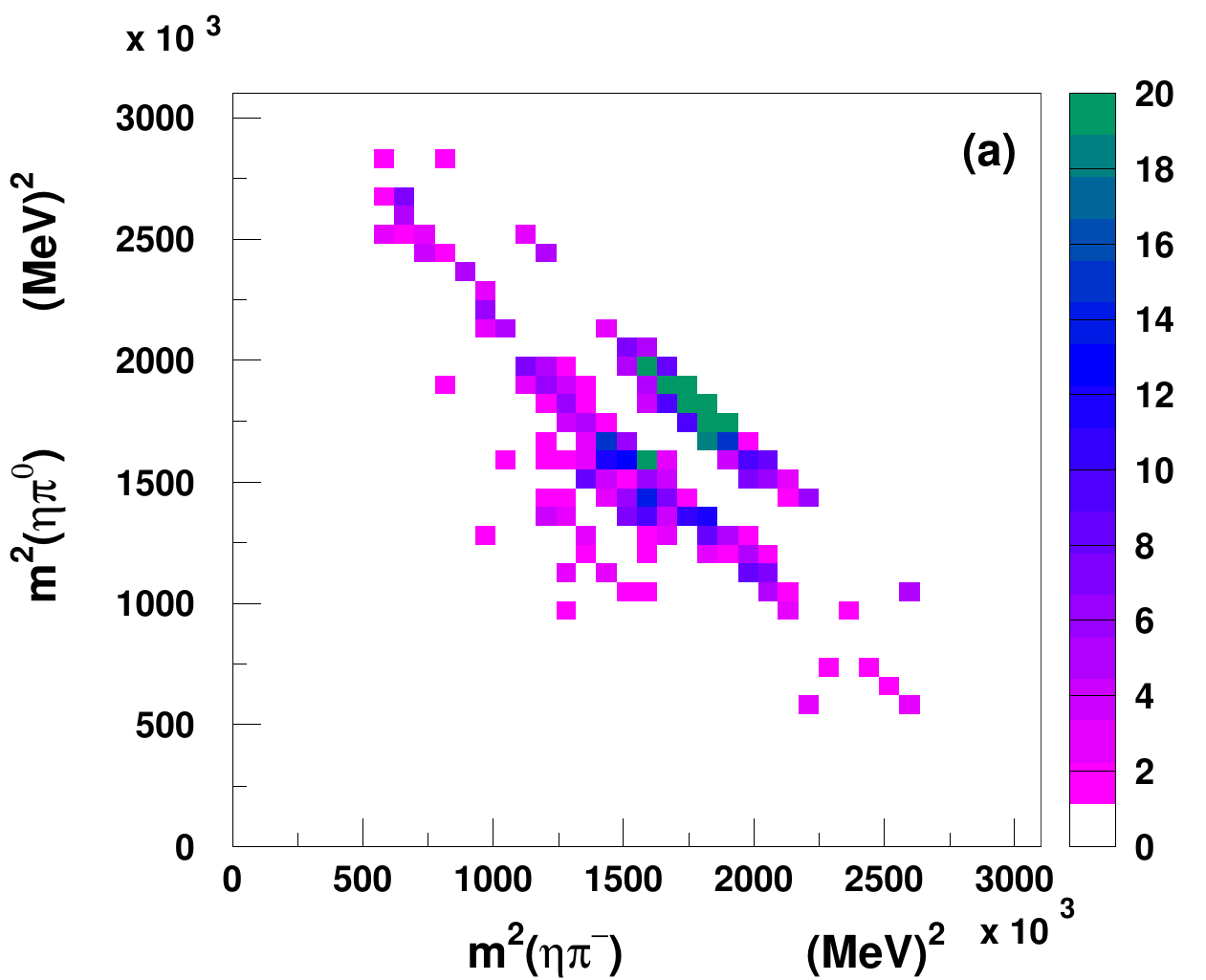} \\
\includegraphics[width=0.5\textwidth]{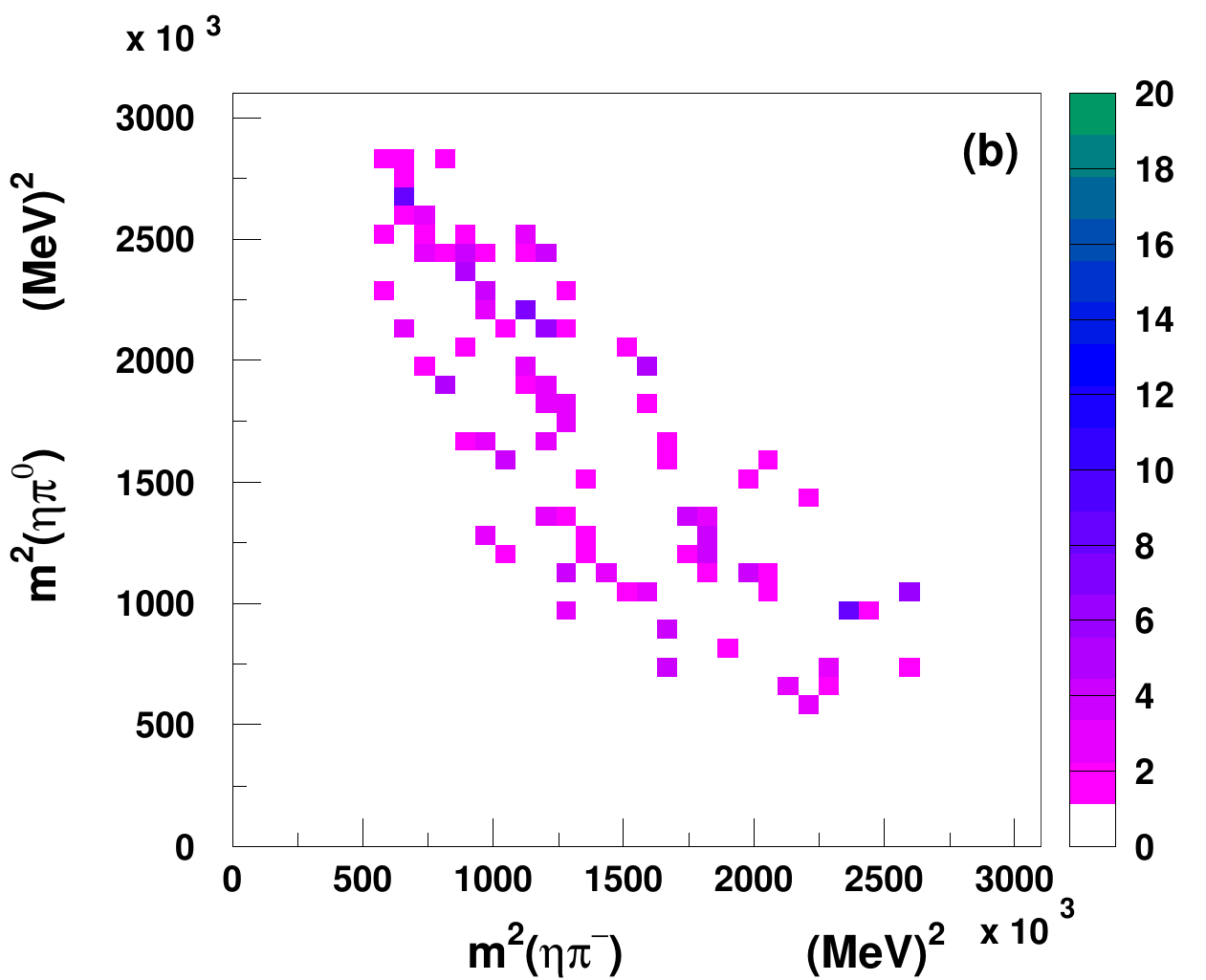} 
\end{tabular}
\caption[]{\label{fig:cbar_chisqr} (Color on line.) 
The difference between the fit and the data in the Dalitz plot 
of $m^{2}(\eta\pi^{0}$ versus $\eta\pi^{-}$ for the reaction $\bar{p}n\rightarrow \eta\pi^{-}\pi^{0}$ 
from the Crystal Barrel experiment~\cite{cbar98}. (a) Does not include the $\pi_{1}(1400)$ while
(b) does include the $\pi_{1}(1400)$. There are clear systematic discrepancies present in (a) that
are not present when the $\pi_{1}(1400)$ is included.}
\end{figure}

The Crystal Barrel Experiment studied antiproton-neutron annihilation at rest in the reaction 
$\bar{p}n\rightarrow \eta \pi^{-} \pi^{0}$~\cite{cbar98}. The Dalitz plot for this final
state is shown in Fig.~\ref{fig:cbar_p1_1400} where bands for the $a_{2}(1320)$ and 
$\rho(770)$ are clearly seen. They reported an exotic $J^{PC}=1^{-+}$ state, the $\pi_{1}(1400)$, 
with a mass of $1.40\pm 0.020 \pm 0.020$~GeV and a width of $0.310\pm 0.050^{+0.050}_{-0.030}$~GeV.
While the signal is not obvious in the Dalitz plot, if one compares the difference between 
a fit to the data without and with the $\pi_{1}(1400)$, a clear discrepancy is seen when
the $\pi_{1}(1400)$ is not included (see Fig.~\ref{fig:cbar_chisqr}). While the $\pi_{1}(1400)$ 
was only a small fraction of the $a_{2}(1320)$ in the E852 measurement~\cite{Thompson:1997bs}, 
Crystal Barrel observed the two states produced with comparable strength. 
The Crystal Barrel group also looked at the reaction $\bar{p}p\rightarrow \eta \pi^{0} \pi^{0}$~\cite{cbar99}. 
Here, a weak signal was observed for the $\pi_{1}(1400)$ (relative to the $a_{2}(1320)$) with 
a mass of $1.360\pm 0.025$~GeV and a width of $0.220\pm 0.090$~GeV. In $I=0$ $\bar{p}p$
annihilations, the $a_{2}(1320)$ is produced strongly from the $^{1}S_{0}$ atomic state. 
However, the $\bar{p}n$ is isospin $1$ and $^{1}S_{0}$ state is forbidden. Thus, the strong $a_{2}$ 
production from $\bar{p}p$ is suppressed in $\bar{p}d$ annihilations---making the $\pi_{1}(1400)$
production appear enhanced relative to the $a_{2}(1320)$ in the latter reaction. 

\subsubsection{OBELIX results on the $\pi_{1}(1400)$}
Both the OBELIX~\cite{Salvini-04} experiment reported on the observation of the $\pi_{1}(1400)$ in 
the reaction $\bar{p}p\rightarrow 4\pi$. They both claimed the observation of the $\pi_{1}(1400)$ 
decaying to $\rho\pi$ final states, however there is some concern about the production mechanism. 
This same result was reported by the Crystal Barrel Experiment~\cite{Duenweber-04} in conference 
only. The $\eta\pi$ signal arises from annihilation from the p-wave initial state, while the signal in 
$\rho\pi$ come from the $^{1}S_{0}$ initial state. Thus, it is unlikely that the exotic state seen 
in $\eta\pi$ and that seen in $\rho\pi$ are the same. The origin of these may not be due to an exotic 
resonance, but rather some re-scattering mechanism that has not been properly accounted for.

\subsubsection{COMPASS results on the $\pi_{1}(1400)$}
The COMPASS experiment used a 191 GeV/$c$ $\pi^{-}$ beam to study the production of both the 
$\eta\pi$ and $\eta^{\prime}\pi$ final states~\cite{Adolph-2014}. They note that in the $\eta\pi$ system,
the exotic $1^{-+}$ wave ``shows a compact peak of 400 MeV/$c^{2}$ width, centered at a mass 
of 1.4 GeV/$c^{2}$''. They make no comments on the resonant nature of this structure.  

\subsubsection{Interpretation of the $\pi_{1}(1400)$}
Interpretation of the $\pi_{1}(1400)$ has been problematic. Its mass is lower than most
predicted values for hybrid mesons, and its observation in only a single decay mode ($\eta\pi$) is not
consistent with models of hybrid decays. Donnachie and Page showed that the $\pi_{1}(1400)$
could be an artifact of the production dynamics. They demonstrated that is possible to 
understand the $\pi_{1}(1400)$ peak as a consequence of the $\pi_{1}(1600)$ (see 
Section~\ref{sec:pi1_1600}) interfering with a non-resonant Deck-type background with 
an appropriate relative phase~\cite{Donnachie:1998ya}.
Zhang~\cite{Zhang:2001sb} considered a molecular picture where the $\pi_{1}(1400)$ was 
an $\eta(1295)\pi$ molecule. However, the predicted decays were inconsistent with
the observations of the $\pi_{1}(1400)$.

Szczepaniak \emph{et al.}~\cite{Szczepaniak:2003vg} considered a model in which $t$-channel forces
could give rise to a background amplitude which could be responsible for the observed
$\pi_{1}(1400)$. In their model, meson-meson interactions which respected chiral symmetry
were used to construct the $\eta\pi$ $p$-wave interaction much like the $\pi\pi$ $s$-wave
interaction gives rise to the $\sigma$ meson. They claimed that the $\pi_{1}(1400)$ 
was not a QCD bound state, but rather dynamically generated by meson exchange forces. 

Close and Lipkin noted that because the SU(3) multiplets to which a hybrid 
and a multiquark state belong are different,  $\eta\pi$ and $\eta^{\prime}\pi$ decays
might be a good way to distinguish them. They found that for a multiquark state, the 
$\eta\pi$ decay should be larger than $\eta^{\prime}\pi$, while the reverse is true for a
hybrid meson~\cite{Close:1987aw}. A similar observation was made by Chung~\cite{Chung:2002fz}
who noted that in the limit of the $\eta$ being a pure octet state,
the decay of an octet $1^{-+}$ state to an $\eta\pi$ $p$-wave is forbidden. Such a 
decay can only come from a decuplet state. Given that the pseudoscalar mixing angle for
the  $\eta$ and $\eta^{\prime}$ are close to this assumption, they argue that the $\pi_{1}(1400)$
is $qq\bar{q}\bar{q}$ in nature. 

While the interpretation of the $\pi_{1}(1400)$ is not clear, most analyses agree that
there is intensity in the $1^{-+}$ wave near this mass. A summary of all reported 
masses and widths for the $\pi_{1}(1400)$ is given in Table~\ref{tab:pi1400}. All 
are reasonably consistent, and even the null observations of VES and E852-IU all concur 
that there is strength near $1.4$~GeV in the $J^{PC}$ exotic wave.  However, the E852 and 
VES results can be explained as either non-resonant background~\cite{Szczepaniak:2003vg}, 
or non-resonant Deck amplitudes~\cite{Donnachie:1998ya}. Another possibility is the opening 
of meson-meson thresholds, such as $f_{1}(1285)\pi$. Unfortunately, no comparisons of these 
hypothesis have been made with the $\bar{p}N$ data (owing to the lack of general availability of
 the data sets), so it is not possible to conclude that they would also explain those data. 
\begin{table}[h!]\centering
\begin{tabular}{crrcc} \hline\hline
Mode & Mass (GeV) & Width (GeV) & Experiment & Reference \\ 
\hline
$\eta\pi^{-}$  & $1.405\pm 0.020$ & $0.18\pm 0.02$ & GAMS & \cite{alde-88} \\
$\eta\pi^{-}$  & $1.343\pm 0.0046$ & $0.1432\pm 0.0125$ & KEK & \cite{aoyagi-93} \\
$\eta\pi^{-}$ & $1.37\pm 0.016$ & $0.385\pm 0.040$ & E852 &\cite{Thompson:1997bs} \\
$\eta\pi^{0}$ & $1.257\pm0.020$ &$0.354\pm 0.064$ & E852 &\cite{Adams:2006sa} \\
$\eta\pi$      & $1.40\pm 0.020$ & $0.310\pm 0.050$ & CBAR & ~\cite{cbar98} \\
$\eta\pi^{0}$  & $1.36\pm 0.025$ & $0.220\pm 0.090$ & CBAR & ~\cite{cbar99} \\
$\rho\pi$      & $1.384\pm 0.028$ & $0.378\pm 0.058$ & Obelix & \cite{Salvini-04} \\
$\rho\pi$      & $\sim 1.4$ & $\sim 0.4$ & CBAR & \cite{Duenweber-04} \\
$\eta\pi$      & $1.354\pm 0.025$ & $0.330\pm 0.035$ & PDG & \cite{pdg12} \\
\hline\hline
\end{tabular}
\caption[]{\label{tab:pi1400}Reported masses and widths of the $\pi_{1}(1400)$ from the
GAMS, KEK, E852, Crystal Barrel (CBAR) and Obelix experiment. Also reported is the 2014 
PDG average for the state. Not reported here is the COMPASS observation of intensity in
the $1^{-+}$ partial wave near a mass of 1400 MeV\cite{Adolph-2014}.}
\end{table}

\subsection{\label{sec:pi1_1600}The $\pi_{1}(1600)$}
While the issue of only one decay mode ($\eta\pi$)  and the low mass have made interpretation
of $\pi_{1}(1400)$ nature difficult, a second $J^{PC}=1^{-+}$ state has a much richer set of
observations, both in production and decay, and in statistics. This other state, the $\pi_{1}(1600)$,
has been observed in diffractive production using incident $\pi^{-}$ beams where its mass and 
width have been reasonably stable over several experiments and  observed decay modes. There 
is some evidence in $\bar{p}p$ annihilation as well as $\chi_{c1}$ decays. These positive results have 
been reported from VES, E852, COMPASS, Crystal Barrel, CLEO-c and others, and are discussed below. 
Negative results have been reported in photoproduction experiments by CLAS.

\subsubsection{VES Results on the $\pi_{1}(1600)$}
In addition to their study of the $\eta\pi^{-}$ system, the VES collaboration also examined the 
$\eta^{\prime}\pi^{-}$ system. Here they observed a $J^{PC}=1^{-+}$ partial wave with intensity 
peaking at a higher mass than the $\pi_{1}(1400)$~\cite{Beladidze:93}.
However, as with the $\eta\pi^{-}$ system, they did not claim the discovery of an 
exotic-quantum-number resonance. VES later reported a combined study of the $\eta^{\prime}\pi^{-}$, 
$f_{1}\pi^{-}$ and $\rho^{0}\pi^{-}$ final states~\cite{Gouz:1992fu}, and reported a 
``resonance-like structure'' with a mass of $1.62\pm 0.02$~GeV and a width of 
$0.24\pm 0.05$~GeV decaying into $\rho^{0}\pi^{-}$. They also noted that the wave with 
$J^{PC}=1^{-+}$ dominates in the $\eta^{\prime}\pi^{-}$ final state, peaking near $1.6$~GeV and 
observed a small $1^{-+}$ signal in the $f_{1}\pi^{-}$ final state.

VES  also reported on the $\omega\pi^{-}\pi^{0}$ final state~\cite{Khokholov:00,Zaitsev:00}. In 
a combined analysis of the $\eta^{\prime}\pi^{-}$, $b_{1}\pi$ and $\rho^{0}\pi^{-}$ final states, they reported
the $\pi_{1}(1600)$  state with a mass of $1.61\pm 0.02$~GeV and a width of $0.29\pm 0.03$~GeV
that was consistent with all three final states. To the extent that they observed these states, they
also observed all three final state produced in natural parity exchange ($M^{\epsilon}=1^{+}$). 
They were also able to report relative branching ratios for the three final states as given in 
equation~\ref{eq:pi11600_rates}.
\begin{eqnarray}
\label{eq:pi11600_rates}
b_{1}\pi : 
\eta^{\prime}\pi : 
\rho\pi : & = & 
1: 1\pm 0.3 : 1.5\pm 0.5.
\end{eqnarray}
However, there were some issues with the $\rho\pi$ final state. Rather than limiting the rank
of the density matrix as was done in~\cite{Adams:1998ff,Chung:02}, they did not limit it. This allowed
for a more general fit that might be less sensitive to acceptance affects. In this model, they 
did not observe any significant structure in the $1^{-+}$ $\rho\pi$ partial wave above $1.4$~GeV. 
However, by looking at how other resonances were produced, they were able to isolate a coherent part
of the density matrix from which they found a statistically significant $1^{-+}$ partial wave
peaking near $1.6$~GeV. While VES was extremely careful not to claim the existence of
the $\rho\pi$ decay of the the $\pi_{1}(1600)$, in the case that it exists, they were able to 
obtain the rates given in equation~\ref{eq:pi11600_rates}.
\begin{figure}[h!]\centering
\includegraphics[width=0.5\textwidth]{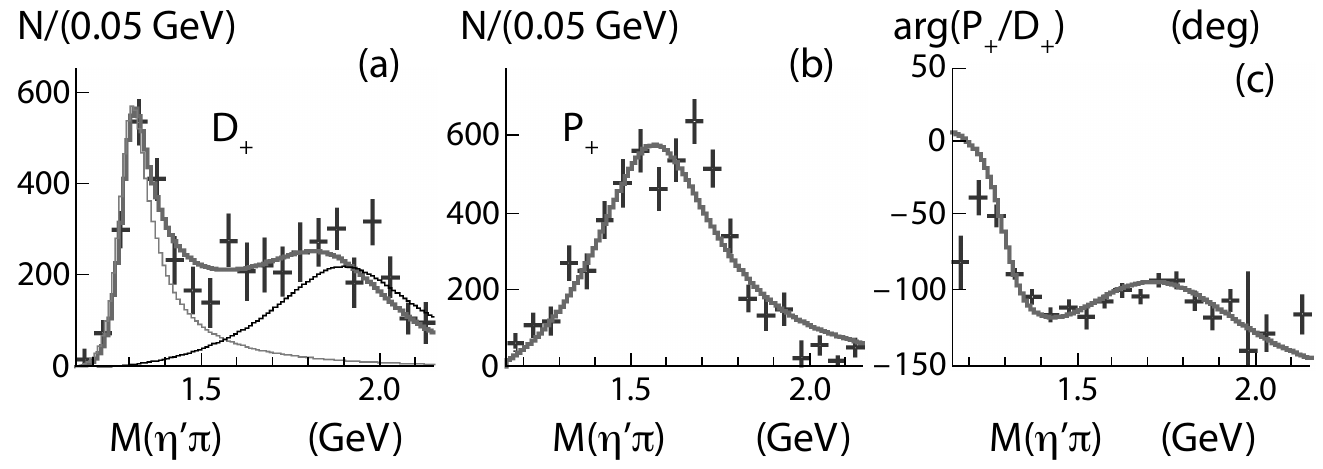}
\caption[]{\label{fig:ves_etaprpi} The results of a partial wave analysis on the $\eta^{\prime}\pi^{-}$
final state from VES. (a) shows he $2^{++}$ partial wave in $\omega\rho$, (b) shows the 
$1^{-+}$ partial wave in $b_{1}\pi$ and (c) shows the interference between them.
(Figure reproduced with permission from reference~\cite{Amelin:2005ry}.)}
\end{figure}
\begin{figure}[h!]\centering
\includegraphics[width=0.5\textwidth]{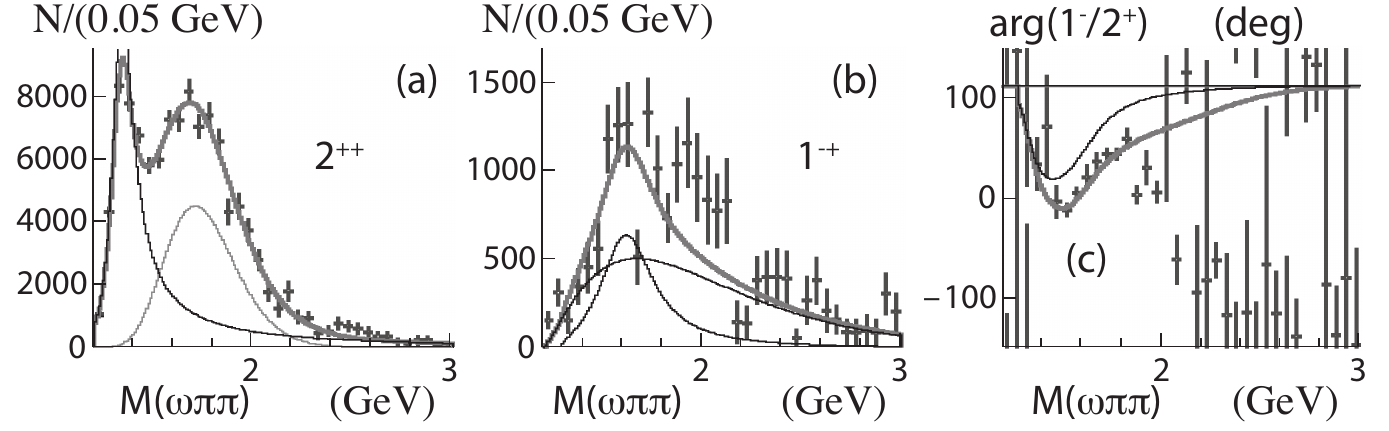}
\caption[]{\label{fig:ves_b1pi}The results of a partial wave analysis on the $b_{1}\pi$
final state from VES. (a) shows he $2^{++}$ partial wave, (b) shows the $1^{-+}$ partial wave and
(c) shows the interference between them.
(Figure reproduced with permission from reference~\cite{Amelin:2005ry}.)}
\end{figure}
\begin{figure}[h!]\centering
\includegraphics[width=0.5\textwidth]{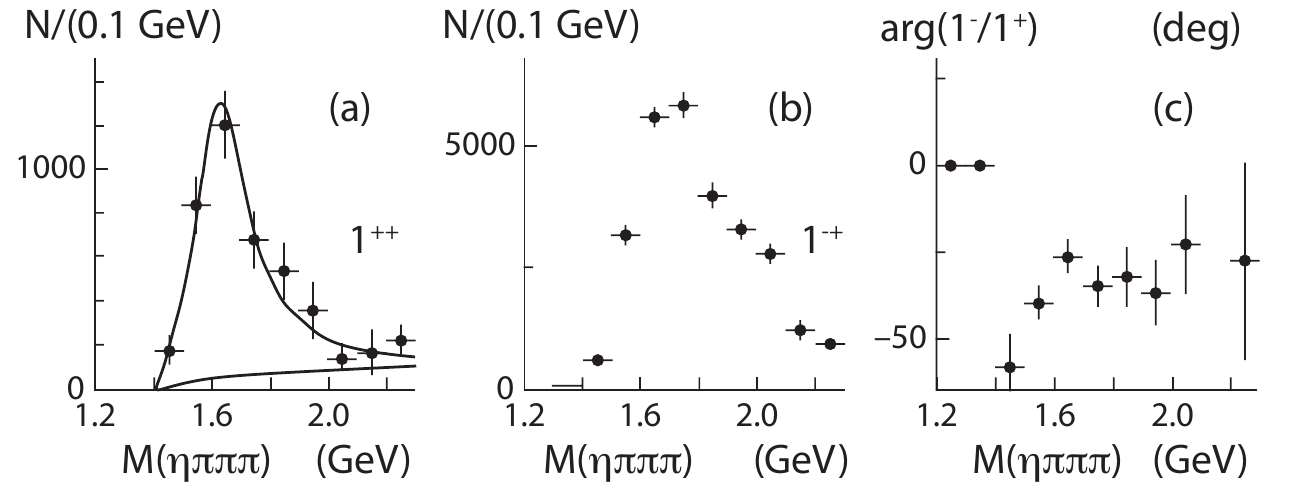}
\caption[]{\label{fig:ves_f1pi}The results of a partial wave analysis on the $f_{1}\pi$
final state from VES. (a) shows he $1^{++}$ partial wave, (b) shows the $1^{-+}$ partial wave and
(c) shows the interference between them.
(Figure reproduced with permission from reference~\cite{Amelin:2005ry}.)}
\end{figure}

The VES results have been summarized in a review of all their work on hybrid 
mesons~\cite{Amelin:2005ry}. This included an updated summary of the $\pi_{1}(1600)$
in all four final states, $\eta^{\prime}\pi$ $\rho\pi$, $b_{1}\pi$ and  $f_{1}\pi$. 
In the $\eta^{\prime}\pi$ final state (Fig.~\ref{fig:ves_etaprpi}), 
they note that the $1^{-+}$
wave is dominant. While they were concerned about the nature of the higher-mass part of
the $2^{++}$ spectrum ($a_{2}(1700)$ or background) they find that a resonant description of 
$\pi_{1}(1600)$ was possible in both cases. For the case of the $b_{1}\pi$ final state
(Fig.~\ref{fig:ves_b1pi}), they found that
the contribution of a $\pi_{1}(1600)$ resonance is required. In a combined fit to both the 
$\eta^{\prime}\pi$ and $b_{1}\pi$ data, they find a mass of $1.56\pm 0.06$~GeV and a width 
of $0.34\pm 0.06$~GeV for the $\pi_{1}(1600)$. In the $f_{1}\pi$ final state
(Fig.~\ref{fig:ves_f1pi}),  they find a
resonant description of the $\pi_{1}(1600)$ with a mass of $1.64\pm 0.03$~GeV and
a width of $0.24\pm 0.06$~GeV which they note is compatible with their measurement
in the previous two final states. They also note, that in contradiction with E852~\cite{Kuhn:2004en},
they find no significant $1^{-+}$ intensity above a mass of $1.9$~GeV (see Section~\ref{sec:pi1_2015}).
For the $\rho\pi$ final state, they are unable to conclude that the $\pi_{1}(1600)$ is present.

They note that the partial-wave analysis of the $\pi^{+}\pi^{-}\pi^{-}$ system finds a significant
contribution from the $J^{PC}=1^{−+}$ wave in the $\rho\pi$ channel ($2$ to $3$\% of the total intensity).
Some of the models in the partial-wave analysis of the exotic wave lead to the appearance of a peak
near a mass of $1.6$~GeV which resembles the $\pi_{1}(1600)$. However, the dependence of the 
size of this peak on the model used is significant~\cite{Zaitsev:00}. They note that because the 
significance of the wave depends very strongly on the assumptions of coherence used in the analysis,
the results for $3\pi$ final states on the nature of the $\pi_{1}(1600)$ are not reliable.

To obtain a limit on the branching fraction of $\pi_{1}(1600)$ decay to $\rho\pi$, they  looked at 
their results of the production of the $\pi_{1}(1600)$ in the charge-exchange reaction to 
$\eta^{\prime}\pi^{0}$ versus that of the $\eta^{\prime}\pi^{-}$ final state. They conclude that the
presence of the $\pi_{1}(1600)$ in $\eta^{\prime}\pi^{-}$ and its absence in $\eta^{\prime}\pi^{0}$
preclude the formation of the $\pi_{1}(1600)$ by $\rho$ exchange. From this, they obtain the 
relative branching ratios for the $\pi_{1}(1600)$ as given in equation~\ref{eq:pi11600_rates3}.
\begin{eqnarray}
\label{eq:pi11600_rates3}
b_{1}\pi : 
f_{1}\pi :
\rho\pi : 
\eta^{\prime}\pi & = & 
1.0 \pm .3: 1.1 \pm .3 : < .3 :1.
\end{eqnarray}

\subsubsection{E852 Results on the $\pi_{1}(1600)$}
\begin{figure}[h!]\centering
\includegraphics[width=0.49\textwidth]{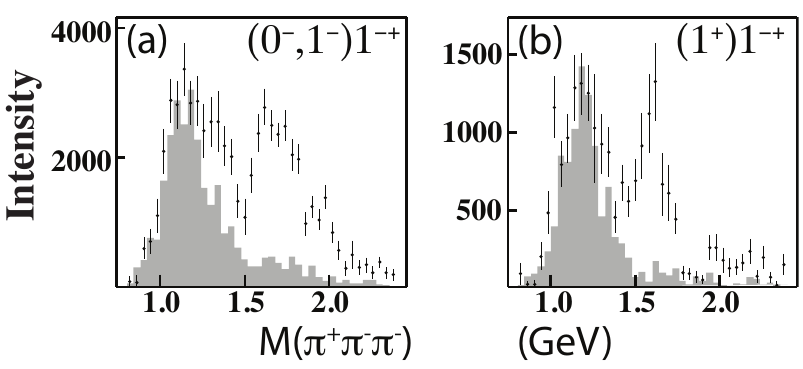}
\caption[]{\label{fig:e852ex}The production of the $1^{-+}$ partial wave as seen in the 
$\pi^{+}\pi^{-}\pi^{-}$ final state by E852. (a) shows the unnatural parity exchange 
($M^{\epsilon}=0^{-}$,$1^{-}$) while (b) shows the natural parity exchange ($M^{\epsilon}=1^{+}$).
(Figure reproduced with permission from reference ~\cite{Adams:1998ff}.)}
\end{figure}

\begin{figure}[h!]\centering
\includegraphics[width=0.47\textwidth]{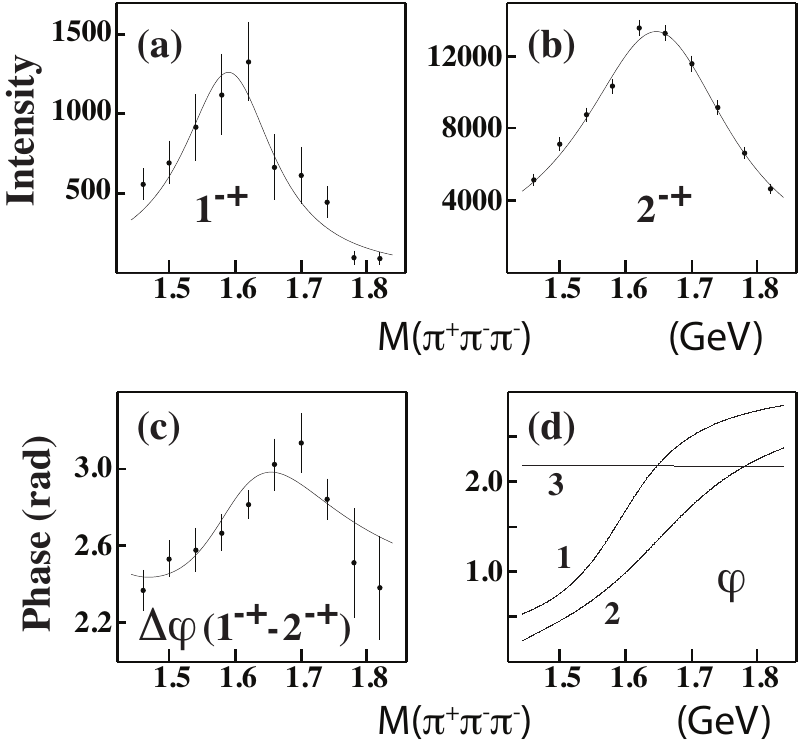}
\caption[]{\label{fig:e8523pi} The results of a PWA to the $\pi^{+}\pi^{-}\pi^{-}$ final
state from E852. (a) shows the intensity of the $J^{PC}=1^{-+}$ wave, (b) shows the 
$2^{-+}$ and (c) shows the phase difference between the two. The solid curves are fits
to two interfering Breit-Wigner distributions. In (d) are shown the phases of the two
Breit-Wigner distributions and (1,2) and a flat background phase (3) that combine to
make the curve in (c).
(Figure reproduced with permission from reference ~\cite{Adams:1998ff}.)}
\end{figure}

Using an $18$~GeV/c $\pi^{-}$ beam incident on a proton target, the E852 collaboration 
carried out a partial wave analysis of the $\pi^{+}\pi^{-}\pi^{-}$ final 
state~\cite{Adams:1998ff,Chung:02}. They saw both the $\rho^{0}\pi^{-}$ and $f_{2}(1270)\pi^{-}$ 
intermediate states and observed a $J^{PC}=1^{-+}$ state which decayed to $\rho\pi$, the 
$\pi_{1}(1600)$. The $\pi_{1}(1600)$ was produced in both natural and unnatural parity 
exchange ( $M^{\epsilon}=1^{+}$ and $M^{\epsilon}=0^{-}$, $1^{-}$) with apparent similar strengths 
in all three exchange mechanisms (see Fig.~\ref{fig:e852ex}). In Ref.~\cite{Chung:02}, they
noted that there were issues with the unnatural exchange production. The signal in the 
$M^{\epsilon}=1^{-}$ wave exhibited very strong model dependence and nearly vanished when
larger numbers of partial waves were included. The signal in the $M^{\epsilon}=0^{-}$
partial wave was stable, but its peak was above $1.7$~GeV. They noted that the 
unnatural-parity exchange is expected to die off at higher energies, so their results are 
not at odds with those of VES, where natural parity exchange dominates. 
In the E852 data, the unnatural parity exchange waves make up a small fraction of the total signal. 
In unnatural parity exchange, they found no significant waves, which made a study of phase 
motion of the $1^{-+}$ in this sector problematic. Thus, in their analysis, they only considered 
the natural parity exchange. There, they found the $\pi_{1}(1600)$ to have a mass of 
$1.593\pm 0.08^{+0.029}_{-0.047}$~GeV and a width of $0.168\pm 0.020^{+0.150}_{-0.012}$~GeV. 
In Fig.~\ref{fig:e8523pi} are shown the intensity
of the $1^{-+}$ and $2^{-+}$ ($\pi_{2}(1670)$) partial waves as well as their phase difference.
The phase difference can be reproduced by two interfering Breit-Wigner distributions and
a flat background.

\begin{figure}[h!]\centering
\includegraphics[width=0.5\textwidth]{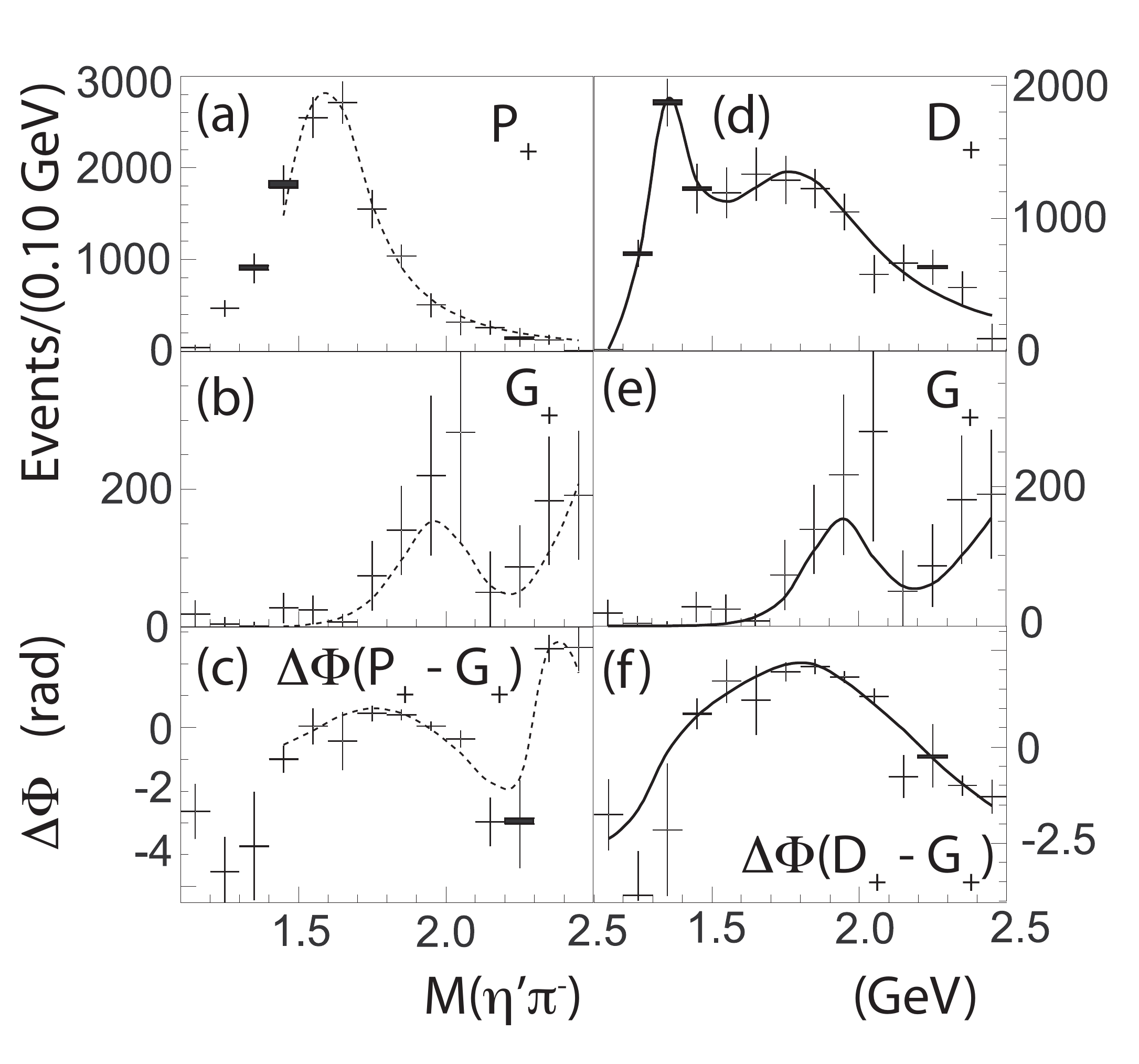}
\caption[]{\label{fig:e852_etaprime} Results from E852 on the $\eta^{\prime}\pi^{-}$ final 
state. (a) shows the $1^{-+}$ partial wave, (b) shows the $4^{++}$ partial wave (an $a_{4}$) 
and (c) shows the phase difference between these. (d) shows the  $2^{++}$ partial wave 
($a_{2}(1320)$), while (e) shows the $a_{4}$ and (f) is the phase difference.
(Figure reproduced with permission from reference~\cite{Ivanov:2001rv}.)}
\end{figure}
 
In a follow-up analysis, E852 also studied the reaction $\pi^{-}p\rightarrow p\eta^{\prime}\pi^{-}$ to 
examine the  $\eta^{\prime}\pi^{-}$ final state~\cite{Ivanov:2001rv}. They observed, consistent with 
VES~\cite{Beladidze:93}, that the dominant signal was the $1^{-+}$ exotic wave produced 
dominantly in the $M^{\epsilon}=1^{+}$ channel, implying only natural parity exchange. 
They found the signal to be consistent with a resonance, the $\pi_{1}(1600)$ and found
a mass of $1.597\pm 0.010^{+0.045}_{-0.010}$~GeV and a width of $0.340\pm 0.040\pm  
0.050$~GeV. The results of the E852 PWA are shown in Fig.~\ref{fig:e852_etaprime}
where the $P_{+}$ wave is the $1^{-+}$, the $D_{+}$ corresponds to the $2^{++}$ $a_{2}$ and
the $G_{+}$ corresponds to the $4^{++}$ $a_{4}$. Clear phase motion is observed between both 
the $2^{++}$ and $4^{++}$ wave and the $1^{-+}$ and the $4^{++}$ wave.

E852 also looked for the decays of the $\pi_{1}(1600)$ to $b_{1}\pi$ and $f_{1}\pi$. The latter
was studied in the reaction $\pi^{-}p\rightarrow p \eta \pi^{+}\pi^{-}\pi^{-}$ with the $f_{1}$
being reconstructed in its $\eta\pi^{+}\pi^{-}$ decay mode~\cite{Kuhn:2004en}.
The $\pi_{1}(1600)$ was seen via interference with both the $1^{++}$ and $2^{-+}$ partial
waves. It was produced via natural parity exchange ($M^{\epsilon}=1^{+}$) and found
to have a mass of $1.709\pm 0.024\pm 0.041$~GeV and a width of 
$0.403\pm 0.080\pm 0.115$~GeV. A second $\pi_{1}$ state was also observed in this
reaction (see Section~\ref{sec:pi1_2015}). 

The $b_{1}\pi$ final state was studied by looking
at the reaction $\pi^{-}p\rightarrow \omega \pi^{-}\pi^{0}p$, with the $b_{1}$ reconstructed in
its  $\omega\pi$  decay mode~\cite{Lu:2004yn}. The $\pi_{1}(1600)$ was seen interfering
with the $2^{++}$ and $4^{++}$ partial waves. In $b_{1}\pi$, they measured a mass of 
$1.664\pm 0.008\pm 0.010$~GeV and a width of $0.185\pm 0.025\pm 0.028$~GeV
for the $\pi_{1}(1600)$. However, the production mechanism was seen to be a mixture
of both natural and unnatural parity exchange, with the unnatural being stronger. As
with the $f_{1}\pi$, they also observed a second $\pi_{1}$ state decaying to $b_{1}\pi$
(see Section~\ref{sec:pi1_2015}).
\begin{table}[h!]\centering
\begin{tabular}{crc} \hline\hline
final state & production ($M^{\epsilon}$) & dominant \\ \hline
$\rho\pi$               & $0^{-}$, $1^{-}$, $1^{+}$ & npe $\sim$ upe \\
$\eta^{\prime}\pi$    & $1^{+}$ & npe \\
$f_{1}\pi$                & $1^{+}$ & npe \\
$b_{1}\pi$               & $0^{-}$, $1^{-}$, $1^{+}$ & upe $>$ npe \\
\hline\hline
\end{tabular}
\caption[]{\label{tab:e852_prod}The production mechanisms for the $\pi_{1}(1600)$ as seen 
in the E852 experiment. Also shown is whether the natural parity exchange (npe) or the 
unnatural parity exchange (upe) is stronger.}
\end{table}

The fact that E852 observed the $\pi_{1}(1600)$ produced in different production mechanisms,
depending on the final state, is somewhat confusing. A summary of the observed mechanisms 
is given in Table~\ref{tab:e852_prod}. In order to understand the variations in production, 
there either needs to be two nearly-degenerate $\pi_{1}(1600)$s, or there is some 
unaccounted-for systematic problem in some of the analyses.
\begin{figure}[h!]\centering
\includegraphics[width=0.5\textwidth]{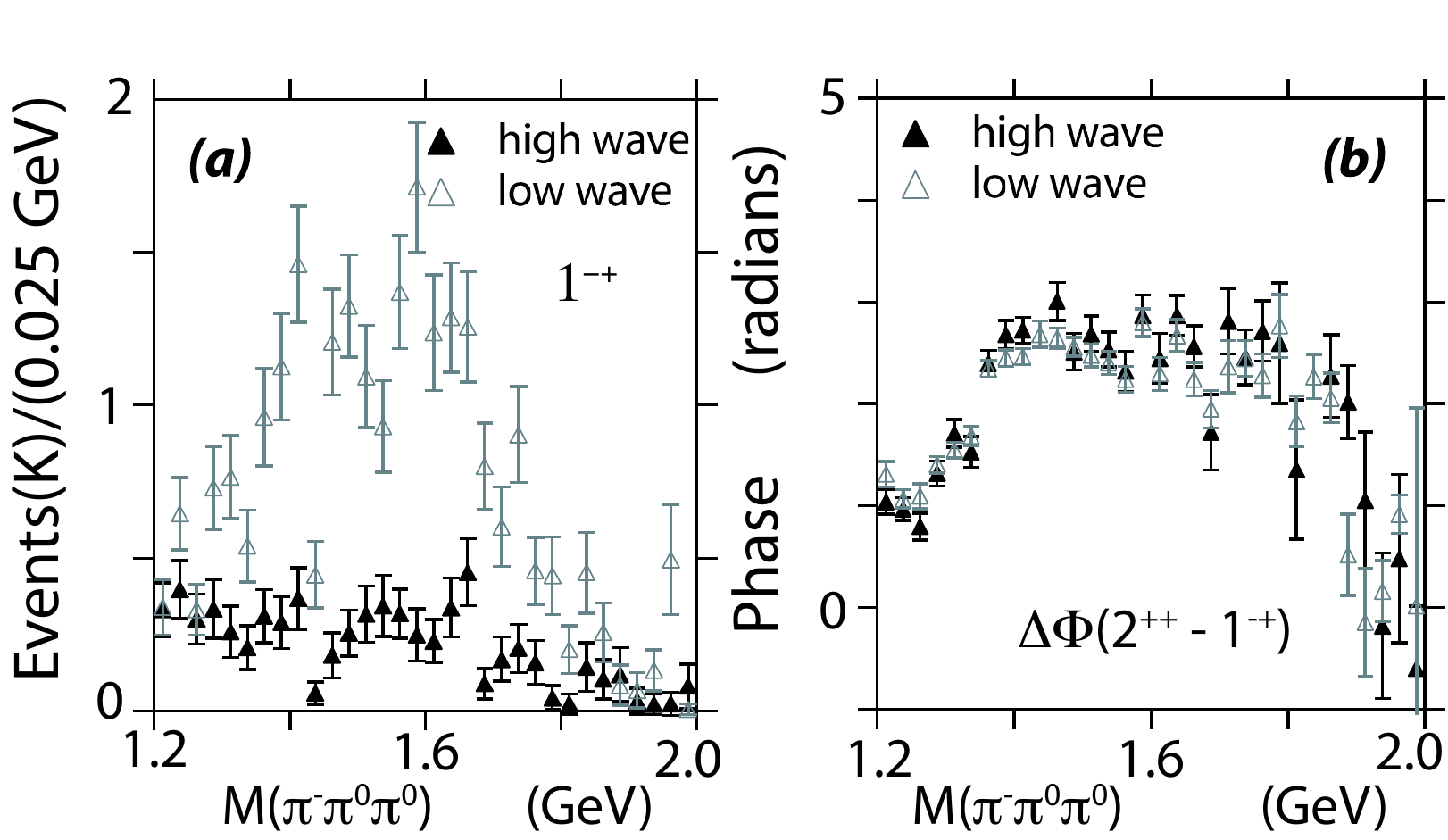}
\caption[]{\label{fig:e852_dz_m00}(Color on line) The E852-IU PWA solutions for the $1^{-+}$ partial wave for the
$\pi^{-}\pi^{0}\pi^{0}$ final state (a) and its interference with the $2^{++}$ partial wave (b). See text 
for an explanation of the labels. (Figure reproduced with permission from reference~\cite{Dzierba:2005jg}.)}
\end{figure}
\begin{figure}[h!]\centering
\includegraphics[width=0.5\textwidth]{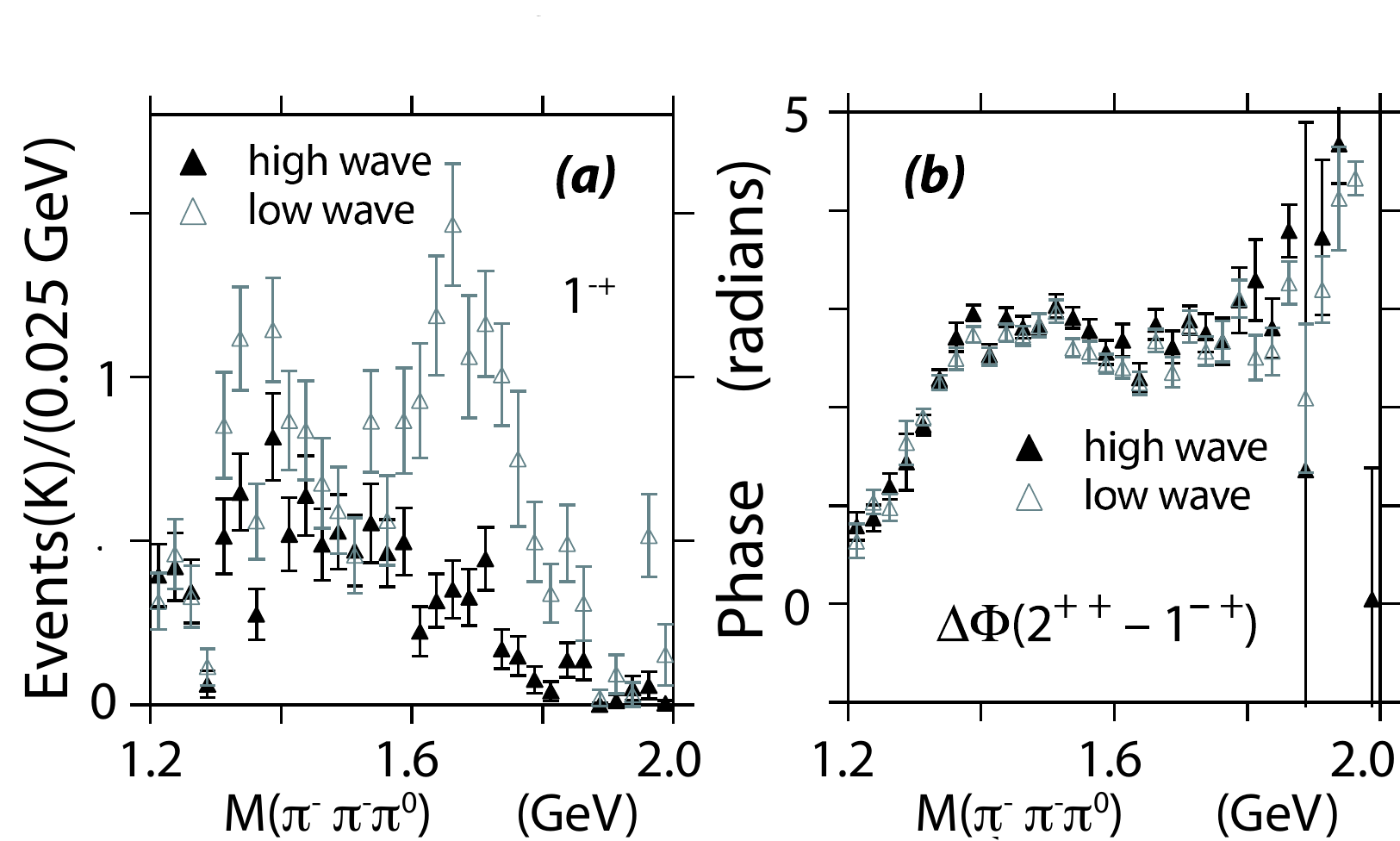}
\caption[]{\label{fig:e852_dz_pmm}(Color on line) The E852-IU PWA solutions for the $1^{-+}$ partial 
wave for the $\pi^{+}\pi^{-}\pi^{-}$ final state (a) and its interference with the $2^{++}$ partial wave (b). 
See text for an explanation of the labels. (Figure reproduced with permission from reference~\cite{Dzierba:2005jg}.)}
\end{figure}

The E852-IU group analyzed an E852 data set that was an order of magnitude larger than that 
used by E852 in Refs.~\cite{Adams:1998ff,Chung:02}. In this larger data set, they looked at the
reactions $\pi^{-}p\rightarrow n \pi^{+}\pi^{-}\pi^{-}$ and $\pi^{-}p\rightarrow n \pi^{-}\pi^{0}\pi^{0}$
and carried out a partial wave analysis for both the $\pi^{+}\pi^{-}\pi^{-}$ and the $\pi^{-}\pi^{0}\pi^{0}$ 
final states. This yielded solutions that were consistent with both final states~\cite{Dzierba:2005jg}.
In this analysis, they carried out a systematic study of which partial waves were important
in the fit. When they used the same wave set as in the E852 analysis~\cite{Adams:1998ff,Chung:02}, 
they found the same solution showing a signal for the $\pi_{1}(1600)$ in both final states. However,
when they allowed for more partial waves, they found that the signal for the $\pi_{1}(1600)$ vanished.
Fig.~\ref{fig:e852_dz_m00} shows these results for the $\pi^{-}\pi^{0}\pi^{0}$ final state,
while Fig.~\ref{fig:e852_dz_pmm} shows the results for the $\pi^{+}\pi^{-}\pi^{-}$ final
state. In both figures, the ``low wave'' solution matches that from E852, while their ``high wave'' 
solution shows no intensity for the $\pi_{1}(1600)$ in either channel. An important point is that
in both their high-wave and low-wave analyses, the phase difference between the exotic $1^{-+}$
wave and the $2^{++}$ wave are the same (and thus the same as in the E852 analysis). While not
shown here, the same is also true for the $1^{-+}$ and $2^{-+}$ waves. 
\begin{table}[h!]\centering
\begin{tabular}{ccccccc}\hline\hline
$\pi_{2}(1670)$ & \multicolumn{2}{c}{$M^{\epsilon}=0^{+}$}
                        & \multicolumn{2}{c}{$M^{\epsilon}=1^{+}$}
                        & \multicolumn{2}{c}{$M^{\epsilon}=1^{-}$} \\
Decay               & L & H & L & H & L & H \\ \hline
$(f_{2}\pi)_{S}$      & $\times$ & $\times$ & $\times$ & $\times$ & $\times$ &     \\
$(f_{2}\pi)_{D}$      & $\times$ & $\times$ & $\times$ & $\times$ &  &  \\
$[(\pi\pi)_{S}]_{D}$ &$\times$ & $\times$ &                 & $\times$ &  &  \\
$(\rho\pi)_{P}$      &$\times$ & $\times$ &                 & $\times$ &  &  \\
$(\rho\pi)_{F}$      &                & $\times$ &                 & $\times$ &  &  \\
$(f_{0}\pi)_{D}$      &                 & $\times$ &                 & $\times$ &  &  \\
\hline\hline
\end{tabular}
\caption[]{\label{tab:p2-decays}The included decays of the $\pi_{2}(1670)$ in two analyses
of the $3\pi$ final state. ``L'' is the wave set used in the E852 analysis~\cite{Adams:1998ff,Chung:02}.
``H'' is the wave set used in the higher statistics analysis~\cite{Dzierba:2005jg}.}
\end{table}

\begin{figure}[h!]\centering
\includegraphics[width=0.35\textwidth]{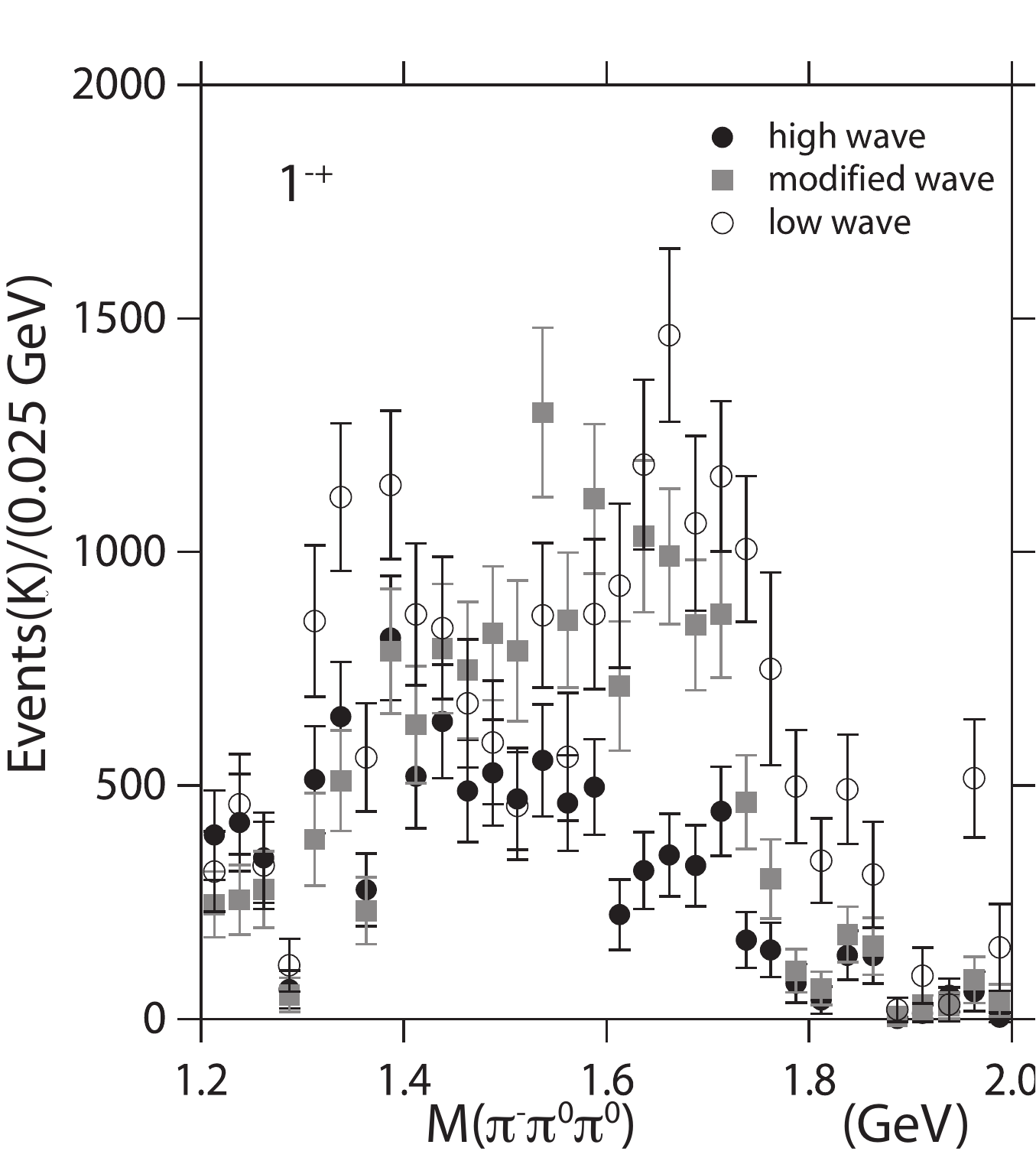}
\caption[]{\label{fig:e852_dz_pi2} Results from the E852-IU analysis showing the $1^{-+}$ intensity 
for the charged mode for the high-wave 
set (filled circles), the modified high-wave set (filled squares), and the low-wave set (open circles). 
In the modified high-wave set the two $\rho\pi$ decays of the $\pi_{2}(1670)$ were removed
from the fit. (Figure reproduced with permission from reference~\cite{Dzierba:2005jg}.)}
\end{figure}

E852-IU carried out a study to determine which of the additional waves in their ``high wave'' 
set were absorbing the intensity of the $\pi_{1}(1600)$. They found that the majority of this was 
due to the inclusion of the $\rho\pi$ decay of the $\pi_{2}(1670)$. The partial waves associated with the 
$\pi_{2}(1670)$ in both analyses are listed in Table~\ref{tab:p2-decays}. While the production
from $M^{\epsilon}=0^{+}$ is similar for both analyses, the E852 analysis only included 
the $\pi_{2}(1670)$ decaying to $f_{2}\pi$ in the $M^{\epsilon}=1^{+}$ production.
The high-statistics analysis included both $f_{2}\pi$ and $\rho\pi$ in both production mechanisms.
The PDG~\cite{pdg12} lists the two main decays of the $\pi_{2}(1670)$ as $f_{2}\pi$ (56\%) and
$\rho\pi$ (31\%), so it seems odd to not include this latter decay in an analysis
including the $\pi_{2}(1670)$. Fig.~\ref{fig:e852_dz_pi2} shows the results of removing the 
$\rho\pi$ decay from the ``high wave'' set for the $\pi^{+}\pi^{-}\pi^{-}$ final state. This
decay absorbs a significant portion of the $\pi_{1}(1600)$ partial wave.

In the E852-IU analyses, the fact that the phase motion of the $1^{-+}$ exotic wave relative 
to other partial waves agrees with those differences as measured by E852, and are the same 
in both the high-wave and low-wave analyses is intriguing. This could be interpreted as a 
$\pi_{1}(1600)$ state which is simply absorbed by the stronger $\pi_{2}(1670)$ as more partial 
waves are added. However, given the small actual phase difference between the $1^{-+}$ and 
$2^{-+}$ partial waves (see Fig.~\ref{fig:e8523pi}), the opposite conclusion is also possible, 
particularly if some small non-zero background phase were present. Here, the $1^{-+}$ signal 
is due to leakage from the stronger $\pi_{2}$ and no $\pi_{1}(1600)$ is needed in the $\rho\pi$ 
final state. 

While the results on $\rho\pi$ between E852 and VES seem at odds, we believe that these 
discrepancies are the result of the assumptions made in the analyses. These assumptions
then manifest themselves in the interpretation of the results. The VES analyses fit both the 
real and imaginary parts of their amplitudes independently. However, for analytic functions, 
the two parts are not independent. Not using these constraints can lead to results that may 
be unphysical, and at a minimum, are discarding important constraints on the amplitudes.
 In E852, many of their results 
rely on the assumption of a flat background phase, but there are many examples where 
this is not true. Thus, their results are biased toward a purely resonant description of the 
data, rather than a combination of resonant and non-resonant parts. It is also somewhat
disappointing that E852 is unable to make statements about relative decay rates, or carry
out a coupled channel analysis of their many data sets. Our understanding is that this 
is due to issues in modelling the rather tight trigger used in collecting their data.

\subsubsection{Crystal Barrel results on the $\pi_{1}(1600)$}
An analysis of Crystal Barrel data at rest for the reaction $\bar{p}p\rightarrow\omega\pi^{+}\pi^{-}\pi^{0}$
was carried by some members of the collaboration~\cite{Baker:2003jh}. They reported evidence
for the $\pi_{1}(1600)$ decaying to $b_{1}\pi$ from both the $^{1}S_{0}$ and $^{3}S_{1}$ initial
states, with the signal being stronger from the former. The total signal including
both initial states, as well as decays with $0$ and $2$ units of angular momentum accounted
for less than $10$\% of the total reaction channel. The mass and width were found consistent
(within large errors) of the PDG value, and only results with the mass and width fixed to the
PDG values were reported. Accounting for the large rate of annihilation to $\omega\pi^{+}\pi^{-}\pi^{0}$
of $13$\%, this would imply that $\bar{p}p\rightarrow \pi_{1}(1600)\pi$ accounts for several
percent of all $\bar{p}p$ annihilations. 

\subsubsection{CLEO-c results on the $\pi_{1}(1600)$}
\begin{figure}[h!]\centering
\includegraphics[width=0.4\textwidth]{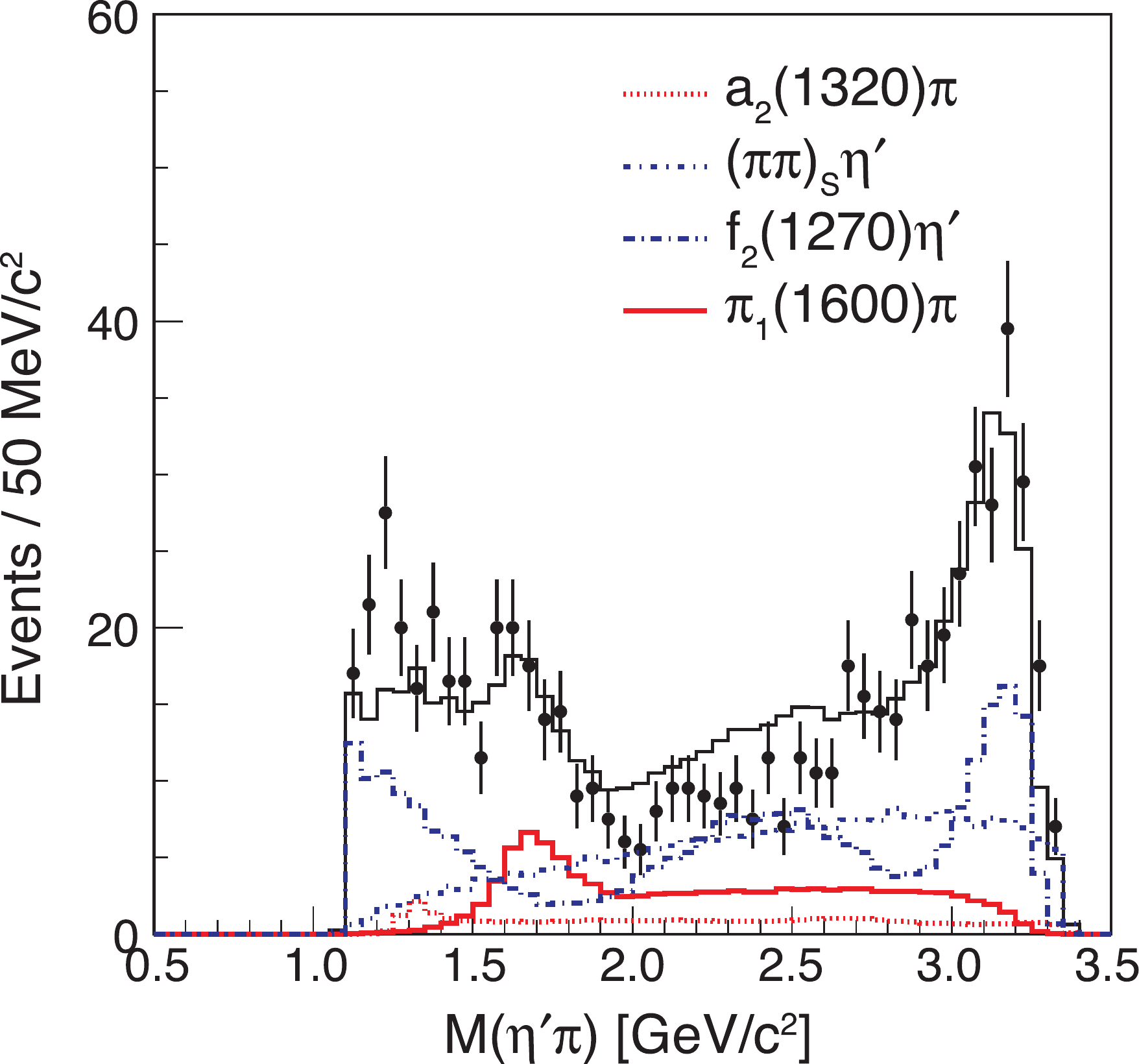}
\caption[]{\label{fig:cleo_c_pi1_1600}(Color on line.) Results from CLEO-c for the decay of
the $\chi_{c1}$ to $\eta^{\prime}\pi^{+}\pi^{-}$. The figure shows the fit intensity in $\eta^{\prime}\pi$
system. The solid (red) fit curve shows the intensity in the exotic $1^{-+}$ partial wave. There is
clear peaking near the mass of the $\pi_{1}(1600)$. (Figure reproduced with permission from 
reference~\cite{Adams:2011sq}.)}
\end{figure}

In CLEO-c,  the decays of the $\chi_{c1}$ to both $\eta\pi^{+}\pi^{-}$ and $\eta^{\prime}\pi\pi$ were 
studied. An amplitude analysis of the $\eta^{\prime}\pi\pi$ final state showed a $4\sigma$ signal 
for an exotic p-wave in the $\eta^{\prime}\pi$ system. Results from this analysis are shown in 
Fig.~\ref{fig:cleo_c_pi1_1600}. While the analysis could not conclude if the 
P-wave is resonant, if it is fit by a Breit-Wigner amplitude, the data can be well described by a 
$\pi_{1}(1600)$ with a mass of $1670\pm 30\pm 20$  MeV/$c^{2}$ and a width of 
$240\pm 50\pm 60$ MeV/$c^{2}$~\cite{Adams:2011sq}.

\subsubsection{CLAS results on the $\pi_{1}(1600)$}
The CLAS experiment at Jefferson Lab studied the reaction 
$\gamma p\rightarrow \pi^{+}\pi^{+}\pi^{-} (n)_{miss}$ to look for the production of the 
$\pi_{1}(1600)$~\cite{Nozar:09}. The photons were produced by bremsstrahlung from a 
$5.7$~GeV electron beam. While there were significant contributions from baryon resonances
in their data, they attempted to remove this by selective cuts on various kinematic regions.
The results of their partial-wave analysis show clear signals for the $a_{1}(1270)$, the 
$a_{2}(1320)$ and the $\pi_{2}(1670)$, but show no evidence for the $\pi_{1}(1600)$ decaying 
into three pions. They place an upper limit of the production and subsequent decay of
the $\pi_{1}(1600)$ to be less than 2\% of the $a_{2}(1320)$. Their results imply (i) the
$\pi_{1}(1600)$ is not strongly produced in photoproduction, (ii) the $\pi_{1}(1600)$ does not
decay to $3\pi$, or (iii) both. Using a much larger data sample, CLAS has continued to search
for the $\pi_{1}(1600)$ in photoproduction through the $3\pi$ decay mode, however the 
preliminary results are still consistent with the earlier work\cite{Eugenio:2013xua}. They propose
 that  either there is a preference for production through mechanisms other
than pion exchange, or that the $\pi_{1}(1600)$ does not decay to $3\pi$.

\begin{figure}[h!]\centering
\includegraphics[width=0.45\textwidth]{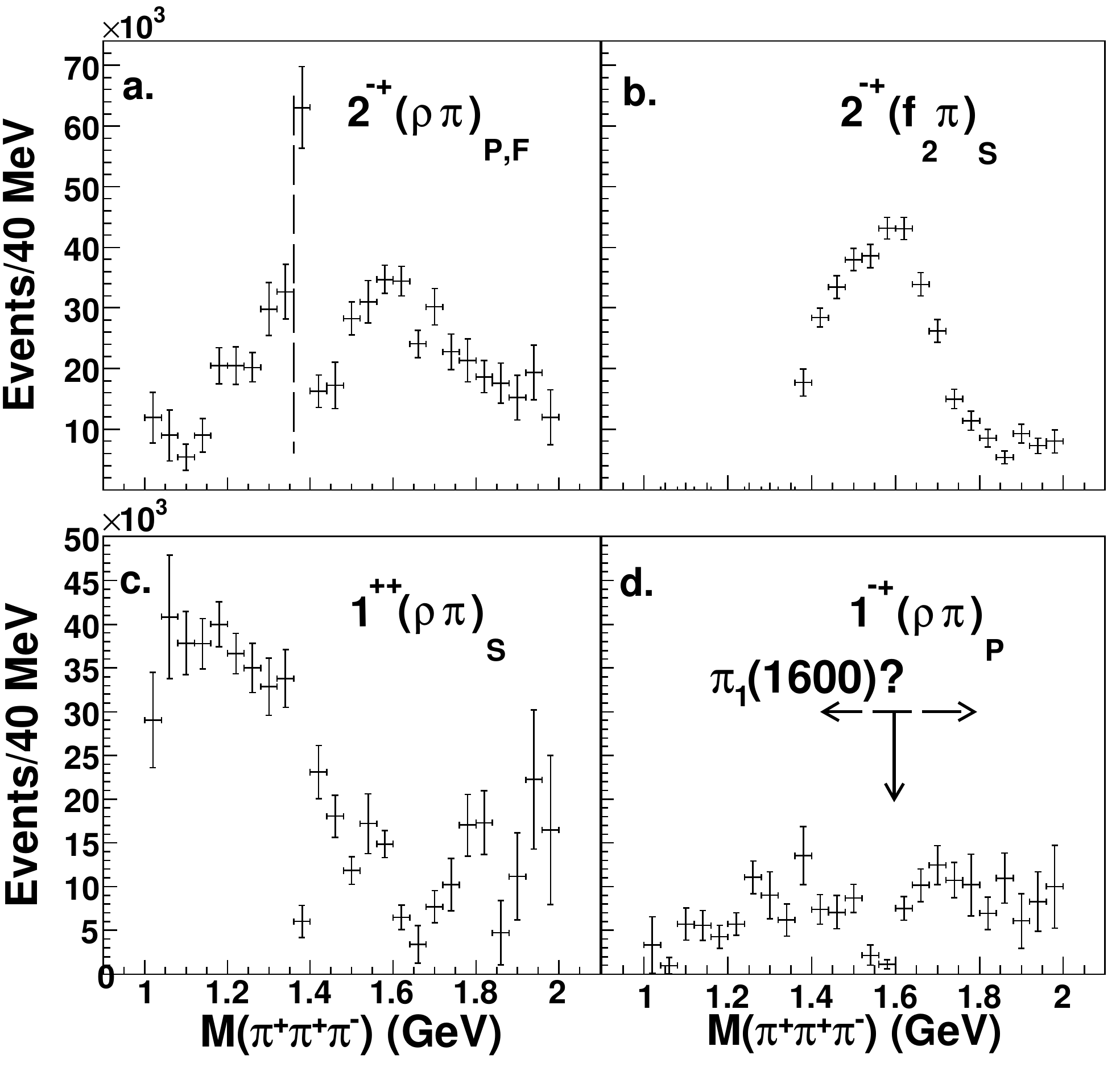}
\caption[]{\label{fig:clas3pi} The results from CLAS of a partial-wave analysis photoproduction
data of the $3\pi$ final state. Intensity is seen in the $2^{-+}$ partial wave, (a) and (b), as
well as the $1^{++}$ partial wave (c). In the $1^{-+}$ exotic wave, (d), no intensity is observed. 
(Figure reproduced with permission from reference~\cite{Nozar:09}.)}
\end{figure}

\subsubsection{COMPASS results on the $\pi_{1}(1600)$}
\begin{figure}[h!]\centering
\includegraphics[width=0.45\textwidth]{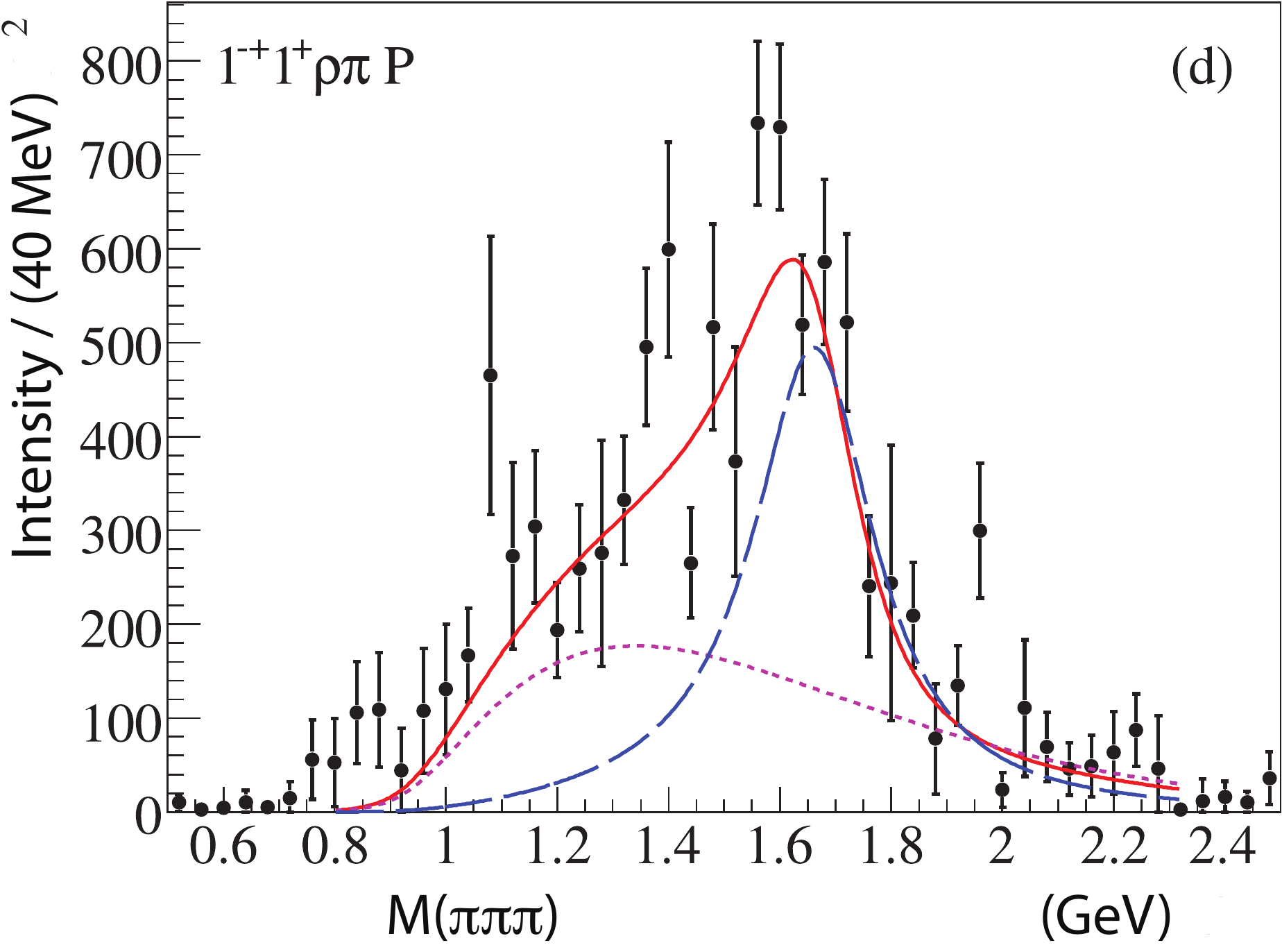}
\caption[]{\label{fig:compass_int} (Color on line.) COMPASS results showing the intensity of 
the exotic $1^{-+}$ wave. The solid (red) curve shows a fit to the corresponding resonances. 
The dashed (blue) curve is the $\pi_{1}(1600)$ while the dotted (magenta) curve is background.
(Figure reproduced with permission from reference~\cite{Alekseev:2009xt}.)}
\end{figure}

\begin{figure}[h!]\centering
\includegraphics[width=0.50\textwidth]{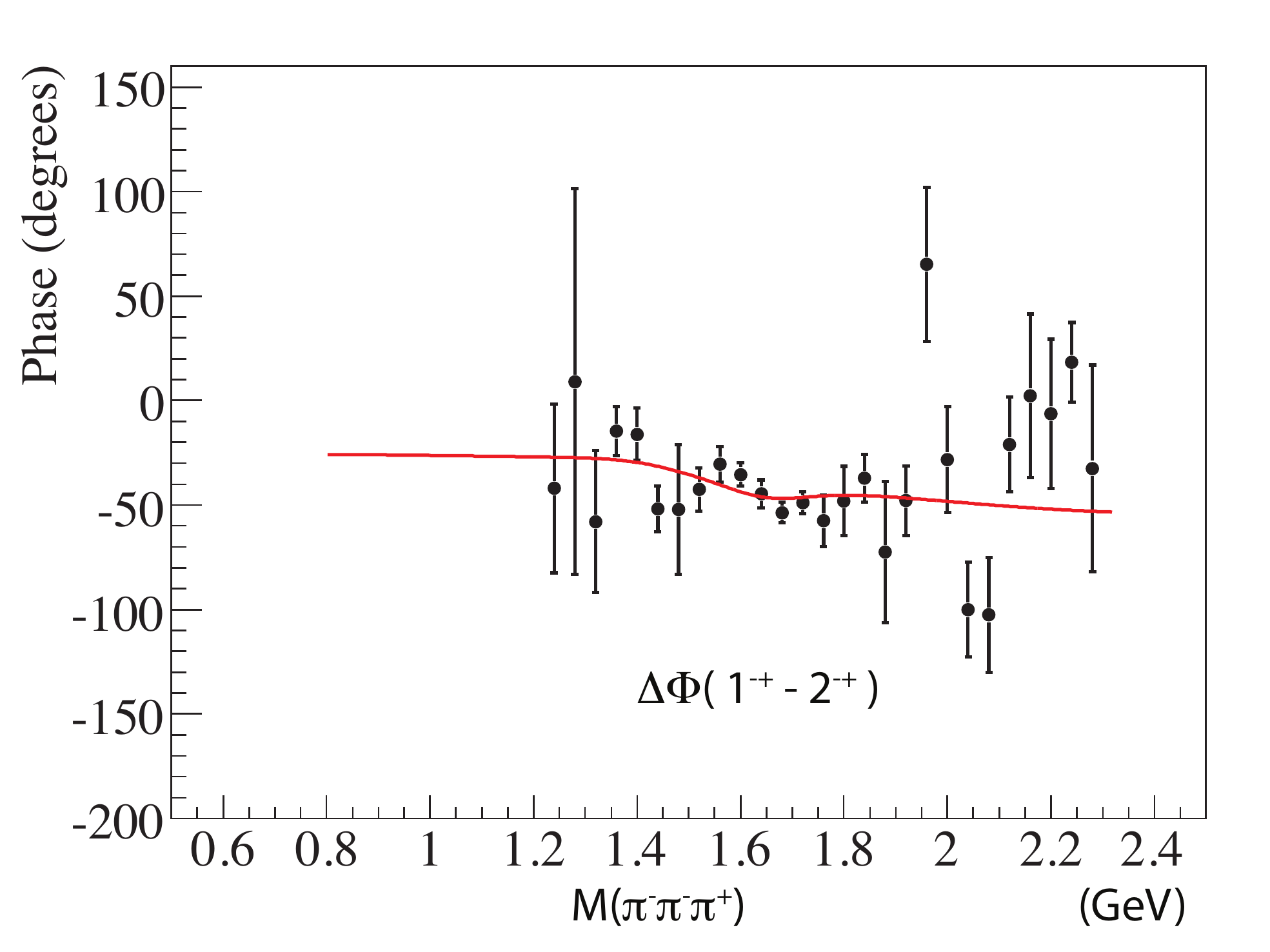}
\caption[]{\label{fig:compass_phs} (Color on line.) COMPASS results showing the phase  
difference between the exotic $1^{-+}$ wave and the $2^{-+}$ wave. The solid (red) curve 
shows a fit to the corresponding resonances.
(Figure reproduced with permission from reference~\cite{Alekseev:2009xt}.)}
\end{figure}

The COMPASS experiment has reported their first study of the diffractively produced 
$3\pi$ final state~\cite{Alekseev:2009xt,Grube:2010ux}.  They used a $190$~GeV/c beam of pions 
to study the reaction $\pi^{-} Pb \rightarrow \pi^{-}\pi^{-}\pi^{+} X$. In their partial-wave
analysis of the $3\pi$ final state, they observed the $\pi_{1}(1600)$ with a mass of 
$1.660\pm 0.010^{+0}_{-0.064}$~GeV and a width of $0.269\pm 0.021^{+0.042}_{-0.064}$~GeV.
The $\pi_{1}(1600)$ was produced dominantly in natural parity exchange ($M^{\epsilon}=1^{+}$)
although unnatural parity exchange also seemed to be required. However, the level was not
reported. The wave set (in reference~\cite{Grube:2010ux}) used appears to be somewhat larger 
than that used in the high-statistics study of E852-IU~\cite{Dzierba:2005jg}.  Thus, in
the COMPASS analysis, the $\rho\pi$ decay of the $\pi_{2}(1670)$ does not appear to absorb 
the exotic intensity in their analysis. They also report on varying the rank of the fit with
the $\pi_{1}(1600)$, with the results being robust against these changes. One point of small concern is that 
the mass and width that they extract for the $\pi_{1}(1600)$ are essentially identical to those 
for the $\pi_{2}(1670)$. For the latter, they observed  a mass of $1.658\pm 0.002^{+0.024}_{-0.008}$~GeV 
and a width of $0.271\pm 0.009^{+0.022}_{-0.024}$~GeV. However, the strength of the exotic wave
appears to be about $20$\% of the $\pi_{2}$, thus feed through seems unlikely. Results from their 
partial-wave analysis are shown in Figs.~\ref{fig:compass_int} and~\ref{fig:compass_phs}. 
These show the $1^{-+}$ partial wave and the phase difference between the $1^{-+}$ and $2^{-+}$
waves. The solid curves are the results of mass-dependent fits to the $\pi_{1}(1600)$ and 
$\pi_{2}(1670)$. 

\begin{figure}[h!]\centering
\includegraphics[width=0.45\textwidth]{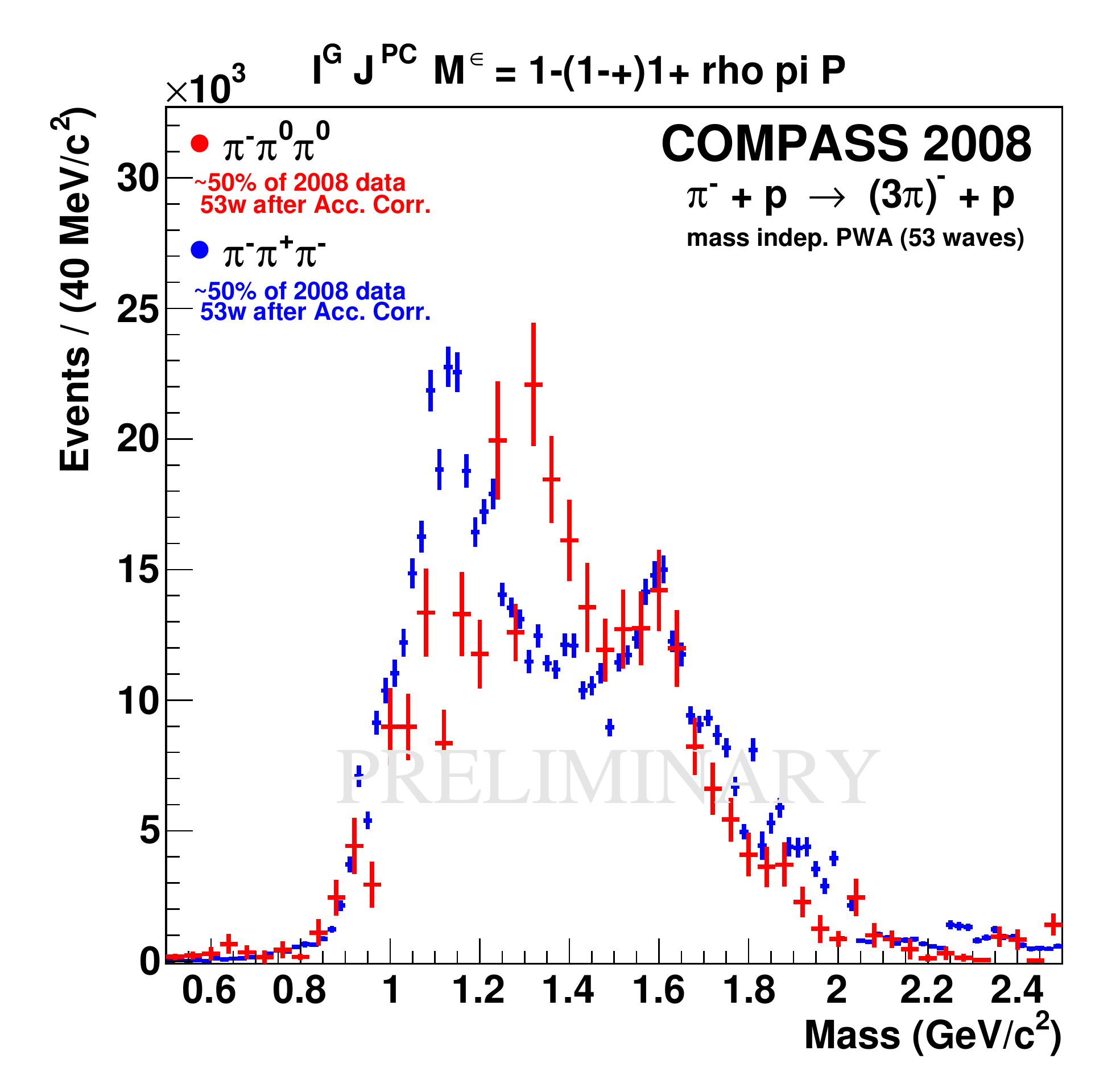}
\includegraphics[width=0.45\textwidth]{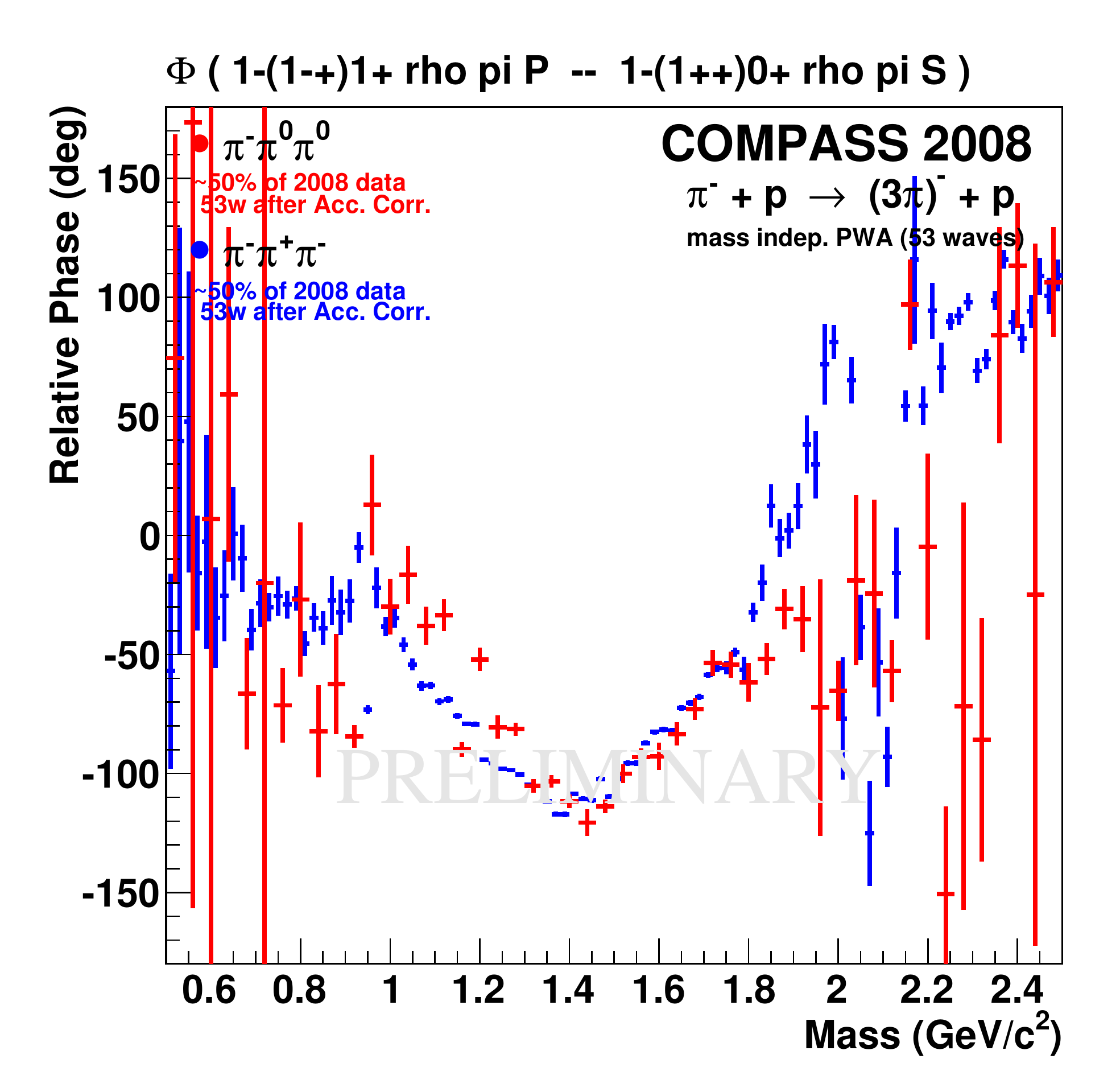}
\caption[]{\label{fig:compass_3pi_minus}(Color on line) (left) The intensity $J^{PC}=1^{-+}$ 
partial wave in the two $(3\pi^{-})$ final states. In the $1.6$~GeV mass region, both the 
$\pi^{-}\pi^{+}\pi^{-}$ and the $\pi^{-}\pi^{0}\pi^{0}$ final states show the same structure.
The peaks in the lower-mass region are unstable and probably artifacts of the analysis.
The phase difference (right) between the  $1^{++}$ and $1^{-+}$ partial waves. 
(Figure reproduced with permission from reference~\cite{Nerling_2011}.)}
\end{figure}

In follow-up studies, COMPASS has looked at three-pion final states produced
from a liquid hydrogen target. They have presented results on both 
$\pi^{-}p\rightarrow p\pi^{-}\pi^{+}\pi^{-}$ and 
$\pi^{-}p\rightarrow \pi^{-}\pi^{0}\pi^{0}$~\cite{Nerling_2011}. These studies confirmed the 
$1^{-+}$ exotic signal found in the Pb running, but the production rates on Pb were found to 
be much higher than on hydrogen. Amplitude analysis showed that the exotic wave was produced
with an $M^{\epsilon}=1^{+}$ exchange in both reactions, but that the $M=1$ in hydrogen is suppressed 
relative to the $M=0$.  Fig.~\ref{fig:compass_3pi_minus} show the intensity of the $1^{-+}$ wave
in both the $\pi^{-}\pi^{+}\pi^{-}$ and $\pi^{-}\pi^{0}\pi^{0}$ systems. Essentially the same
information on the $1^{-+}$ exotic signal in the $(3\pi)^{-}$ system can be found in 
references~\cite{Nerling_2011b,Nerling_2012,Grube:2013ux}.

In order to more fully understand the three-pion system, COMPASS has studied their very large data 
set on three pions. In particular, the non-exotic $1^{++}$ partial wave which has historically
been modeled using both the broad $a_{1}(1260)$ as well as non-resonant Deck-effect terms. 
They have reported on a new, narrow structure in this partial wave, the $a_{1}(1420)$, which
decays dominantly to $f_{0}(980)\pi$~\cite{Paul-2013},\cite{Uhl-2014},\cite{Nerling-2014},
\cite{Uhl-2014-2},\cite{Hass-2014}. They report a mass of $1414^{+15}_{-13}$ MeV and
a width of $153^{+8}_{-23}$ MeV~\cite{Adolph-2015}. Since their discovery of this
new state, results of the exotic $\pi_{1}(1600)$ have been limited until a more detailed 
analysis can be completed. They report that an exotic signal in the $1^{−+}$ partial wave at 
1.6 GeV/$c^{2}$ is observed and that it shows a clean phase motion with respect to 
well-known resonances. Their results are consistent for both the $\pi^{-}\pi^{0}\pi^{0}$ and
$\pi^{+}\pi^{-}\pi^{-}$ final states. The state is observed in the $\rho\pi$ decay mode for both
charged and neutral $\rho$s. However they exclude that a narrow (150 MeV to 200 MeV width) state exists~\cite{Nerling-2014},\cite{Krinner-2015},\cite{Adolph-2015}.

\begin{figure}[h!]\centering
\includegraphics[width=0.45\textwidth]{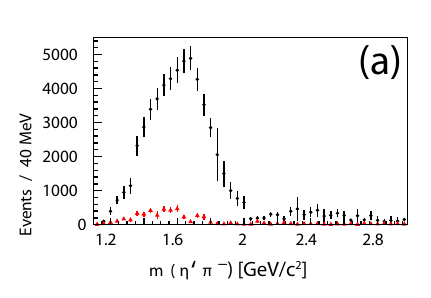}
\includegraphics[width=0.45\textwidth]{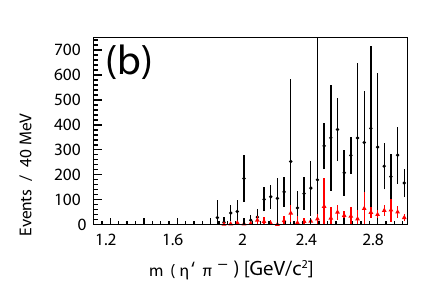}
\caption[]{\label{fig:compass_eta_etap}(Color on line) The intensity of the $\eta^{\prime}\pi^{-}$
(black) and $\eta\pi^{-}$ (red) signals in the $J^{PC}=1^{-+}$ partial wave (a) and the $3^{-+}$
partial wave (b). The signal for the $\eta\pi^{-}$ is much smaller than that for the $\eta^{\prime}\pi^{-}$
for all odd partial waves. (Figure reproduced with permission from reference~\cite{Adolph-2014}.)}
\end{figure}

The COMPASS Collaboration has also examined the 
$\pi^{-}p\rightarrow \eta \pi^{-} p$ and $\pi^{-}p\rightarrow \eta^{\prime} \pi^{-} p$ 
reactions~\cite{Schluter_2011,Adolph-2014}. Amplitude analyses of both systems 
shows that the even partial waves are produced with similar strengths in both
system, while the odd partial waves are suppressed in the $\eta\pi$ system relative 
to the $\eta^{\prime}\pi$ system. The $a_{2}(1320)$ and $a_{4}(2040)$ are produced 
in both systems with a relative ratio given by phase-space factor and the pseudoscalar 
mixing angle. The odd partial waves, $J^{PC}=(1,3,5)^{-+}$ all correspond to exotic quantum 
numbers, and are produced much more strongly in the $\eta^{\prime}$ system. This
is shown for the spin $1$ and $3$ partial waves in Fig.~\ref{fig:compass_eta_etap}.
In terms
of quark structure, the $\eta$ is predominantly an octet state, while the $\eta^{\prime}$ is
predominantly a singlet state. Both VES and E852 reported that the $1^{-+}$ exotic wave
is the dominant feature in the $\eta^{\prime}\pi$ final state, while it is strongly suppressed in 
the $\eta$ channel. In addition, the relative phase motion between the
exotic wave and the $a_{2}$ wave are different between the $\eta$ and $\eta^{\prime}$ channels.  
Even though COMPASS saw the exotic $1^{-+}$ wave in $\eta^{\prime}\pi$ as the dominant wave, they 
were unable to confirm the resonant nature of the signal. Large non-resonant contributions appear
to be required to describe both the intensity and the phase simultaneously. Essentially the same
information on the $1^{-+}$ exotic signal in the $\eta^{\prime}\pi^{-}$ system can be found in 
references~\cite{Nerling_2011b,Schluter_2012,Nerling_2012}.

\subsubsection{Summary of $\pi_{1}(1600)$ results}
Table~\ref{tab:p1_1600} summarizes the masses and widths found for the $\pi_{1}(1600)$
in the four decay modes and from the experiments which have seen a positive result.
While the $\eta^{\prime}\pi$, $f_{1}\pi$ and $b_{1}\pi$ decay modes appear to be robust in
the observation of a resonant $\pi_{1}(1600)$, there are concerns about the $3\pi$ final 
states. While we report these in the table, the results should be taken with some caution.

\begin{table}[h!]\centering
\begin{tabular}{cllcc} \hline\hline
Mode & Mass (GeV) & Width (GeV) & Experiment & Reference \\ 
\hline
$b_{1}\pi$             & $1.58\pm 0.03$ & $0.30\pm 0.03$ & VES & \cite{Dorofeev:99} \\
$b_{1}\pi$             & $1.61\pm 0.02$ & $0.290\pm 0.03$ & VES & \cite{Khokholov:00} \\
$b_{1}\pi$             & $\sim 1.6$ & $\sim 0.33$ & VES & \cite{Dorofeev:02} \\
$b_{1}\pi$             & $1.56\pm 0.06$ & $0.34\pm 0.06$ & VES & \cite{Amelin:2005ry} \\
$f_{1}\pi$             & $1.64\pm 0.03$ & $0.24\pm 0.06$ & VES & \cite{Amelin:2005ry} \\
$\eta^{\prime}\pi$  & $1.58\pm 0.03$ & $0.30\pm 0.03$ & VES & \cite{Dorofeev:99} \\
$\eta^{\prime}\pi$  & $1.61\pm 0.02$ & $0.290\pm 0.03$ & VES & \cite{Khokholov:00} \\
$\eta^{\prime}\pi$    & $1.56\pm 0.06$ & $0.34\pm 0.06$ & VES & \cite{Amelin:2005ry} \\
$\rho\pi$            & $1.593\pm 0.08$& $0.168\pm 0.020$ & E852 & \cite{Adams:1998ff} \\
$\eta^{\prime}\pi$ & $1.597\pm 0.010$ & $0.340\pm 0.040$ & E852 & \cite{Ivanov:2001rv} \\
$f_{1}\pi$              & $1.709\pm 0.024$ & $0.403\pm 0.080$ & E852 & \cite{Kuhn:2004en} \\
$b_{1}\pi$             & $1.664\pm 0.008$ & $0.185\pm 0.025$ & E852 & \cite{Lu:2004yn} \\
$b_{1}\pi$  & $\sim 1.6$ & $\sim 0.23$ & CBAR & \cite{Baker:2003jh} \\
$\rho\pi$ & $1.660\pm 0.010$ & $0.269\pm 0.021$ & COMPASS & \cite{Alekseev:2009xt} \\
$\eta^{\prime}\pi$ & $1.670\pm 0.030$ & $0.240\pm 0.050$ & CLEO-c & \cite{Adams:2011sq} \\
all              & $1.662^{+0.008}_{-0.009}$ & $0.241\pm 0.040$ & PDG & \cite{pdg12} \\
\hline\hline
\end{tabular}
\caption[]{\label{tab:p1_1600}Reported masses and widths of the $\pi_{1}(1600)$ along
with the 2014 PDG average.}
\end{table}

Models for hybrid decays predict rates for the decay of the $\pi_{1}$.
Equation~\ref{eq:decay_rates} gives the predictions from reference~\cite{Close:1995}.
\begin{eqnarray}
\nonumber
& \pi b_{1} : \pi f_{1} : \pi\rho : \eta\pi : \pi\eta^{\prime} &  \\ 
\nonumber
& = & \\
& 170 : 60 : 5-20 : 0-10 : 0-10 & 
\label{eq:decay_rates}
\end{eqnarray}
A second model from reference~\cite{PSS-1} predicted the following rates for a
$\pi_{1}(1600)$.
\begin{center}
\begin{tabular}{lccccc}
      & $\pi b_{1}$ &  $\rho\pi$ & $\pi f_{1}$ &  $\eta(1295)\pi$ & $K^{*}K$  \\ 
PSS & $24$ & $9$ & $5$ & $2$ & $0.8$ \\ 
IKP  & $59$ & $8$ & $14$ & $1$ & $0.4$ 
\end{tabular}
\end{center}
These can be compared to the results from VES in equation~\ref{eq:pi11600_rates3},
which are in moderate agreement. The real identification of the $\pi_{1}(1600)$ as a
hybrid will almost certainly involve the identification of other members of the nonet:
the $\eta_{1}$ and/or the $\eta^{\prime}_{1}$, both of which are expected to have widths
that are similar to the $\pi_{1}$. For the case of the $\eta_{1}$, the most promising 
decay mode may be the $f_{1}\eta$ as it involves reasonably narrow daughters.

We believe that the current data support the existence of a resonant $\pi_{1}(1600)$ which
decays into $b_{1}\pi$, $f_{1}\pi$ and $\eta^{\prime}\pi$, however, confirmation of the $b_{1}\pi$
and $f_{1}\pi$ modes by COMPASS would be useful. For the $\rho\pi$ decay, things are somewhat
uncertain. As noted earlier, the phase motion results observed by both E852 and E852-IU are can 
be interpreted as either the $\pi_{2}(1670)$ absorbing the $\pi_{1}(1600)$, or leakage from the 
$\pi_{2}(1670)$ generating a spurious signal in the $1^{-+}$ channel. While COMPASS does not
yet have final results, they seem to confirm that the exotic partial wave near 1600 MeV does couple to the $\rho\pi$ decay mode and they exclude a narrow-resonance interpretation
of the $\pi_{1}(1600)$. The careful follow-on studies from COMPASS to more broadly explore 
the model space and production mechanisms have started to yield very interesting new 
results and we look forward to new information in the near future.
\subsection{\label{sec:pi1_2015}The $\pi_{1}(2015)$}
\subsubsection{E852 Results on the $\pi_{1}(2015)$}
\begin{figure}[h!]\centering
\includegraphics[width=0.35\textwidth,angle=270]{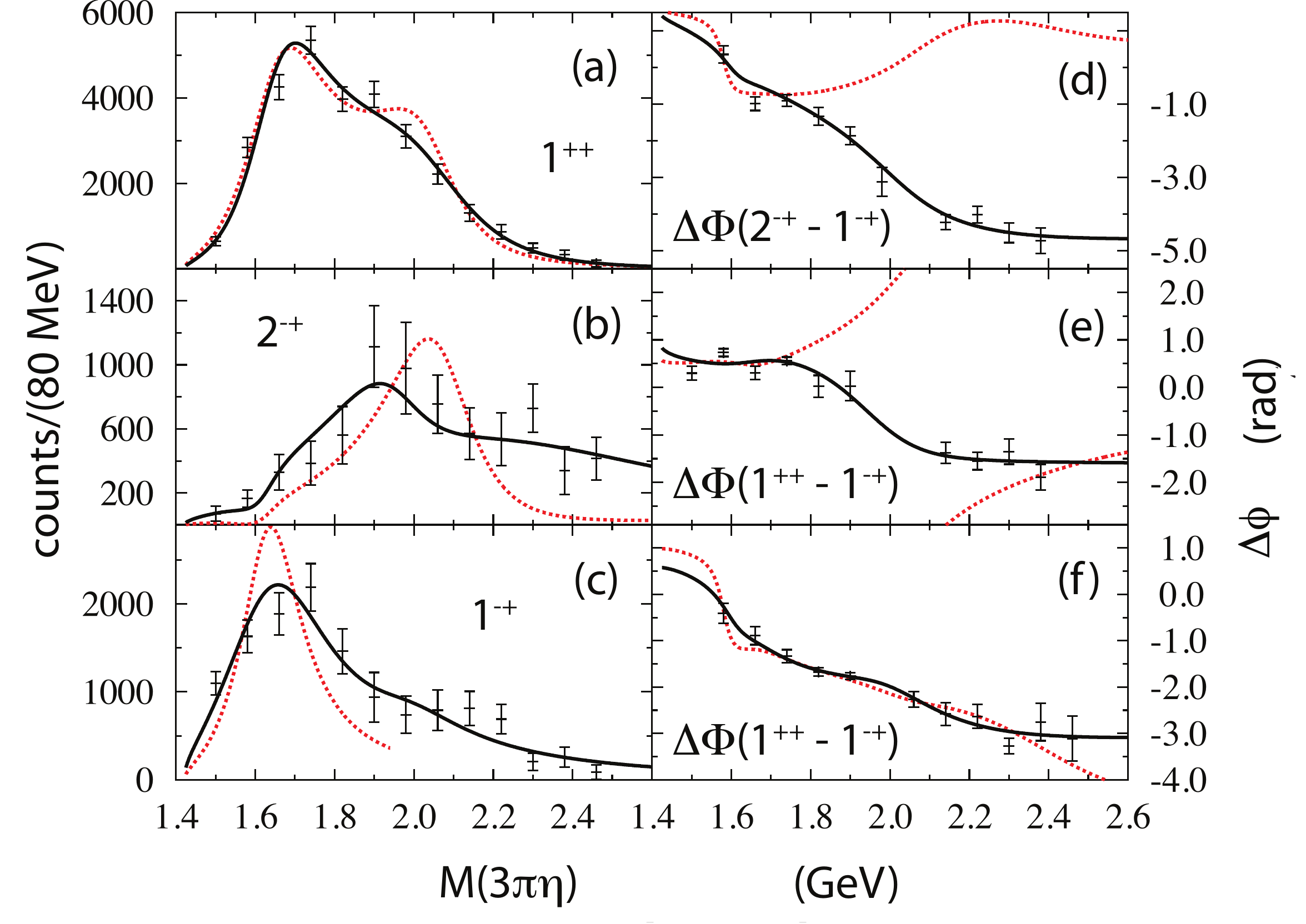}
\caption[]{\label{fig:e852f1pi} (Color on line.) The $f_{1}\pi$ invariant mass from E852~\cite{Kuhn:2004en}.
(a) The $1^{++}$ partial wave ($a_{1}(1270)$), (b) the $2^{-+}$ partial wave ($\pi_{2}(1670)$) and
(c) the exotic $1^{-+}$ partial wave. The dotted (red) curves show the fits of Breit-Wigner
distributions to the partial waves. (d) shows the phase difference between the $2^{-+}$ and 
$1^{-+}$ partial waves, while (e) shows the difference between the $1^{++}$ and $1^{-+}$
partial waves. The dotted (red) curves show the results for a single $\pi_{1}$ state, the $\pi_{1}(1600)$.
(f) shows the same phase difference as in (d), but the dotted (red) curve shows a fit with two
poles in the $1^{-+}$ partial wave, the $\pi_{1}(1600)$ and the $\pi_{1}(2015)$.
(Figure reproduced with permission from reference~\cite{Kuhn:2004en}.)}
\end{figure}

The E852 experiment has also reported a third $\pi_{1}$ state seen decaying to both $f_{1}\pi$
~\cite{Kuhn:2004en} and to $b_{1}\pi$~\cite{Lu:2004yn}. In the $f_{1}\pi$ final state, the
$\pi_{1}(2015)$ is produced with $M^{\epsilon}=1^{+}$ in conjunction with the $\pi_{1}(1600)$.
The description of the $1^{-+}$ partial wave requires two poles. They report a  mass of
$2.001\pm 0.030\pm 0.092$~GeV and a width of $0.333\pm 0.052\pm 0.049$~GeV.
Fig.~\ref{fig:e852f1pi} shows the E852 data from this final state. Parts $e$ and $f$ of
this show the need for the two-pole solution.
VES also examined the $f_{1}\pi$ final state, and their intensity of the $1^{-+}$ partial
wave above $1.9$~GeV (see Fig.~\ref{fig:ves_f1pi}) is not inconsistent with that of 
E852~\cite{Amelin:2005ry}. However, VES made no comment on this, nor have they 
claimed the existence of the $\pi_{1}(2015)$.

\begin{figure}[h!]\centering
\includegraphics[width=0.5\textwidth]{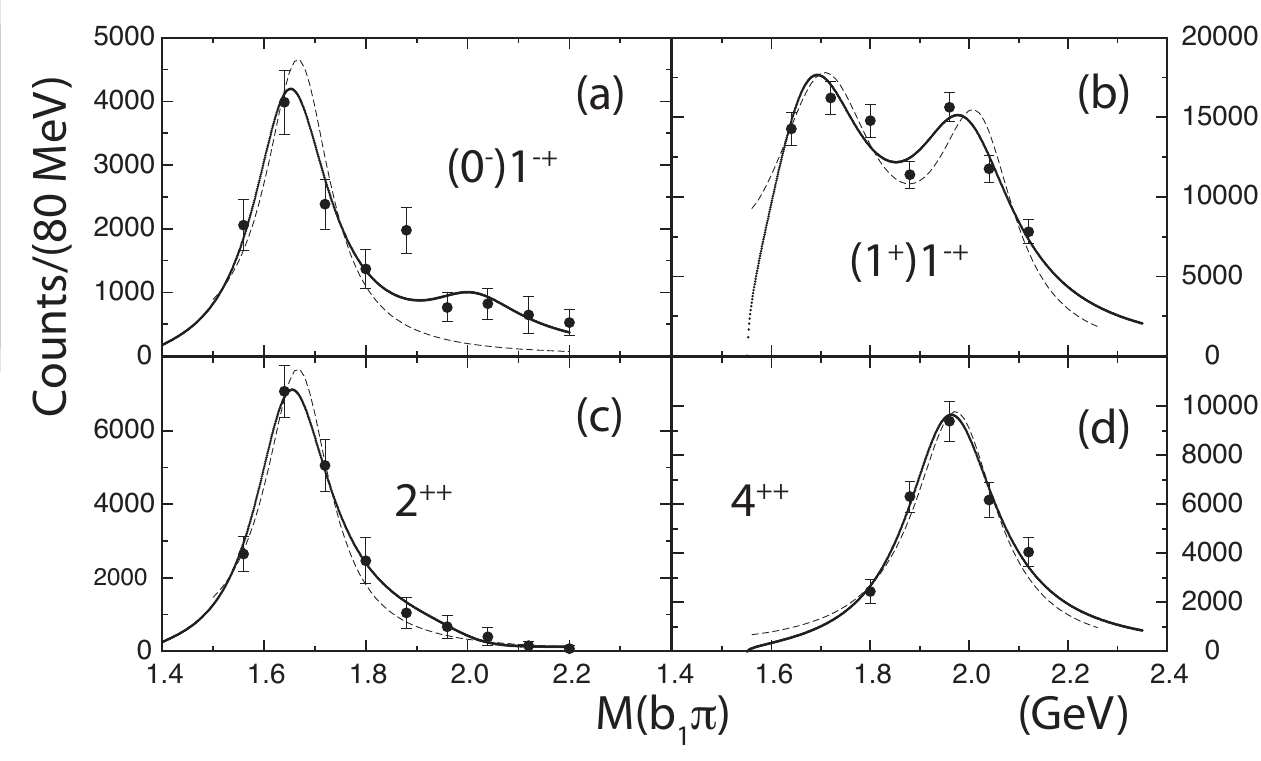}
\caption[]{\label{fig:e852b1pi}
The $b_{1}\pi$ invariant mass from the E852 experiment. (a) shows the $1^{-+}$ $b_{1}\pi$ partial
 wave produced in natural parity exchange ($M^{\epsilon}=1^{+}$) while (b) shows the $1^{-+}$ 
$b_{1}\pi$ partial wave produced in unnatural parity exchange  ($M^{\epsilon}=0^{-}$). In (c) 
is shown the $2^{++}$ $\omega\rho$ partial wave, while (d) shows the $4^{++}$ $\omega\rho$
partial wave. The curves are fits to the $\pi_{1}(1600)$ and $\pi_{1}(2015)$ (a and b), the $a_{2}(1700)$
in (c) and the $a_{4}(2040)$ in (d). (Figure reproduced with permission from reference~\cite{Lu:2004yn}.)}
\end{figure}
In the $b_{1}\pi$ final state, the $\pi_{1}(2015)$ is produced dominantly through natural parity
exchange ($M^{\epsilon}=1^{+}$) while the $\pi_{1}(1600)$ was reported in both natural and 
unnatural parity exchange, where the unnatural exchange dominated. They observe a mass 
of $2.014\pm 0.020\pm 0.016$~GeV  and a width of $0.230\pm 0.032\pm 0.073$~GeV,
which are consistent with that observed in the $f_{1}\pi$ final state. Fig.~\ref{fig:e852b1pi}
shows the intensity distributions for several partial waves in this final states. The need for
two states is most clearly seen in panel $(b)$. VES also looked at the $b_{1}\pi$ final state, but did 
not observe $1^{-+}$ intensity above $1.9$~GeV~\cite{Amelin:2005ry}. However, the intensity
shown in Fig~\ref{fig:ves_b1pi} may be consistent with that observed by E852.
The reported masses and widths are summarized in Table~\ref{tab:p1_2000}. We note that this 
state does not appear in the summary tables of the PDG~\cite{pdg12}. 
\begin{table}[h!]\centering
\begin{tabular}{crrcc} \hline\hline
Mode & Mass (GeV) & Width (GeV) & Experiment & Reference \\ 
\hline
$f_{1}\pi$ & $2.001\pm 0.030$ & $0.333\pm 0.052$ & E852 & \cite{Kuhn:2004en} \\
$b_{1}\pi$ & $2.014\pm 0.020$ & $0.230\pm 0.032$ & E852 & \cite{Lu:2004yn} \\
\hline\hline
\end{tabular}
\caption[]{\label{tab:p1_2000} Reported masses and widths of the $\pi_{1}(2015)$ as observed
in the E852 experiment. The PDG does not report an average for this state.}
\end{table}
\subsubsection{Interpretation of the $\pi_{1}(2015)$}
With so little experimental evidence for this high-mass state, it is difficult to say much. We note
that the observed decays, $f_{1}\pi$ and $b_{1}\pi$ are those expected for a hybrid meson.
We also note that the production of this state is consistent (natural parity exchange) for both
of the observed final states. In the case that the $\pi_{1}(1600)$ is associated with the lowest-mass
hybrid state, one possible interpretation of the $\pi_{1}(2015)$ would be an excited state (as suggested
by recent LQCD calculations~\cite{Dudek:2010wm}). The mass
splitting is typical of radial excitations observed in the normal mesons. In the case that the  
$\pi_{1}(1600)$ is identified as something else, the $\pi_{1}(2015)$ would be a prime candidate 
for the lightest mass hybrid.

\section{Outlook}
Hybrid mesons should be observed in nonets, and lattice QCD has recently made quite clear
predictions that several of these nonets should have exotic quantum numbers. The lightest
supermultiplet of hybrids, $0^{-+},1^{-+}, 1^{--}, 2^{-+}$, contains one exotic quantum number nonet
($1^{-+}$). Excitations of these should contain another $1^{-+}$ nonet, in addition to a single spin-0
($0^{+-}$) and two spin-2 ($2^{+-}$) nonets. Experimentally, we have seen evidence for up to 
three isospin-1, spin-1 exotic states: $\pi_{1}(1400)$, $\pi_{1}(1600)$ and $\pi_{1}(2105)$. As we
noted it seems unlikely that the lightest of these is a hybrid, while the heavier two states could
map onto the isospin-1 members of the two $1^{-+}$ nonets predicted by lattice QCD. Unfortunately,
lacking experimental evidence for other members of these nonets, it is difficult to draw
solid conclusions. In particular, observation of one or both isospin-0 members of either nonet
is needed. Beyond the spin-1 exotic hybrids, observation of states with quantum numbers 
$2^{+-}$ and or $0^{+-}$ is needed. 

As we move forward over the next several years, new experimental efforts will join COMPASS
in searching for these states. At Jefferson Lab, the GlueX experiment expects to take first
physics data in late 2015. GlueX will use linearly polarized photons incident on a hydrogen target to produce these states. In the next five years or so, PANDA at FAIR will join these efforts with antiproton beams --  hopefully dramatically new information will be gleaned from these experimental efforts.

Keeping pace with experiment will require a number improvements in theory. It is hoped, for example, that lattice gauge theory will be able to provide predictions for light and charmonium hybrid masses on unquenched lattices with physical pion masses and multihadron operators relatively soon. Conclusive and comprehensive computations of strong and electromagnetic transitions of light and charmonium hybrids would also be most welcome.

A new breed of models that is capable of reproducing central lattice results is also required. Ideally these models will reproduce the gluonic adiabatic potentials and the spectrum of heavy and light hybrids reasonably well. Presumably this will require a formalism that captures short range and long range dynamics in an approximate fashion without double counting or other conceptual issues. Perhaps a promising approach would be a model based on Coulomb gauge QCD with a variable number of constituent gluons and many-body interactions. Such a model should also be able to describe strong and electromagnetic decays reasonably accurately. These are demanding criteria, but new experimental and lattice data should provide many valuable clues, which will hopefully lead to a quantitative and qualitative understanding of this enigmatic sector of the Standard Model.

\section*{Acknowledgments}
The authors would like to thank many people for useful discussions, especially  
J.~J.~Dudek and C.~J.~Morningstar. This work was supported in part by the U.S. Department 
of Energy under grant No. DE-FG02-87ER40315.

\end{document}